\documentclass[aip,jcp,reprint,superscriptaddress,footinbib,floatfix]{revtex4-2}

\usepackage{graphicx}
\usepackage{bm}
\usepackage{amssymb}
\usepackage{amsbsy}
\usepackage{amsmath}
\usepackage{mathtools}
\usepackage{xcolor}
\usepackage{stmaryrd}
\usepackage{mathrsfs}
\usepackage{tikz}
\usetikzlibrary{shapes}
\usepackage{cancel}
\usepackage[normalem]{ulem}
\usepackage{hyperref}
\usepackage{multirow}

\usepackage{titlesec}
\titlespacing\section{0pt}{12pt plus 4pt minus 2pt}{0pt plus 2pt minus 2pt}
\titlespacing\subsection{0pt}{12pt plus 4pt minus 2pt}{0pt plus 2pt minus 2pt}
\titlespacing\subsubsection{0pt}{12pt plus 4pt minus 2pt}{0pt plus 2pt minus 2pt}

\newcommand{\fref}[1]{Fig.~\ref{#1}}
\newcommand{\frefs}[1]{Figs.~\ref{#1}}
\newcommand{\tref}[1]{Table~\ref{#1}}
\newcommand{\eref}[1]{Eq.~(\ref{#1})}
\newcommand{\sref}[1]{Sec.~\ref{#1}}

\makeatletter
\let\cat@comma@active\@empty
\makeatother

\makeatletter 
\gdef\@ptsize{2}
\let\@currsize\normalsize 
\makeatother 

\begin{document}
\title{A mesoscopic model for the rheology of dilute and semidilute solutions of wormlike micelles}
\author{Avishek Kumar}
\affiliation{IITB-Monash Research Academy, Mumbai, Maharashtra 400076, India}
\affiliation{Department of Chemical Engineering, Indian Institute of Technology Bombay, Mumbai, Maharashtra 400076, India}
\affiliation{Department of Chemical and Biological Engineering, Monash University, Melbourne, VIC 3800, Australia}
\author{Rico F. Tabor}
\affiliation{School of Chemistry, Monash University, Melbourne, VIC 3800, Australia}
\author{P.Sunthar}
\affiliation{Department of Chemical Engineering, Indian Institute of Technology Bombay, Mumbai, Maharashtra 400076, India}
\author{J. Ravi Prakash}
\email{ravi.jagadeeshan@monash.edu}
\affiliation{Department of Chemical and Biological Engineering, Monash University, Melbourne, VIC 3800, Australia}
\homepage{https://ravijagadeeshan.ac}

\begin{abstract}
The concept of a `persistent worm' is introduced, representing the smallest possible length of a wormlike micelle, and modelled by a bead-spring chain with sticky beads at the ends. Persistent worms are allowed to combine with each other at their sticky ends to form wormlike micelles with a distribution of lengths, and the semiflexibility of a wormlike micelle is captured with a bending potential between springs, both within and across persistent worms that stick to each other. Multi-particle Brownian dynamics simulations of such polydisperse and `polyflexible' wormlike micelles, with hydrodynamic interactions included and coupled with reversible scission/fusion of persistent worms, are used to investigate the static and dynamic properties of wormlike micellar solutions in the dilute and unentangled semidilute concentration regimes.  The influence of the sticker energy and persistent worm concentration are examined and simulations are shown to validate theoretical mean-field predictions of the universal scaling with concentration of the chain length distribution of linear wormlike micelles, independent of the sticker energy. The presence of wormlike micelles that form rings is shown not to affect the static properties of linear wormlike micelles, and mean-field predictions of ring length distributions are validated. Linear viscoelastic storage and loss moduli are computed and the unique features in the intermediate frequency regime compared to those of homopolymer solutions are highlighted. The inclusion of hydrodynamic interactions enables the distinction between Rouse and Zimm dynamics in wormlike micelle solutions to be elucidated. While the neglect of  hydrodynamic interactions is justified in the case of entangled wormlike micelle solutions, here the scaled concentration $c/c^*$ at which the onset of the screening of hydrodynamic interactions occurs with increasing concentration, as a function of the scission energy, is clearly identified.
\end{abstract}

\maketitle

\section{\label{sec:intro} Introduction}

Beyond a critical micelle concentration, surfactant molecules in solution self-assemble into a variety of aggregate structures, ranging from spherical, wormlike and rodlike micelles, to vesicles and bilayers depending on the molecular geometry, concentration, temperature, charge and salt concentration~\cite{Cates2006,Lerouge2010,Israel2011}. Amongst these manifold structures, attention is focussed here on the behaviour of dilute and semidilute solutions of unentangled wormlike micelles. While wormlike micelles exhibit characteristics similar to those of semiflexible homopolymers, their continuous breakage and reformation has led to their being described as `living' or `equilibrium' polymers~\cite{Dreiss2007,Dreiss2013}. This dynamic aspect of their instantaneous length leads to intricate viscoelastic properties and complex flow behaviour, including both shear-thickening and shear-thinning, the formation of banded structures in shear flow, and the occurrence of various flow instabilities~\cite{Pine1996,Pine1998,Berret1999,Berret2006,Rothstein2007,Dhont2008,Shibaev2015}. Changes in the microstructure of wormlike micelles and their mutual interactions significantly impact the behaviour of their solutions. This has made it possible to finely modulate their rheology through careful solution preparation and has contributed to their wide-ranging use in a number of different industrial applications~\cite{Ezrahi2007,Nicolas2007,Sullivan2007,Zakin2007,Raghavan2017}. The design of efficient and targeted formulations of wormlike micelle solutions clearly requires a detailed understanding of how wormlike micelle structure and dynamics dictate the bulk rheological behaviour of their solutions. In this paper, a mesoscopic model for wormlike micelle solutions is introduced that enables some insight to be drawn into the connection between dynamics on the molecular scale and the observed behaviour on the macroscopic scale.

Following the seminal work of Cates~\cite{Cates1987}, a number of analytical theories based on mean-field, scaling and kinetic theory arguments have been developed to describe the static properties of wormlike micelle solutions~\cite{Cates1990,Kroger1998,Wittmer1998,Kroger2004,Huang2006}, and predictions of various quantities such as the distribution of wormlike micelle lengths and the dependence of the mean length on concentration and scission energy have been made. These predictions have been extensively validated by simulations using a variety of different methods, such dynamic Monte Carlo based on the bond fluctuation model~\cite{Rouault1995,Wittmer1998}, non-equilibrium molecular dynamics~\cite{Kroger1995,Kroger1998,Kroger2004,PaddingPRE2004} and Brownian dynamics~\cite{Rouault1999,Huang2006,Ryckaert2006}. While early theories did not account for the possibility of ring formation through the fusion of the ends of linear wormlike micelles~\cite{Cates1990,Wittmer1998}, subsequent developments explicitly accounted for the presence of rings and predictions for quantities such as the distribution of the lengths of rings have been made, which have, in turn, been validated by numerical simulations~\cite{Kroger1998,Milchev2000,Wittmer2000,PaddingPRE2004}. The results of these early theories provide an excellent benchmark with which to test the validity of any new mesoscopic model, and it is shown here that the model introduced in this work is able to reproduce with great fidelity the analytical predictions of a variety of static properties.

Viscoelastic properties of wormlike micellar solutions that are most relevant to industrial applications are typically observed in  the concentrated entangled regime. A majority of experimental and theoretical studies of the flow behaviour of wormlike solutions are consequently focussed on this regime. Constitutive models that are commonly used to describe concentrated entangled wormlike micellar solutions,  and that can be integrated into computational fluid dynamics (CFD) simulations, are not yet able to predict many of the features observed in flows that involve both shear and extensional components~\cite{Shen2022}. Nevertheless, in recent years significant progress has been made in the development of nearly quantitatively accurate models for their linear and nonlinear rheological behaviour in rheometric flows~\cite{Zou2019,Peterson2021,Sato2022,Peterson2023,Hudson2024a,Hudson2024b}. Since the subject of this work is the rheological behaviour of wormlike micelle solutions in the dilute and unentangled semidilute regimes, the concentrated regime is not discussed here further. As is appropriate, however, a brief summary of prior work relevant to the goals of the current work, in the dilute and semidilute regimes (and of studies that are agnostic to concentration regime), is given below.

Attempts to understand, at the continuum level, the strikingly different behaviour exhibited by wormlike micellar solutions in shear flow compared to the behaviour of polymer solutions, such as the complex dependence of viscosity on shear rate, and the occurrence of macroscopic bands of different viscosities, have been based on analysing the nature of constitutive equations for wormlike micelle solutions~\cite{Bautista1999,Boek2005,olmsted2008,Dhont2008,Landazuri2016,Yamada2020} and on stability analyses~\cite{Fielding2007,FieldingPRE2007,Morozov2012,Moorcroft2014,FieldingPRL2018,Graham2024}. Significant progress in capturing the behaviour of dilute wormlike micelle solutions in a variety of flow fields has been made recently with the help of refined models that combine a constitutive equation for the fluid with a kinetic equation for the destruction and creation of structure~\cite{Manero2018,Yamada2020,Manero2022}. These approaches are very convenient for integration with exisiting CFD simulators that have been designed for use with similar viscoelastic constitutive equations. However, the procedure for determining model parameters by fitting experimental measurements and the physical connection of the parameters to molecular aspects of wormlike micelle structure and dynamics is not clear. Additionally, the shear stress predicted by these models is not a multi-valued function of shear rate over any range of shear rates and as a consequence, they are not expected to predict the existence of vorticity banding, for which a reentrant flow curve is a necessary condition~\cite{Shinnar1970,olmsted2008,FieldingPRL2018,Graham2024}.

Between the two extremes of continuum models and coarse-grained mesoscopic models for wormlike micelle solutions lie phenomenological models that are `microstructurally' inspired, and which aim to predict both the shapes of flow curves and the existence of banded structures in shear flow~\cite{Turner1992,Vasquez2007,Zhou2014,Adams2018,Graham2021,Graham2024}. In contrast to vorticity banding, which is typically observed in shear flows of dilute wormlike micellar solutions and requires the existence of a reentrant flow curve, shear banding is observed in semidilute entangled and concentrated solutions, and requires the shear rate to be a multi-valued function of shear stress over a range of shear stresses~\cite{olmsted2008,Dhont2008}. 

The non-monotonic dependence of shear stress on shear rate is captured by phenomenological models for semidilute entangled and concentrated solutions, such as the VCM model~\cite{Vasquez2007,Zhou2014} (where entanglement effects are treated implicitly through an empirically tuneable parameter) and Brownian dynamics simulations~\cite{Cook2020} of refined versions of these models~\cite{Adams2018}. These studies, which are built on the representation of micelles by two species of dumbbells, short and long, where two short dumbbells can combine to form a long one and the long dumbbell can break into two short ones, suggest that the high shear rate band consists of the short micelle species while the low shear rate band consists of the longer micelle species. This is consistent with experimental observations of differences in birefringence in the two banded regions in Couette flow that are at different shear rates but at the same shear stress~\cite{decruppe1995}. Though examination of the semidilute entangled and concentrated regimes is outside the scope of this work, it is worth pointing out that this insight into the molecular composition of the shear bands obtained with models that are highly coarse-grained, is already quite valuable, and suggests that much greater understanding could be gained by the development of more fine-grained models which are capable of capturing  wormlike micellar structure and dynamics on the molecular scale more accurately and that can predict the occurrence of banding. Such models are, however, currently lacking.

 In the case of dilute wormlike micellar solutions, the reactive rod model was originally introduced by Dutta and Graham~\cite{Dutta2018} and further refined to better capture experimental results in two recent groundbreaking papers by Hommel and Graham~\cite{Graham2021,Graham2024}. This model belongs to the class of phenomenological models that are strongly motivated by physics on the microscopic scale, and which is capable of capturing several of the key experimentally observed rheological signatures in these solutions, such as shear thickening and thinning in simple shear flow~\cite{Graham2021}, and vorticity banding and the presence of finger-like instabilities in circular Couette flow~\cite{Graham2024}. In this model, wormlike micelles are represented as rods that can combine together by fusion into longer rods, which in turn can breakdown into shorter rods by scission. Both the fusion and scission can happen spontaneously or be induced by flow. By solving coupled evolution equations for the average orientation of rods and the collective length of the micelles, the model computes the micelle contribution to solution stress, which is then combined with the equations for the conservation of mass and momentum to solve for macroscopic observables in any geometry. These papers represent a significant advance in the description of dilute wormlike micellar solutions and the reactive rod model will undoubtedly provide greater insight into experimental observations in other flow fields. Nevertheless, there are some aspects of wormlike micelle structure and dynamics that the model does not take into account. 

By modeling wormlike micelles as rods and by using a single representative length and orientation, the reactive rod model does not take into consideration the semiflexibility of micelles and their distribution of lengths and orientations. One can anticipate that micelles of different lengths and flexibility will align differently in flow. By neglecting all forms of micelle-micelle interactions, such as excluded volume interactions, the reactive rod model is strictly valid only in the dilute regime. Additionally, it is well known for dilute and semidilute polymer solutions that it is vitally important to account for hydrodynamic interactions in order to accurately predict dynamic properties~\cite{Bird1987,Rubinstein2003}. While mesoscopic models such as the model introduced in the present work cannot be easily integrated into CFD solvers (an aspect which is a significant advantage of models such as the reactive rod model), it is important to recognise the extent of inaccuracy that might be present in predictions by the reactive rod model through the neglect of some of the key aspects of wormlike micelle structure and dynamics that have been identified above.

Given the vast range of length and time scales involved, carrying out simulations at the level of individual surfactant molecules in order to model the rheological behaviour of wormlike micelle solutions is still not computationally feasible, and scales that are relevant to rheological properties can only be probed meaningfully through coarse-grained simulations at the mesoscopic level~\cite{Padding2005,Padding2008}. There have been many mesoscopic models developed to describe the  rheological behaviour of wormlike micellar solutions based on a variety of different numerical methods ranging from non-equilibrium molecular dynamics~\cite{Kroger1995,Kroger1997,Kroger2004,PaddingPRE2004} to Brownian dynamics simulations~\cite{Kroger1998,Padding2005,Ryckaert2006,Padding2008,Adams2018,Cook2020}, and predictions of a number of linear and nonlinear rheological properties in shear flow have been made. All these approaches have as their key component  the fusion of individual `indivisible' units that represent the shortest possible length of a wormlike micelle into longer linear or ring-like wormlike micelles, which subsequently breakup by scission, possibly all the way back down to the individual indivisible units. These indivisible units are typically beads~\cite{Kroger1995,Kroger1997,Kroger1998,PaddingPRE2004,Ryckaert2006}, dumbbells~\cite{Adams2018,Cook2020} or rods~\cite{Padding2005,Padding2008} and a range of different potentials and algorithms have been used to determine the processes of fusion and scission. In nearly all these cases, the size of the single indivisible unit is equal to one persistence length of the wormlike micelle. The only exceptions are the FENE-CB model of Kr\"oger~\cite{Kroger1998,Kroger2004} and rod model of Padding et.~al.~\cite{Padding2005,Padding2008}, where a bending potential is used between beads in the former case and rods in the latter, to account for semiflexibility. In neither of these cases, however, has the effect of the bending rigidity of wormlike micelles on their rheological behaviour been studied systematically. Models that assume that the length of the indivisible unit is equal to a persistence length do not account for semiflexibility since the wormlike micelles formed by their fusion are intrinsically flexible in nature. 

The single indivisible unit of a wormlike micelle is clearly determined by entropic and enthalpic considerations, and would depend on the chemistry and geometry of the surfactant molecules, the chemistry of the solvent molecules and ions present in solution, and on their mutual interactions.~\cite{Israel2011,Nagarajan2007} It seems reasonable to argue that the resultant shortest possible length need not necessarily be equal to exactly one persistent length. In contrast to the previous work, we introduce the concept of a `persistent worm', which is the shortest possible length of a wormlike micelle, that can in principle be any fraction of the actual persistent length of the wormlike micelle. The persistent worm is modelled as a bead-spring chain with an arbitrary number of beads, and the terminal beads are made `sticky', i.e., they can associate with other sticky beads to form pairs of stuck beads and thus form a wormlike micelle. A bending potential, with which the persistent length can be tuned, is imposed between the springs within a persistent worm, and when two persistent worms combine, the same bending potential is imposed across the two persistent worms in order to maintain a uniform degree of bending stiffness along the entire backbone of a wormlike micelle formed by the fusion of persistent worms. Since the degree of flexibility of a polymer chain is determined by the ratio of the persistent length to the contour length, this implies that the current model leads to a polydisperse and `polyflexible' solution of wormlike micelles.

Additionally, nearly all existing mesoscopic models neglect the presence of hydrodynamic interactions. Early non-equilibrium molecular dynamics simulations did account for the presence of hydrodynamic interactions by including solvent molecules explicitly~\cite{Kroger1995,Kroger1997}. However, in order to examine the influence of hydrodynamic interactions, it is necessary to compare results of simulations with and without their presence, which has not been done so far. While neglecting hydrodynamic interactions is well justified for solutions that are sufficiently concentrated such that they are screened~\cite{Rubinstein2003}, there has not been a meticulous investigation so far to determine the critical concentration beyond which hydrodynamic interactions may be neglected. In the context of the accurate prediction of the rheological behaviour of dilute and semidilute homopolymer solutions, it is now well established that it is vitally  important to include hydrodynamic interactions~\cite{Prakash2019}. 

In the present work, a mesoscopic model is introduced that is valid for finite concentrations, which can account for both semiflexibility and hydrodynamic interactions, and as a consequence, distinguish between Rouse and Zimm dynamics in wormlike micelle solutions and enable the determination of the critical concentration beyond which hydrodynamic interactions are screened. Essentially, hydrodynamic interactions have been included within the framework of HOOMD-Blue, using an efficient algorithm for their fast computation~\cite{PSE2017}.

With this mesoscopic model, Brownian dynamics simulation predictions of equilibrium properties and the linear viscoelasticity of wormlike micelle solutions, across a range of concentrations from the dilute to the unentangled semidilute regime, are made as a function of model parameters. The model is validated by comparison of static property predictions with the classical results of mean-field and scaling theories, and the influence of hydrodynamic interactions on storage and loss moduli is explored for the first time.

In~\sref{sec:pw} and~\sref{sec:wlm}, the persistent worm mesoscopic model and wormlike micelles formed by their assembly are introduced, along with the concomitant notation and defining equations. This is followed by the description of the governing equations for the simulation algorithm and the various properties computed in this work in~\sref{sec:MCBD} and~\sref{sec:propertydefs}, respectively. The validity of the procedure used to maintain a uniform semiflexibility along a wormlike micelle's backbone is demonstrated in~\sref{sec:valid}, and the overlap concentration for a wormlike micellar solution is determined in~\sref{sec:cstar}. Static property predictions of the present model and a  comparison with earlier mean-field theory results is carried out in \sref{sec:staticprop}, with the length distributions and mean lengths  of linear wormlike micelles discussed in~\sref{sec:lengths}, and the length distributions and the fraction of rings discussed in~\sref{sec:rings}. Two different definitions of the radius of gyration are discussed in~\sref{sec:radius_of_gyration}, while the influence of finite size effects on the prediction of static properties is discussed in~\sref{sec:Fse}. \sref{sec:gprime} presents the new results of this work on the storage and loss moduli $G^{\prime}$ and $G^{\prime\prime}$ for wormlike micelle solutions. Material functions obtained using the Green-Kubo formulation  are compared with those obtained with a direct simulation of small amplitude shear flow in~\sref{sec:gprimevalid}. In~\sref{sec:homopolymer}, the moduli for linear wormlike micellar solutions are compared with those for a monodisperse homopolymer solution, with chains having the same length as the mean length of wormlike micelles, and at the same scaled concentration. In~\sref{sec:HI}, the effect of hydrodynamic interactions on the predicted shapes of $G^{\prime}$ and $G^{\prime\prime}$ is examined in the dilute and semidilute concentration regimes. A summary of the key conclusions of this work is provided in~\sref{sec:conclusions}.

\section{\label{sec:goveqn} The mesoscopic model and governing equations}

\subsection{\label{sec:pw} The persistent worm as an indivisible unit}

\begin{figure}[b]
 \begin{center}
\resizebox{8.65cm}{!}{\includegraphics*[width=12.5cm]{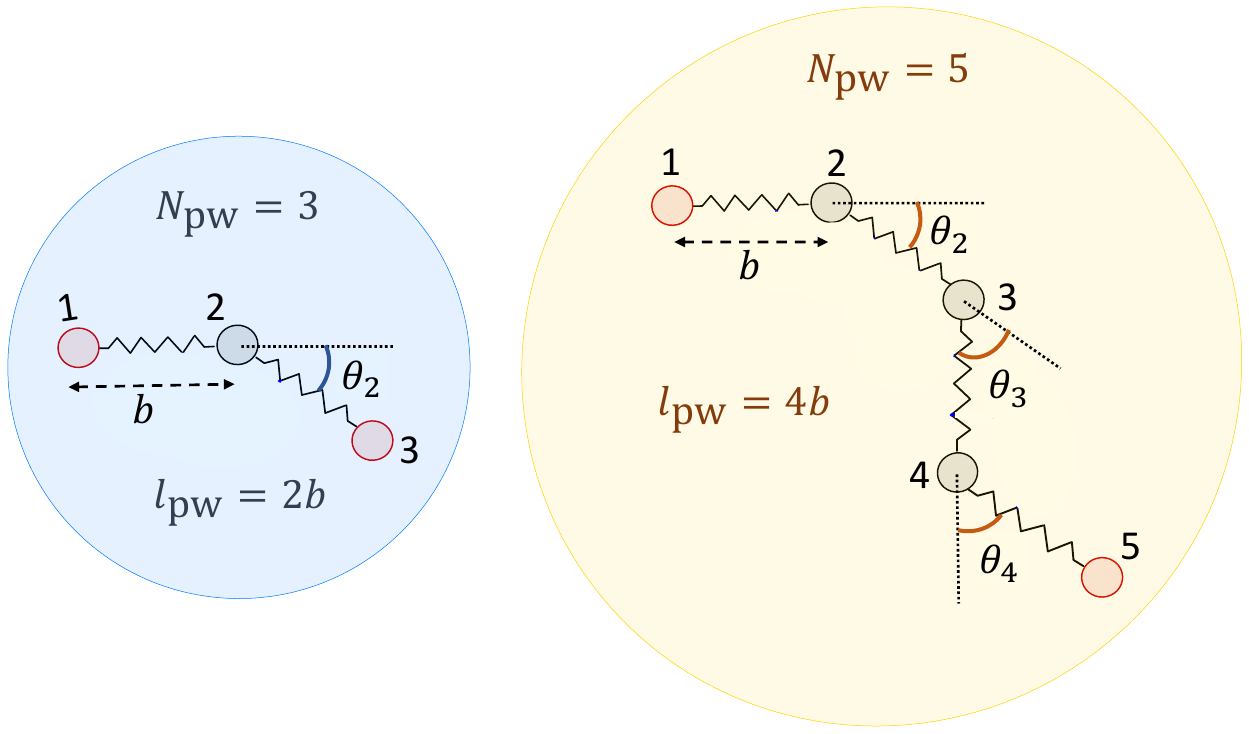}}  
\end{center}
\vspace{-10pt}
\caption{Schematic diagrams of two persistent worms represented by bead-spring chains with $N_{\text{pw}} =3$ and $N_{\text{pw}} =5$ beads, respectively. The natural length of the Fraenkel spring is denoted by $b$ and $\ell_{\text{pw}}$ is the natural length of a persistent worm. The angle $\theta$ between successive connector vectors is used in the calculation of the bending energy of the chain. The beads at the ends of a chain are `sticky' and can associate with other sticky beads to form pairs of stuck beads, and consequently, wormlike micelles. }
\label{fig:pw}
\end{figure}

A persistent worm represents the shortest possible length of wormlike micelle and is modelled by a bead-spring chain with $N_{\text{pw}}$ beads, as shown schematically in \fref{fig:pw} for two sample cases, with $N_{\text{pw}} = 3$ and 5 beads, respectively. While the spring force law that governs the spring can be chosen arbitrarily, the Fraenkel force law is used here,
\begin{equation}
    \bm{F}^{(\text{s})} \left(\bm{r}_{\mu+1} - \bm{r}_{\mu}\right) = H \left[1 - \frac{b}{\left\vert\bm{r}_{\mu+1} - \bm{r}_{\mu}\right\vert}\right] \left(\bm{r}_{\mu+1} - \bm{r}_{\mu}\right)
\label{eq:Fraenkel}
\end{equation}
where $\bm{r}_{\mu}$ is the position of bead $\mu$ with respect to an arbitrarily chosen origin, and the spring force acts along the spring connector vector $ \left(\bm{r}_{\mu+1} - \bm{r}_{\mu}\right)$ connecting beads $\mu$ and $\mu +1$, with magnitude $\left\vert\bm{r}_{\mu+1} - \bm{r}_{\mu}\right\vert$, $H$ is the spring constant, and $b$ is the natural length of the spring. With an appropriate choice of the magnitude of the spring constant $H$, the Fraenkel spring has been shown to mimic a rigid rod of length $b$~\cite{Indranil2022}. The natural length $\ell_{\text{pw}}$ of a persistent worm is clearly, $\ell_{\text{pw}} = (N_{\text{pw}} - 1) b$, since there are $(N_{\text{pw}} - 1)$ springs in a persistent worm.

In order to model the semiflexibility of wormlike micelles, which is a measure of the energetic resistance to bending along the backbone~\cite{yamakawa2016}, both intra- and inter-persistent-worm bending potentials are used. In the case of the intra-persistent-worm bending potential, a bending cost is imposed based on the angle $\theta_{\mu}$,
\begin{equation}
\label{eq:bending potential}
    \frac{U^{\text{(b)}}_{\mu}}{k_{\text{B}} T} = C \left(1-\cos{\theta_\mu} \right)
\end{equation}
where $k_{\text{B}}$ is Boltzmann's constant, $T$ is the absolute temperature, $C$ is the bending stiffness and $\theta_\mu$ is the included angle between adjacent springs in the persistent worm, represented by the vectors $ \left(\bm{r}_{\mu} - \bm{r}_{\mu-1}\right)$ and $ \left(\bm{r}_{\mu+1} - \bm{r}_{\mu}\right)$, respectively. Details of the inter-persistent-worm bending potential, which is imposed across two persistent worms that are stuck together, along with expressions for the stiffness parameter $C$ and the bending force $\bm{F}^{(\text{b})}_{\mu}$ on a bead $\mu$, are discussed in the next section.

If the number of persistent worms in a simulation box is $n_{\text{pw}}^{\text{T}}$, then there are a total of $N^{\text{T}} = N_{\text{pw}} \times n_{\text{pw}}^{\text{T}}$ beads (or monomers) in the box, and the monomer concentration is consequently, $c = N^{\text{T}}/V$, where $V$ is the volume of the simulation box. Note that both $n_{\text{pw}}^{\text{T}}$ and $c$ are parameters that can be controlled in a simulation, and the monomer concentration $c$ includes monomers that could belong to either linear or closed loop (ring) wormlike micelles.

The terminal beads in a persistent worm can associate with terminal beads on other persistent worms and as a consequence, are `sticky'. In this work, sticker beads are only allowed to associate in pairs. Thus, while the formation of linear and ring-like wormlike micelles is permitted, for simplicity, branching is prohibited (though it is possible in principle). Equations governing wormlike micelles formed by the fusion of persistent worms are discussed below.

\subsection{\label{sec:wlm} Wormlike micelles assembled from persistent worms}

\subsubsection{\label{sec:lcd} Lengths, concentrations and distributions}

It is appropriate to first introduce some notation related to wormlike micelles before providing details of pairwise bead-bead interaction and bending potentials. If a wormlike micelle of length $L$ (either a linear chain or a ring) is composed of $m_{\text{pw}}^{\text{L}}$ persistent worms, then,
\begin{equation}
L = m_{\text{pw}}^{\text{L}}  \ell_{\text{pw}} = m_{\text{pw}}^{\text{L}} (N_{\text{pw}} - 1) b
\label{eq:wlmlength}
\end{equation}
Since wormlike micelles increase or decrease in length in units of persistent worm lengths ($\ell_{\text{pw}}$), $L$ is a discrete quantity. It follows that the shortest wormlike micelle is exactly one persistent worm long, while the longest wormlike micelle is composed of all the persistent worms in the simulation box,
\begin{equation}
    \begin{split}
        L_{\text{min}} & =  \ell_{\text{pw}}  \\
         L_{\text{max}} & = n_{\text{pw}}^T  \ell_{\text{pw}} = n_{\text{pw}}^T (N_{\text{pw}} - 1) \, b           
               \end{split}
 \label{eq:Lminmax}   
\end{equation}
If we denote the number of wormlike micelles of length $L$ that are linear by $n^{\text{lin}}_{\text{L}}$ and the number 
 that are rings by $n^{\text{R}}_{\text{L}}$ (which are both functions of $L$), then the total number of wormlike micelles in a simulation box (of all lengths) that are either linear or rings are, respectively,
\begin{align}
& \mathcal{N}^{\text{lin}}_{\text{wlm}}   = \sum_{L_{\text{min}}}^{ L_{\text{max}}}  n^{\text{lin}}_{\text{L}} (L) ,  \,  \, \text{and} \, \,  
\mathcal{N}^{\text{R}}_{\text{wlm}} = \sum_{L_{\text{min}}}^{ L_{\text{max}}}  n^{\text{R}}_{\text{L}} (L) \nonumber \\
 \text{and}, \quad & \mathcal{N}_{\text{wlm}}= \mathcal{N}^{\text{lin}}_{\text{wlm}} + \mathcal{N}^{\text{R}}_{\text{wlm}} 
\label{eq:Nwlm}
\end{align}
where $\mathcal{N}_{\text{wlm}}$ is the total number of wormlike micelles in the solution at any instant of time. Since paired terminal beads of persistent worms in a wormlike micelle are counted as a single bead, the number of `effective' monomers $N_{\text{L,lin}}^{\text{eff}}$ in a linear wormlike micelle of length $L$ is,
\begin{equation}
        N_{\text{L,lin}}^{\text{eff}} =  (N_{\text{pw}} - 1) \,  m_{\text{pw,lin}}^{\text{L}} +1      
 \label{eq:Nefflin}   
\end{equation}
where $m_{\text{pw,lin}}^{\text{L}}$ is the number of persistent worms in a linear wormlike micelle  of length $L$, and the number of `effective' monomers $N_{\text{L,R}}^{\text{eff}}$  in a ring wormlike micelle of length $L$ is,
\begin{equation}
        N_{\text{L,R}}^{\text{eff}}   = (N_{\text{pw}} - 1)  \, m_{\text{pw,R}}^{\text{L}}
\label{eq:NeffR}   
\end{equation}
where $m_{\text{pw,R}}^{\text{L}}$ is the number of persistent worms in a ring wormlike micelle of length $L$.
As a result, the total number of effective monomers in the simulation box that belong to either linear or ring wormlike micelles are, respectively,
\begin{equation}
N_{\text{T,lin}}^{\text{eff}} = \sum_{L_{\text{min}}}^{ L_{\text{max}}}  n^{\text{lin}}_{\text{L}} (L) N_{\text{L,lin}}^{\text{eff}} ,  \,  \, \text{and} \, \, 
N_{\text{T,R}}^{\text{eff}} = \sum_{L_{\text{min}}}^{ L_{\text{max}}}  n^{\text{R}}_{\text{L}} (L) N_{\text{L,R}}^{\text{eff}} 
\label{eq:Nefftotal}
\end{equation}
and the respective effective monomer concentrations are, 
\begin{equation}
c^{\text{eff}} = \frac{N_{\text{T,lin}}^{\text{eff}}}{V}  , \,  \, \text{and} \, \, c_{\text{R}}^{\text{eff}} = \frac{N_{\text{T,R}}^{\text{eff}}}{V} 
\label{eq:ceff}
\end{equation}
Note that the subscript `lin' has been dropped for the sake of notational simplicity in the symbol for the effective concentration of  monomers in linear wormlike micelles ($c^{\text{eff}}$) since a majority of the results in this work are presented in terms of this concentration. It follows that the probabilities of finding either linear or ring wormlike micelles of length between $L$ and $L + \Delta L$ are,  respectively, 
\begin{align}
\psi_{\text{lin}}(L) \Delta L &= \frac{n^{\text{lin}}_{L + \Delta L} - n^{\text{lin}}_{L}}{\mathcal{N}^{\text{lin}}_{\text{wlm}}}, \nonumber \\[5pt]
\text{and} \quad 
\psi_{\text{R}}(L) \Delta L &= \frac{n^{\text{R}}_{L + \Delta L} - n^{\text{R}}_{L}}{\mathcal{N}^{\text{R}}_{\text{wlm}}}
\label{eq:psiL}
\end{align}
and the mean lengths of linear and ring wormlike micelle are defined, respectively, by,
\begin{equation}
\bar{L} = \sum_{L_{\text{min}}}^{ L_{\text{max}}}  L \, \psi_{\text{lin}} (L) \Delta L , \,  \, \text{and} \, \, \bar{L}_{\text{R}} = \sum_{L_{\text{min}}}^{ L_{\text{max}}}  L \, \psi_{\text{R}} (L) \Delta L 
\label{eq:meanL}
\end{equation}
where the subscript `lin' has been dropped for the sake of notational simplicity in the symbol for the mean length of linear wormlike micelles ($\bar{L}$). It is common in the literature on the static properties of wormlike micelle solutions to define the scaled length of linear wormlike micelles, $x = L/\bar{L}$, and to calculate the probability distribution $p(x)$ of scaled linear wormlike micelle lengths, both analytically and through simulations. It is straightforward to show that,
\begin{equation}
p(x) = \bar{L} \, \psi_{\text{lin}}  (L) 
\label{eq:pxpsiL}
\end{equation}

\subsubsection{\label{sec:SDK} The pairwise bead-bead interaction potential}

There are three possible pairwise bead-bead interaction scenarios between two beads that are either on the same persistent worm or on different persistent worms: (i) a backbone (non-sticky) bead interacts with another backbone bead, (ii) a backbone bead interacts with a sticky bead, and (iii) a sticky bead interacts with a sticky bead. All these three cases are modelled here with a common potential, the Soddemann-D\"unweg-Kremer (SDK)~\cite{SDK2001} potential, that acts between any two interacting beads $\mu$ and $\nu$ separated by a distance $r_{\mu\nu}= \left\vert\bm{r}_{\mu} - \bm{r}_{\nu}\right\vert$, 
{\small
\begin{align}\label{eq:SDK} 
\frac{U_{\mu\nu}^{\text{SDK}}}{k_{\text{B}} T} = \left\{
\begin{array}{l l}
 \hspace{-1mm} 4 \left[ \left( \dfrac{\sigma}{  r_{\mu\nu}} \right)^{12} - \left( \dfrac{\sigma}{  r_{\mu\nu}} \right)^6 + \dfrac{1}{4} \right] - \epsilon \, ;& r_{\mu\nu}\leq 2^{1/6}\sigma \\ [15pt]
\hspace{-1mm} \dfrac{1}{2}\,  \epsilon  \left[ \cos \left(\alpha \left( \dfrac{ r_{\mu\nu}}{\sigma} \right)^2+ \beta\right) - 1 \right] ;& \hspace{-5mm} 2^{1/6}\sigma \leq  r_{\mu\nu} \leq r_{\rm c} \\ [15pt]
\hspace{-1mm} 0. &   r_{\mu\nu} \geq  r_{\rm c}
\end{array}\right.
\end{align}
}
The repulsive part of the SDK potential is identical to a truncated Lennard-Jones potential, while the attractive part is represented by a cosine function. The minimum of the attractive well-depth occurs at a distance of $r_{\mu\nu}=2^{1/6}\sigma$  and $\epsilon$ is the well-depth of the potential in units of $k_BT $. In terms of the length scale $l_H = \sqrt{ k_{\text{B}} T/H }$ (used in this study for non-dimensionalising lengths), $\sigma / l_H$ is set equal to 1 in all the simulations reported here. Following the arguments laid out in the recent work by~\citet{Santra2019}, the cutoff radius $r_c$ is set equal to $1.82 \, \sigma$. The constants $\alpha$ and $\beta$ are determined by the boundary conditions: $\text{U}_{\mu\nu}^{\text{SDK}} = 0$ at $r_{\mu\nu} = r_c$ and $\text{U}_{\mu\nu}^{\text{SDK}} = -\epsilon$ at $r_{\mu\nu} = 2^{1/6}\sigma$. Using these conditions, the values of $\alpha$ and $\beta$ are computed to be 1.530633312 and 1.213115524, respectively. 
\begin{figure}[t]
\centering
\resizebox{8.5cm}{!}{\includegraphics*[width=9cm]{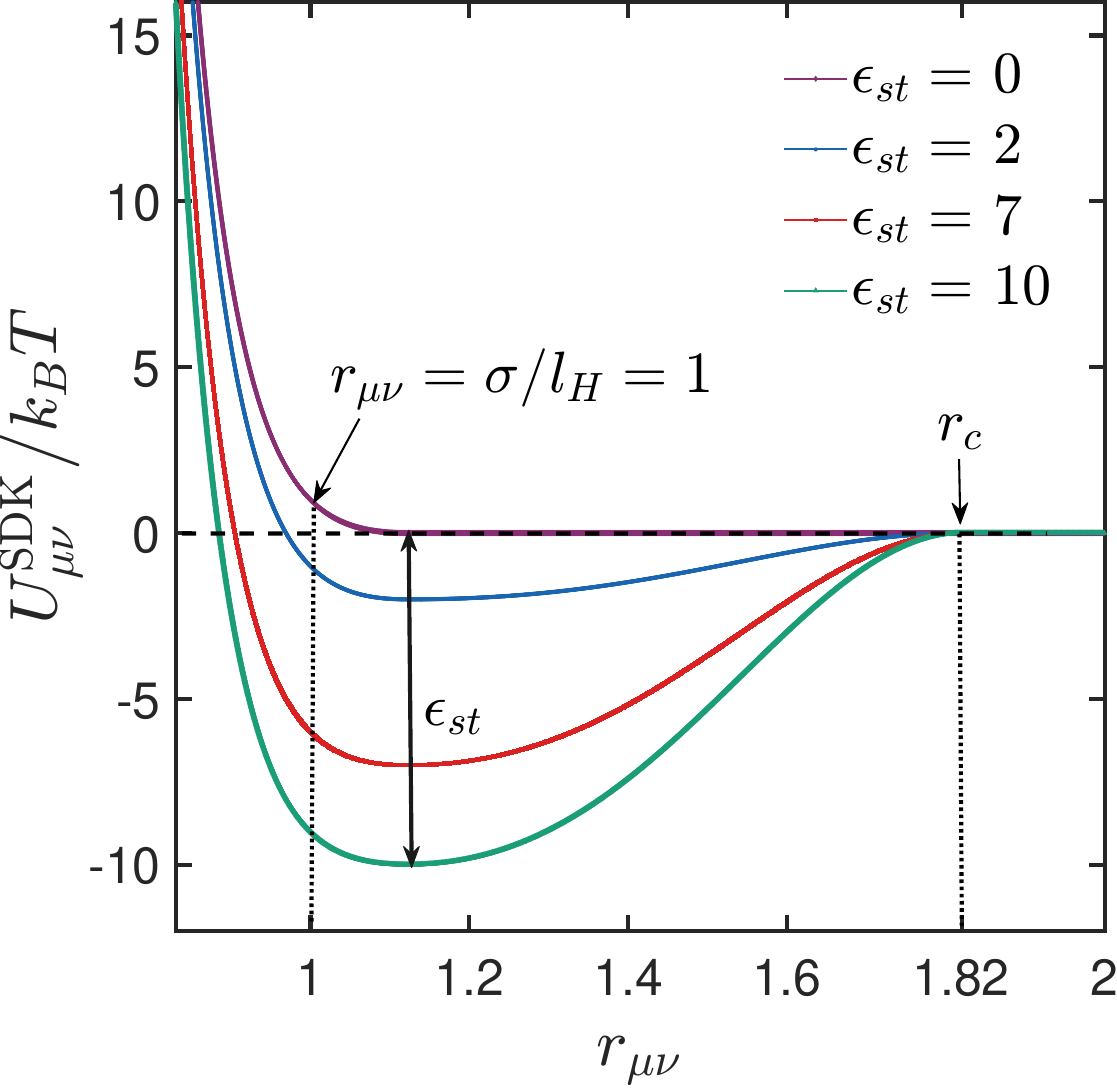}}  
\vspace{-10pt}
\caption{The Soddemann-D\"unweg-Kremer (SDK) potential ($U_{\mu\nu}^{\mathrm{SDK}}/k_{\text{B}} T$), defined in~\eref{eq:SDK}, plotted as a function of the radial distance $r_{\mu\nu}$ (in units of $l_H$), for different well depths $\epsilon_{st}$.}
\label{fig2}
\end{figure}
Amongst the many advantages of the SDK potential is its ability to interpolate between good solvent and poor solvent conditions by varying the well depth \text{$\epsilon$}~\cite{Santra2019}. Here, the well-depth corresponding to interactions between backbone-backbone and backbone-sticker bead pairs is denoted by $\epsilon_{bb}$, and both are considered to be purely repulsive, i.e.,  $\epsilon_{bb} = 0$, corresponding to the good solvent limit. The well-depth corresponding to associative interactions between sticker-sticker beads is denoted by $\epsilon_{st}$, and is typically chosen to be strongly attractive, with $\epsilon_{st} \geq 1$. The shape of the SDK potential for various values of the well depth $\epsilon_{st}$  is illustrated in~\fref{fig2}.

In reality, the distribution of wormlike micelle lengths is governed by a balance between entropic and enthalpic factors, with the equilibrium micelle length arising from a trade-off between the enthalpic cost of forming hemispherical endcaps and the entropic penalty of maintaining long micelles. Forming an endcap requires a distortion in the packing of surfactant molecules and a loss of favourable tail–tail interactions, which introduces an energy penalty, typically termed the endcap energy~\cite{Nagarajan2007,May2007,Israel2011}. In the present model, where the complex energetics of scission and fusion is captured with a phenomenological SDK interaction potential between the ``sticky” ends of persistent worms, the well depth $\epsilon_{st}$ acts as a surrogate for the endcap energy. It quantifies the energy lost when two sticky ends fuse to form a longer chain, thereby eliminating two endcaps, and conversely, the creation of two new endcaps through scission leads to an increase in the micelle energy by $\epsilon_{st}$. This approach is consistent with previous coarse-grained models, where similar short-range potentials have been used to capture complex fusion–scission dynamics in micellar systems~\cite{Cates1990,Wittmer1998,Wittmer2000,Padding2005,Huang2006,Padding2008}. Within this coarse-grained framework, as discussed in greater detail in ~\sref{sec:anal}, using arguments based on minimizing the Helmholtz free energy, mean-field theories predict the distribution of chain lengths and the mean micelle length's dependence on concentration and scission energy. As demonstrated subsequently in ~\sref{sec:lengths}, the use of the SDK potential effectively captures the thermodynamic cost of endcap formation and removal, and recovers the theoretical predictions of equilibrium static properties based on free energy minimization arguments.

Further, the choice of using the SDK potential to describe associative interactions implies that while there is a free energy barrier to the scission of two sticky beads that are stuck together, there is no barrier to their fusion if they are within cut-off distance of each other. Several previous studies have included a free energy barrier to fusion, arguing that the recombination of chains maybe a complex activated process~\cite{Huang2006,Ryckaert2006,PaddingPRE2004,Padding2008}. For simplicity, in this preliminary study that aims to introduce a new mesoscopic model for wormlike micelles, an activation barrier for fusion has been neglected. Depending on the performance of the model in its ability to capture rheological signatures of wormlike micelle solutions, an activation barrier can be introduced subsequently if this is considered necessary to improve its predictions. The occurrence of a scission/fusion event is determined by a Monte Carlo scheme, which is described in greater detail below. 

\subsubsection{\label{sec:interbp} The inter-persistent-worm bending potential}

\begin{figure}[tbhp]
\begin{center}
 \hspace{0cm}
\resizebox{8.5cm}{!}{\includegraphics*[width=16.5cm]{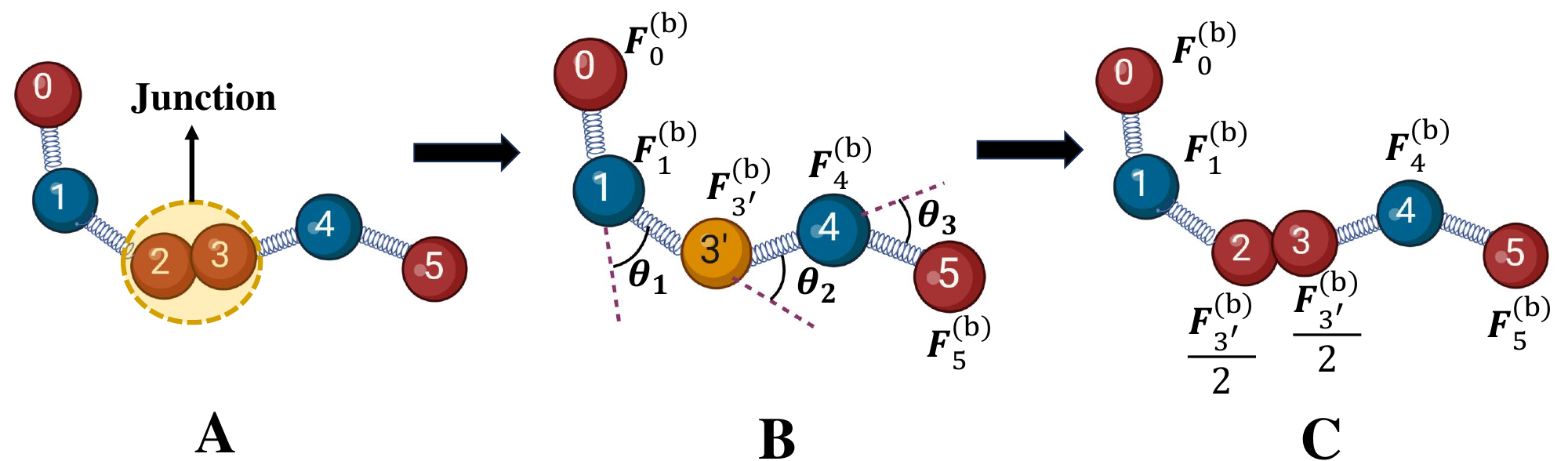}}  
\end{center}    
\vspace{-10pt}
\caption {Three step process (A $\longrightarrow$ B $\longrightarrow$ C) for implementing an inter-persistent-worm bending potential between two persistent worms, represented here by trumbbells. The terminal beads (red) of the persistent worms are sticky monomers connected to each other by backbone monomers (blue). The hypothetical bead $3^\prime$ is positioned at the centre of mass of beads 2 and 3. The labels $\bm{F}_\mu^{(\text{b})}$ ($\mu = 1,2,3^{\prime},4,5$) represent the bending forces on each of the beads $\mu$.} 
    \label{fig3}
\end{figure} 

A three step process is followed in order to ensure a constant degree of semi-flexibility along a wormlike micelle's backbone when two stickers at the ends of either the same persistent worm, or on different persistent worms, come within cutoff distance of each other. Basically, a bending potential which is identical to that used between springs adjacent to a backbone monomer within a persistent worm, i.e., \eref{eq:bending potential}, is imposed across the springs that are adjacent to the two neighbouring stickers. This is illustrated in \fref{fig3} and described below:
\begin{description}
    \item[A] The pairs of stickers within the cutoff radius of each other are identified (the junction points), shown as beads 2 and 3 in \fref{fig3}A.
    \item[B] The pairs of stickers are replaced with a hypothetical bead positioned at their centre of mass and the bending force on each bead in the updated configuration is calculated using \eref{eq:bendingforce} below, as illustrated by bead $3^\prime$ and the forces $\bm{F}_\mu^{(\text{b})} \, (\mu=0,1,3^{\prime},4,5)$ in \fref{fig3}B.
    \item[C] The bending force on the hypothetical bead is split into two halves, with one half allocated to each of the original sticky beads, while keeping the bending force on all the other beads the same as in the previous step. This is illustrated in \fref{fig3}C, where $\bm{F}_{3^{\prime}}^{(\text{b})}$ is divided equally between beads 2 and 3. 
\end{description}
Note that these steps are carried out after a Monte Carlo scheme (described below) has determined which sticky bead pairs will fuse to become part of a wormlike micelle, and before each timestep of the Brownian dynamics integration, in order to obtain the updated set of bending forces on each bead. The simulation then proceeds with the newly calculated bending forces. 

The bending force on a bead $\mu$ subjected to a bending potential given by~\eref{eq:bending potential}, is given by the following expression~\cite{Isaac2023},
\begin{multline}
\label{eq:bendingforce}
    \frac{\bm{F}^{(\text{b})}_\mu}{k_{\text{B}} T} = C \Bigg\lbrace \left[\frac{1}{Q_\mu} \left(  \bm{u}_\mu \cos{\theta_\mu}  - \bm{u}_{\mu-1} \right) \right. \\
    + \left. \frac{1}{Q_{\mu-1}} \left( - \bm{u}_{\mu-1} \cos{\theta_\mu}   + \bm{u}_{\mu} \right) \right] \\
    + \left. \left[\frac{1}{Q_{\mu-1}} \left( - \bm{u}_{\mu-1} \cos{\theta_{\mu-1}}  + \bm{u}_{\mu-2} \right) \right] \right. \\
    + \left[\frac{1}{Q_{\mu}} \left( \bm{u}_{\mu} \cos{\theta_{\mu+1}}  - \bm{u}_{\mu+1} \right) \right] \Bigg\rbrace
\end{multline}
where the segment from bead $\mu$ to $\mu + 1$ has unit vector $\bm{u}_\mu$ with length $Q_\mu = \left\vert \bm{Q}_\mu \right\vert = \left\vert \bm{r}_{\mu+1} - \bm{r}_{\mu} \right\vert$, and the bending stiffness $C$ for a wormlike micelle of length $L$ can be determined from,
\begin{equation}
\label{bending potential expression SK}
C = \frac{1+p_{\text{b,1}}(2N_{\text{K,s}}) + p_{\text{b,2}}(2N_{\text{K,s}})^2}{2N_{\text{K,s}}+p_{\text{b,3}}(2N_{\text{K,s}})^2 + p_{\text{b,4}}(2N_{\text{K,s}})^3}
\end{equation}
Here, $N_{\text{K,s}}$ is the ratio of $L$ to the persistence length $l_p$, given by, 
\[ N_{\text{K,s}} = \frac{L}{ 2N^{\text{L}}_{s} l_p } = \frac{\ell_{\text{pw}}}{ 2 \left(N_{\text{pw}} - 1\right)\, l_p } \]
where $N^{\text{L}}_{s} = m_{\text{pw}}^{\text{L}} \left(N_{\text{pw}} - 1\right)$ is the number of springs in a wormlike micelle (linear or ring) of length $L$, and the quantities $p_{\text{b},i} = -1.237, 0.8105, -1.0243, 0.4595$ for $i=1,2,3,4$, respectively, are constants in the Pad\'e approximation used to calculate $C$~\cite{Khomami2016}. This enables the expression of chain semiflexibility in terms of the physically meaningful $l_p$, rather than as a function of the parameter $C$. The validity of the implementation of the intra- and inter-persistent-worm bending potentials described above is demonstrated in \sref{sec:valid}, where it is shown that a wormlike micelle formed by the association of two persistent worms has  similar properties to a homopolymer of the same length when the bending stiffness is identical in both cases.

Having set out the key features of the model developed here for wormlike micelle solutions, it is appropriate now to discuss the mesoscopic simulation algorithm that has been used to explore their static and dynamic properties.

\subsection{\label{sec:MCBD} Monte Carlo \& Brownian dynamics} 

\subsubsection{\label{sec:MC} Fusion and scission rules}

The breakage and reformation of wormlike micelles is implemented here with a modified version of the Monte Carlo scheme that has been used previously to determine the binding and unbinding of sticker pairs in the context of flexible associative polymers~\cite{Robe2024}. The modification accounts for the fact that in addition to the energetics of sticking/unsticking, the energetics of persistent worm alignment must also be taken into account at each Monte Carlo step. Basically, when two stickers are within a distance $2^{1/6}\sigma \le r \le r_c$ of each other, the change in energy $\Delta E$ if the bond state were changed, is calculated according to,
\begin{equation}
\label{DeltaE}
 \Delta E(r,\theta) =  \frac{1}{2}\left[\cos \left(\alpha \, {\left( \frac{r}{\sigma} \right) }^2+\beta \right)-1 \right] \,  \epsilon_{st} + C \Big( 1-\cos \theta \Big) 
 \end{equation}
where $\theta$ is the angle between the adjacent springs at junction points (for instance, $\theta_2$ in Fig. \ref{fig3}B). Note that the function, $f(r) = ({1}/{2})\left[\cos \left(\alpha \, {\left( {r}/{\sigma} \right) }^2+\beta \right)-1 \right]$, which is the prefactor to the sticker energy term in~\eref{DeltaE}, has the values $f(r) = -1$ for $r= 2^{1/6}\sigma$ and $f(r) = 0$ for $r = r_c$. A pseudo-random number is drawn from a uniform distribution between 0 and 1, and if the random number is less than  $\exp \left(- \Delta E /k_{\text{B}}T \right)$, the change of the state is accepted and the bond is formed. In an update
sweep, each existing bond attempts to break in this manner, and then bond formations are attempted. Further, if two stickers are within the cutoff distance but at least one of them already has as many bonds as its functionality allows, then bond formation for the new pair is not attempted. Here, as mentioned earlier, the functionality of a sticker is set to one, which avoids branching. The use of \eref{DeltaE} to calculate the change in energy ensures that persistent worms are more likely to stick when they are highly aligned and that detailed balance is satisfied, meaning all bond states are reversible and have complementary probabilities. As noted earlier, the breakage and reforming of wormlike micelles is queried once before each Brownian dynamics timestep. Simulations have been carried out at various decreasing timesteps to ensure convergence of dynamic property predictions, and reported values are independent of the choice of timestep.

\subsubsection{\label{sec:BD} Euler integration algorithm}

The position of each bead, $\bm{r}_{\mu}(t)$ $(\mu = 1,2,3, \ldots, N_{\text{pw}})$, in each of the $n_{\text{pw}}^{\text{T}}$ persistent worms in the simulation box, is evolved in time according to the following first-order Euler integration scheme for the numerical solution of the It\^o stochastic differential equation that governs its motion~\cite{JainPRE2012},
\begin{equation}\label{eq:BD}
    \bm{r}_{\mu}(t+\Delta t) = \bm{r}_{\mu}(t) + \frac{\Delta t}{4} \sum_{\nu=1}^{N^{\text{T}}} \bm{D}_{\mu\nu} \cdot \bm{F}_\nu +  \frac{1}{\sqrt{2}} \sum_{\nu=1}^{N^{\text{T}}}\bm{B}_{\mu\nu} \cdot \bm{\Delta W}_\nu
    \end{equation}
Here length and time scales have been non-dimensionalized using  $l_H=\sqrt{{k_{\text{B}}T}/{H}}$ and $\lambda_H={\zeta}/{4H}$, with $\zeta=6\pi \eta_{\text{s}} a$ being the Stokes friction coefficient for a spherical bead of radius $a$, and $\eta_{\text{s}}$ represents the solvent viscosity. $\Delta\bm W_\nu$ is a non-dimensional Wiener process, whose components are obtained from a real-valued Gaussian distribution with zero mean and variance $\Delta t$. $\bm{B}_{\mu\nu }$ is a non-dimensional tensor whose evaluation requires the decomposition of the diffusion tensor $\bm D_{\mu\nu}$, defined as $\bm D_{\mu\nu} = \delta_{\mu\nu} \bm{\delta} + \bm{\Omega}_{\mu\nu}$, where $\delta_{\mu\nu}$ is the Kronecker delta, $\bm{\delta}$ is the unit tensor, and $\bm{\Omega}_{\mu\nu}$ is the hydrodynamic interaction tensor. Block matrices $\mathcal{D}$ and $\mathcal{B}$ consisting of $N^{\text{T}} \times N^{\text{T}}$ blocks each having dimensions of $3 \times 3$ are defined such that the $(\mu,\nu)$-th block of $\mathcal{D}$ contains the components of the diffusion tensor $\bm{D}_{\mu\nu }$, whereas, the corresponding block of $\mathcal{B}$ is equal to $\bm{B}_{ \mu\nu}$. The decomposition rule for obtaining $\mathcal{B}$ can be expressed as $\mathcal{B} \cdot {\mathcal{B}}^{\textsc{t}} = \mathcal{D}\label{decomp}$. In the present study, the regularized Rotne-Prager-Yamakawa (RPY) tensor is used to compute hydrodynamic interactions,
\begin{equation}
{\bm{\Omega}_{\mu \nu}} = {\bm{\Omega}} ( {\bm{r}_{\mu}} - {\bm{r}_{\nu}} )
\end{equation}
where 
\begin{equation}
\bm{\Omega}(\bm{r}) =  {\Omega_1\, \bm{\delta} +\Omega_2 \, \frac{\bm{r r}}{{r}^{2}}} \, ; \,\, \text{with} \, \left\vert \bm{r} \right\vert = r
\end{equation}
and
\begin{equation*}
\Omega_1 = \begin{cases} \dfrac{3\sqrt{\pi}}{4} \dfrac{h^*}{r}\left({1+\dfrac{2\pi}{3}\dfrac{{h^*}^2}{{r}^{2}}}\right) & \text{for} \quad r\ge2\sqrt{\pi}h^* \\
 1- \dfrac{9}{32} \dfrac{r}{h^*\sqrt{\pi}} & \text{for} \quad r\leq 2\sqrt{\pi}h^* 
\end{cases}
\end{equation*}
\vspace{5pt}
\begin{equation*}
\Omega_2 = \begin{cases} \dfrac{3\sqrt{\pi}}{4} \dfrac{h^*}{r} \left({1-\dfrac{2\pi}{3}\dfrac{{h^*}^2}{{r}^{2}}}\right) & \text{for} \quad r\ge2\sqrt{\pi}h^* \\
 \dfrac{3}{32} \dfrac{r}{h^*\sqrt{\pi}} & \text{for} \quad r\leq 2\sqrt{\pi}h^* 
\end{cases}
\end{equation*}
The hydrodynamic interaction parameter $h^* = a/(\sqrt{\pi k_BT/H})$ is the dimensionless bead radius. The net nondimensional force $\bm{F}_\nu$ acting on the $\nu$-th bead in \eref{eq:BD} is a sum of the nondimensional forms of the spring force, $\bm{F}_\nu^{(\text{s})} = \bm{F}^{(\text{s})}(\bm{Q}_\nu) - \bm{F}^{(\text{s})}(\bm{Q}_{\nu-1})$ (derived from 
 \eref{eq:Fraenkel}), the force due to the SDK potential, $\bm{F}_\nu^{\text{SDK}}$ (derived from \eref{eq:SDK}), and the bending force $\bm{F}_\nu^{(\text{b})}$ (given by \eref{eq:bendingforce}),
\begin{equation}
    \bm{F}_\nu =  \bm{F}_\nu^{(\text{s})} + \bm{F}_\nu^{\text{SDK}} + \bm{F}_\nu^{(\text{b})} 
\end{equation}
Note that while the spring force and the bending force on bead $\nu$ are associated with springs that are on the same wormlike micelle as the bead, the force due to the SDK potential is a consequence of all the beads that are within a cutoff distance of bead $\nu$, which could belong to the same wormlike micelle as bead $\nu$, or to other wormlike micelles in the neighbourhood. Details of the Brownian dynamics simulation algorithm used to integrate \eref{eq:BD} are given in the next section.

\subsubsection{\label{sec:simdet} Simulation Details}

Brownian dynamics simulations have been carried out here using the HOOMD-Blue simulation toolkit~\cite{Nlist2019,HOOMD2020}, with the decomposition of the diffusion tensor (necessary for the computation of hydrodynamic interactions) performed using the positively split Ewald (PSE) method that has been implemented by Aleks Donev, Jim Swan and co-workers as a plugin to HOOMD-Blue~\cite{PSE2017}. While the original algorithm was developed for colloidal suspensions~\cite{PSE2017}, it has recently been adapted to polymer solutions and applied to the description of gelation in associative polymer solutions~\cite{Robe2024}. The latter algorithm is the version used here.

The simulations were conducted in three distinct phases: an initial pre-equilibration phase, where only backbone monomers were present, followed by a stepwise increase in sticker strength for the sticker beads until the system has equilibrated, and finally, a production phase during which data sampling was carried out. Each of these phases was carried out for different lengths of time measured in terms of a relaxation time, which, given that the wormlike micelle solution is polydisperse, is taken to be the longest relaxation time of a homopolymer chain with length equal to the mean length of the linear wormlike micelles in the system. Initially, of course, the mean length, which is a function of monomer concentration and sticker energy, is unknown. As a result, preliminary simulations are carried out for a sufficiently long time such that properties achieve a stationary state. The mean length is then estimated, and for simulations with hydrodynamic interactions the Zimm relaxation time of a homopolymer chain with this length is calculated, while for those without, the Rouse relaxation time is found~\cite{Bird1987}. Once these relaxation times are determined, in the remaining simulations, the first phase is carried out for typically 5 to 10 relaxation times, the second is usually for 20 relaxation times, while the duration of the final production run, when all the relevant properties are calculated as a function of time, is around 10 relaxation times. The non-dimensional timestep was set to $\Delta t = 10^{-3}$ in all phases of the simulations, while ensemble averages and error estimates were obtained during the production run from 1000 independent simulation instances ($N_{\text{run}}$). For simulations without hydrodynamic interactions,  $h^*$ was set to zero, while simulations with hydrodynamic interactions were conducted with $h^* = 0.2$. 

The estimation of the overlap concentration $c^*$ for a polydisperse wormlike micellar solution is a subtle concept, which is discussed here in some detail in \sref{sec:overlap_conc} below. However, for the purposes of planning simulations, it has been found convenient to define the nondimensional scaled concentration $c/c^*_{\text{pw}}$, where $c^*_{\text{pw}}$ is the overlap concentration of a system consisting purely of persistent worms (i.e., where they have not formed wormlike micelles) defined by,
\begin{equation}\label{c*pw}
    c^*_{\text{pw}} = \frac{N_{\text{pw}}}{\left({4\pi}/{3}\right) R_{{g_0},\text{pw}}^3}
\end{equation}
Here, $R_{{g_0},\text{pw}}$ is the radius of gyration of a single persistent worm in the dilute limit, which is determined by carrying out simulations of a dilute solution of the bead-spring chains that constitute a persistent worm (without stickers). In the present work, these are trumbbells in most instances. The beads interact through a purely repulsive potential and hydrodynamic interactions are switched off since $R_{{g_0},\text{pw}}$ is a static property. Clearly, the persistent worm overlap  concentration $c^*_{\text{pw}}$ is not useful for distinguishing dilute from semidilute solutions, and is just used here as a convenient way to non-dimensionalize the monomer concentration $c$. Note that both $c$ and $c^*_{\text{pw}}$ can be calculated before the full-blown wormlike micelle solution simulations are carried out, unlike the \emph{effective} monomer concentration $c^{\text{eff}}$ (\eref{eq:ceff}), which is an outcome of the simulation and not known a~priori. 

\begin{figure}[t]
\begin{center}
 \hspace{0cm}
\resizebox{9cm}{!}{\includegraphics*[width=9cm]{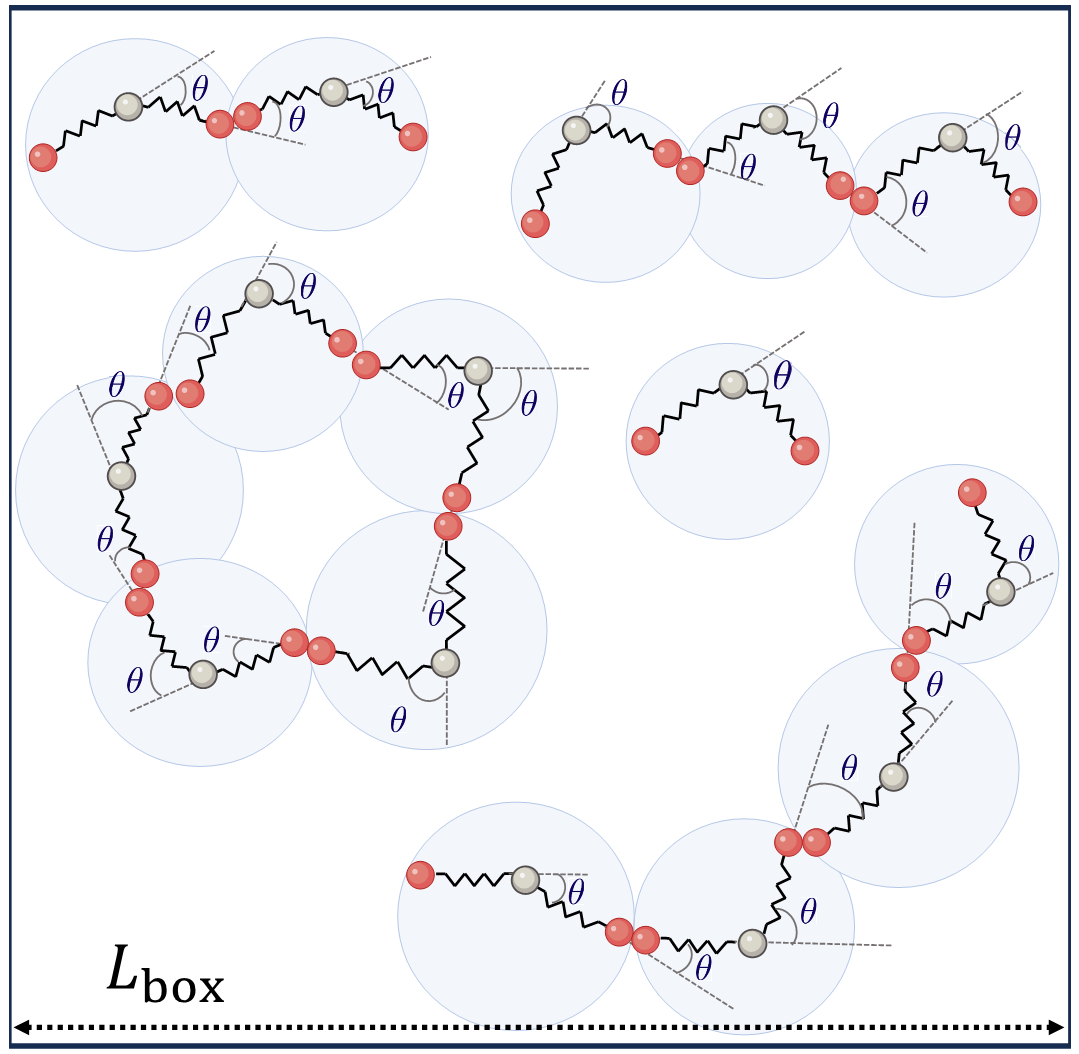}}  
\end{center}   
\vspace{-10pt}
    \caption{A schematic illustratation of a simulation box (of magnitude $L_{\text{box}}$ in each dimension), displaying an ensemble containing a ring and linear wormlike micelles of various lengths formed by the fusion of persistent worms represented by trumbells. The  application of intra- and inter-persistent-worm bending potentials is illustrated by the angles $\theta$.}  
    \label{fig4}
\end{figure} 

As discussed earlier, the semiflexibility of wormlike micelles has been accounted for through the use of intra- and inter-persistent-worm bending potentials, and the key parameter in this context is the persistent length $\ell_p$ (or equivalently, the bending stiffness $C$). While the validity of the implementation of this concept is demonstrated in \sref{sec:valid}, the majority of the simulations in the present work, which is essentially aimed at introducing a new mesoscopic model, have been carried out for flexible wormlike micelles. A systematic investigation of the effect of semiflexibility on the dynamics of wormlike micelle solutions will be carried out in a future study.

In order to examine the influence of the presence of ring wormlike micelles on solution properties, two versions of the algorithm have been developed; one version with rings, with results denoted by ``With Rings" in figure legends, that includes both ring and linear wormlike micelles, and a second version, which only permits the formations of linear micelles and excludes any ring wormlike micelles, results of which are identified as ``No Rings" in the figure legends. 
    
    \begin{table}[t]
    \begin{center}
           \caption{Typical parameter values used in the Brownian dynamics simulations \label{parametervalues}}         
     \vspace{10pt}
     \setlength{\tabcolsep}{0.1em}
     \begin{tabular}{llllll}
     	     \hline
            {} & Parameter & & Symbol  &  & Values \\
            \hline
            1 & Backbone interaction strength  & & $\text{$\epsilon$}_{bb}$   &  & 0 \\
            2 & Sticker strength  & & $\text{$\epsilon$}_{st}$    & & 2 to 10 \\
            3 & Simulation box size  & & $L_{\text{box}} $    & & 24, 40 \\
            4 & Rest length of the spring  & & $b$  & & 3\\
            5 & Hydrodynamic interaction parameter  & & $h^*$  & & 0, 0.2 \\
            6 & Integration time step  & & $\Delta t$ &  & 0.001 \\
            7 & Number of beads in a persistent worm  & & $N_{\text{pw}}$   & & 2, 3 \\
            8 & Persistent worm length & & $\ell_{\text{pw}}$   & & 3, 6 \\
            9 & Persistence length  & & $\ell_{p}$   & & 6 to 600  \\
            10 & Independent simulation instances  & & $N_{\text{run}}$  & & 1000 \\
            11 & Scaled monomer concentration  & & $c/c^*_{\text{pw}}$  & & 0.01 to 1 \\
            \hline
           \end{tabular}
           \end{center}
    \end{table}

A schematic illustration of a simulation box with an ensemble of wormlike micelles is shown in \fref{fig4} and typical parameter values used in the simulations are displayed in \tref{parametervalues}. The various static and dynamic properties computed in this work are defined in the next section.

\subsection{\label{sec:propertydefs} Property definitions} 

In addition to the distribution of lengths of linear and ring wormlike micelles $\psi_{\text{lin}} (L)$ and $\psi_{\text{R}} (L)$, and the mean lengths $\bar{L}$ and $\bar{L}_{\text{R}}$, which are static properties of wormlike micelles solutions, the mean size of wormlike micelles is estimated  by calculating the radius of gyration  in two different ways.

The radius of gyration of linear wormlike micelles of length $L$ is calculated using the expression,
\begin{equation}\label{RgL}
R_{g_L}^2 = \frac{1}{N_{\text{L,lin}}^{\text{eff}} } \left \langle \sum_{\mu=1}^{N_{\text{L,lin}}^{\text{eff}} } \left (\bm{r}_\mu-\bm{r}_{\text{c}} \right)^2 \right \rangle
\end{equation}
where $\bm{r}_{\text{c}}$ is the centre of mass of a linear wormlike micelle of length $L$,
\begin{equation}
    \bm{r}_{\text{c}} = \frac{1}{N_{\text{L,lin}}^{\text{eff}} }  \sum_{\mu=1}^{N_{\text{L,lin}}^{\text{eff}} } \bm{r}_\mu
\end{equation}
and angular brackets $\langle \cdots \rangle$ represent an ensemble average over all stochastic trajectories. The radius of gyration of ring wormlike micelles, $R_{g_R}$, is calculated similarly, with the number of effective monomers $N_{\text{L,R}}^{\text{eff}}$ in a ring wormlike micelle of length $L$, taking the place of $N_{\text{L,lin}}^{\text{eff}}$ in the expressions above.

While $R_{g_L}$ is the mean size of micelles of length $L$, the mean size of all linear wormlike micelles that are present at a given concentration and sticker energy, regardless of their length, is given by the radius of gyration, $R_{g}$, defined by,
\begin{equation}\label{meanradius}
R_{g}^2 = \left \langle \,  \frac{1}{\mathcal{N}_{\text{wlm}}^{\text{lin}}}  \sum_{\alpha=1}^{\mathcal{N}_{\text{wlm}}^{\text{lin}}} \Bigg\lbrace  \frac{1}{N_{\text{L,lin}}^{{\text{eff}},\alpha}} \sum_{\mu=1}^{N_{\text{L,lin}}^{{\text{eff}},\alpha} } \left ( \bm{r}_\mu^{(\alpha)}-\bm{r}_{\text{c}}^{(\alpha)} \right)^2  \Bigg \rbrace \,  \right \rangle
\end{equation}
where $N_{\text{L,lin}}^{{\text{eff}},\alpha}$ is the effective number of monomers in the $\alpha$-th micelle, and the index $\alpha$ varies over all the wormlike micelles (of all lengths) in a simulation box.

The nondimensional contribution to the stress tensor from both linear chains and rings in a polydisperse wormlike micelle solution is given by the Kramers-Kirkwood expression,
\begin{equation}\label{Kramers}
\bm{\tau}^{\text{p}} = \frac{1}{\mathcal{N}_{\text{wlm}}} \left\langle \sum_{\alpha=1}^{\mathcal{N}_{\text{wlm}}} \sum_{\nu=1}^{N_b^{\text{L}}} \left(\bm{r}_\nu^{(\alpha)}-\bm{r}_c^{(\alpha)} \right) \bm{F}_{\alpha\nu} \right\rangle
\end{equation}
where the stress has been nondimensionalised by $\mathcal{N}_{\text{wlm}} \, (k_{\text{B}} T/V)$, the number of beads in a wormlike micelle of length $L$ is $N_b^{\text{L}} = m_{\text{pw}}^{\text{L}}N_{\text{pw}}$, and the force $\bm{F}_{\alpha\nu}$ on the bead $\nu$ in chain $\alpha$ is given by,
\begin{equation}\label{totalforce}
\bm{F}_{\alpha\nu} = \sum_{\beta=1}^{\mathcal{N}_{\text{wlm}}} \sum_{\substack{\mu=1 \\ \mu\ne\nu}}^{N_b^{\text{L}}} \bm{F}_{\alpha\nu,\beta\mu}^{\text{SDK}} + \sum_{\substack{\mu=1 \\ \mu\ne\nu}}^{N_b^{\text{L}}} \bm{F}_{\alpha\nu,\alpha\mu}^{\text{(s)}} + \bm{F}_{\alpha\nu}^{\text{(b)}}
\end{equation}
where, the pair-wise nature of the bead-bead interaction and spring forces, and the fact that both the spring and bending forces are intra-chain forces that only act on beads within the chain $\alpha$, is explicitly recognized in the summations. Note that the bending force, given by \eref{eq:bendingforce}, is not a pair-wise force but rather involves successive triplets of beads. The expression for the dimensionless tensor $\bm{S}$ appearing in \eref{Kramers},
\[ \bm{S} = \sum_{\alpha=1}^{\mathcal{N}_{\text{wlm}}} \sum_{\nu=1}^{N_b^{\text{L}}} \left( \bm{r}_\nu^{(\alpha)}-\bm{r}_c^{(\alpha)} \right) \bm{F}_{\alpha\nu}  \]
which is the total contribution to the stress tensor from all the chains in a simulation box from a single trajectory in the ensemble, can be simplified and shown to be given by,
\begin{multline}
\label{Kramersv2}
\bm{S} = 
\frac{1}{2} \sum_{\nu=1}^{N^{\text{T}}}\sum_{\substack{\mu=1 \\ \mu\ne\nu}}^{N^{\text{T}}} \bm{r}_{\nu\mu} \bm{F}_{\nu\mu}^{\text{SDK}} + \sum_{\alpha=1}^{\mathcal{N}_{\text{wlm}}} \sum_{i=1}^{N_s^{\text{L}}} \bm{Q}_{i}^{(\alpha)} \bm{F}^{{\text{(s)}}}
\left( \bm{Q}_{i}^{(\alpha)} \right)\\
+ C \sum_{\alpha=1}^{\mathcal{N}_{\text{wlm}}} \sum_{i=1}^{N_s^{\text{L}}-1} \Big[\bm{u}_i^{(\alpha)}\bm{u}_{i+1}^{(\alpha)} +\bm{u}_{i+1}^{(\alpha)}\bm{u}_i^{(\alpha)} \\
- \cos \theta_{i+1}^{(\alpha)} \left(\bm{u}_i^{(\alpha)}\bm{u}_i^{(\alpha)} + \bm{u}_{i+1}^{(\alpha)}\bm{u}_{i+1}^{(\alpha)} \right) \Big] 
\end{multline}
where, as previously defined, $N^{\text{T}}$ is the total number of beads in a simulation box and $N_s^{\text{L}}$ is the number of springs in a wormlike micelle of length $L$, regardless of its architecture. The tensor $\bm{S}$ is required for the calculation of the only linear viscoelastic material functions of wormlike micellar solutions that are discussed in this work, namely, the storage ($G^\prime$) and loss ($G^{\prime\prime}$) moduli. These are obtained here from the dimensionless shear relaxation modulus $G(t)$, which is defined (because of isotropic conditions at equilibrium) by,
\begin{equation}\label{Gtiso}
    G(t) = \frac{1}{3} \Big [  G_{xy}(t) + G_{xz}(t) + G_{yz}(t) \Big ]
\end{equation}
where the individual components $G_{ij}(t)$ are estimated from the Green-Kubo expression,
\begin{equation}\label{stressauto}
    G_{ij}(t) =\frac{1}{\mathcal{N}_{\text{wlm}}} \, \Big\langle \! S_{ij}(0) \, S_{ij}(t) \! \Big\rangle 
\end{equation}
The modulus, $G(t)$, is fitted to a sum of exponential functions: $G(t) = \sum_{k=1}^{n}a_k \exp(-t/\tau_k)$, where $a_k$ and $\tau_k$ are fitting parameters and $n$ is the number of exponentials used to fit the auto-correlation function. The functions $G^\prime (\omega)$ and $G^{\prime\prime} (\omega)$ are then obtained by carrying out the Fourier transforms,
\begin{align}  \label{Gprime}
    G^\prime (\omega) = \int_0^\infty \!\! d (\omega t) \, G(t) \sin(\omega t)\\[10pt] \nonumber
    G^{\prime\prime} (\omega) =  \int_0^\infty \!\! d (\omega t) \, G(t) \cos(\omega t)
\end{align}

\subsection{\label{sec:overlap_conc} The overlap concentration for a wormlike micellar solution}

In a monodisperse homopolymer solution, with all the chains having the same length $L$, the overlap concentration $c^*$, which demarcates the end of the dilute concentration regime and the onset of the semidilute regime, is defined by the expression,
\begin{equation}\label{homooverlap}
    c^* = \frac{N}{\dfrac{4\pi}{3} \Big\lbrack R_{g0}(L)\Big\rbrack^3}
\end{equation}
where $N$ is the number of beads on a chain with length $L$, and $R_{g0}(L)$ is its radius of gyration in the dilute limit. This expression cannot be used to define the overlap concentration for a wormlike micellar solution because they are polydisperse in nature, with a distribution of chain lengths that depend on the effective concentration $c^{\text{eff}}$ of monomers and the sticker energy $\epsilon_{st}$. Indeed, the definition of $c^*$ cannot be based on the mean chain length either, since $\bar{L} = \bar{L} (c^{\text{eff}}, \epsilon_{st})$. In this work, a unique overlap concentration, $c^*$, for a fixed value of the sticker energy $\epsilon_{st}$, is defined as described in the sequence of steps below, by adapting an approach introduced by~\citet{Wittmer1998} in their seminal work on the static properties of wormlike micelle solutions.

\begin{figure}[t]
\begin{center}
\resizebox{9cm}{!}{\includegraphics*[width=9cm]{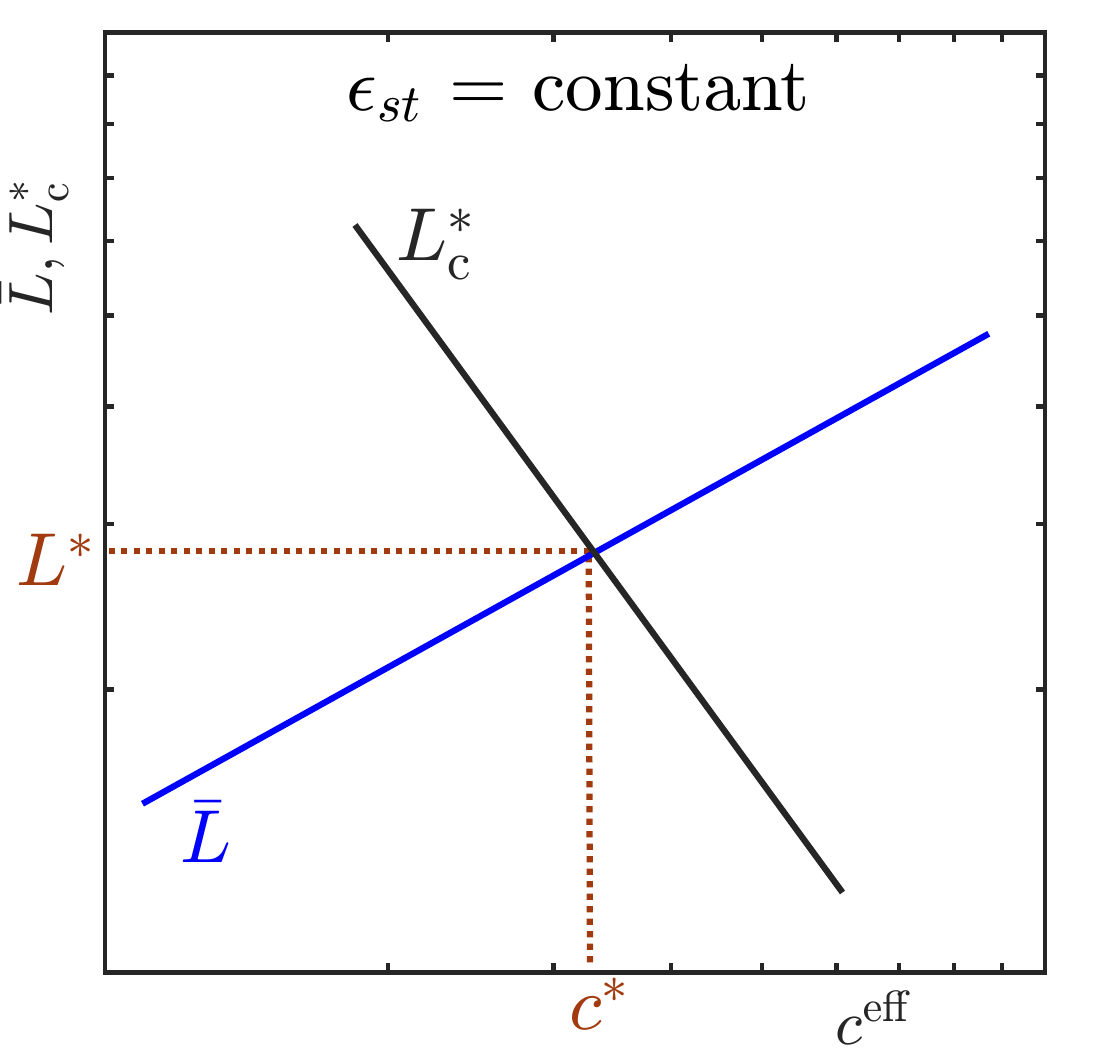}} 
\end{center}
        \vspace{-10pt}
\caption [Overlap concentration for WLMs.] {Schematic diagram describing the procedure for determining the overlap concentration in a wormlike micelle solution for a fixed value of the sticker energy $\epsilon_{st}$. The black line depicts the dependence of the overlap length $L^*_{\text{c}}$ on the effective concentration $c^{\text{eff}}$, while the blue line models the dependence of the mean length $\bar{L}$ on $c^{\text{eff}}$. The coordinates of the intersection point, ($c^*, L^*$), correspond to the overlap concentration and overlap length, respectively, with $c^*$ representing the effective concentration $c^{\text{eff}}$ at which the mean length of wormlike micelles in the system is exactly the length for which $c^{\text{eff}}$ is the overlap concentration.} 
    \label{fig5}
    \vspace{-10pt}
    \end{figure} 
   
\begin{enumerate}\label{ovdescrip}
\item  For each length $L$ of linear wormlike micelles in a typical polydisperse solution, with numbers of effective monomers $N_{\text{L,lin}}^{\text{eff}}$, the radius of gyration $R_{g0} (L)$ can be determined by simulating a dilute homopolymer solution with bead-spring chains having $N_{\text{L,lin}}^{\text{eff}}$ beads, under athermal solvent conditions.
\item The overlap concentration $c_L^* = f (L)$ is then calculated for each of these lengths $L$, using \eref{homooverlap} with $N_{\text{L,lin}}^{\text{eff}}$ in place of $N$. The function $f(L)$ needs to be computed only once for a range of lengths $L$.  
\item For a wormlike micelle solution at an effective concentration $c^{\text{eff}}$, the overlap length $L^*_{\text{c}}$ is determined by inverting the function $f$, i.e., $L^*_{\text{c}} = f^{-1} (c^{\text{eff}})$. This implies that a monodisperse homopolymer solution, all of whose chains are of length $L^*_{\text{c}}$, would have an overlap concentration equal to $c^{\text{eff}}$. We expect $f^{-1} (c^{\text{eff}})$ to be a monotonically decreasing function of $c^{\text{eff}}$, as shown schematically in Fig.~\ref{fig5}, where $L^*_{\text{c}}$ is plotted as a function of $c^{\text{eff}}$.
\item As noted earlier, $\bar{L} = \bar{L} (c^{\text{eff}}, \epsilon_{st})$. For a fixed value of $\epsilon_{st}$, it is possible to compute $c^{\text{eff}}$ and $\bar{L} (c^{\text{eff}})$ by carrying out equilibrium simulations for increasing concentrations $c/c^*_{\text{pw}}$. We expect $\bar{L}$ to be an increasing function of $c^{\text{eff}}$, as shown schematically in Fig.~\ref{fig5}, where $\bar L$ is plotted as a function of $c^{\text{eff}}$.
\item For each value of $\epsilon_{st}$, the two curves for $L^*_{\text{c}}$ and $\bar L$ as functions of $c^{\text{eff}}$, intersect at a unique value of $c^{\text{eff}}$. The $x$ and $y$ coordinates of the intersection point, ($c^*, L^*$), are denoted as the overlap concentration and overlap length, respectively, at the sticker energy $\epsilon_{st}$. The intersection point corresponds to the concentration at which the mean length of wormlike micelles in the system is exactly the length for which $c^{\text{eff}}$ is the overlap concentration.
\end{enumerate}

For all concentrations $c^{\text{eff}} < c^*$, the mean length $\bar{L} < L^*_{\text{c}}$, corresponding to the dilute concentration regime, while for all concentrations $c^{\text{eff}} > c^*$, the mean length $\bar{L} > L^*_{\text{c}}$, and the solution is considered to be semidilute. In solutions where both ring and linear chains are present, the procedure outlined above is followed identically, with the difference that the definition of $c^*$ is based only on the linear chains in the system, and with $c^{\text{eff}}$ calculated based only on the number of effective monomers in linear chains. It is shown subsequently that with this definition, the static properties of linear chains are identical in solutions with and without rings.

An alternative approach to defining an overlap concentration for wormlike micelle solutions is to determine the effective concentration for which the total volume occupied by all the linear wormlike micelles in the solution is equal to the volume of the simulation box. This is in accord with the physical representation of the overlap concentration in monodisperse homopolymer solutions as the concentration at which all the polymer coils just begin to touch each other and the total volume occupied by the chains is equal to the system volume. The total volume occupied by all the linear wormlike micelles in a simulation box is given by,
\begin{equation}
\mathcal{V}_{\text{wlm}} = \sum_{\alpha=1}^{\mathcal{N}_{\text{wlm}}^{\text{lin}}} \left(\frac{4\pi}{3} \right)\Bigg\lbrack \frac{1}{N_{\text{L,lin}}^{\text{eff},\alpha}}\sum_{\mu=1}^{N_{\text{L,lin}}^{\text{eff},\alpha}} \left (\bm{r}_\mu^{(\alpha)}-\bm{r}_{\text{c}}^{(\alpha)} \right)^2 \Bigg\rbrack^{\frac{3}{2}}  
\label{eq:vwlm}   
\end{equation}
In terms of $\mathcal{V}_{\text{wlm}}$, the alternative definition of $c^*$ is the value of $c^{\text{eff}}$ at which,
\[ \langle \mathcal{V}_{\text{wlm}} \rangle = V_{\text{box}} = L_{\text{box}}^3\]
Only the total volume occupied by linear wormlike micelles is considered, since the definition of $c^*$ above (based on the procedure described schematically in~\fref{fig5}), takes into account only linear wormlike micelles. As will be demonstrated in \sref{sec:cstar} below, the two different definitions of $c^*$ turn out to be fairly close to each other.

\section{\label{sec:results} Results and Discussion}

\subsection{\label{sec:valid} Validation of bending potential implementation}

\begin{figure*}[t]
\centering
\begin{tabular}{ccc}
\includegraphics[width=5.4cm]{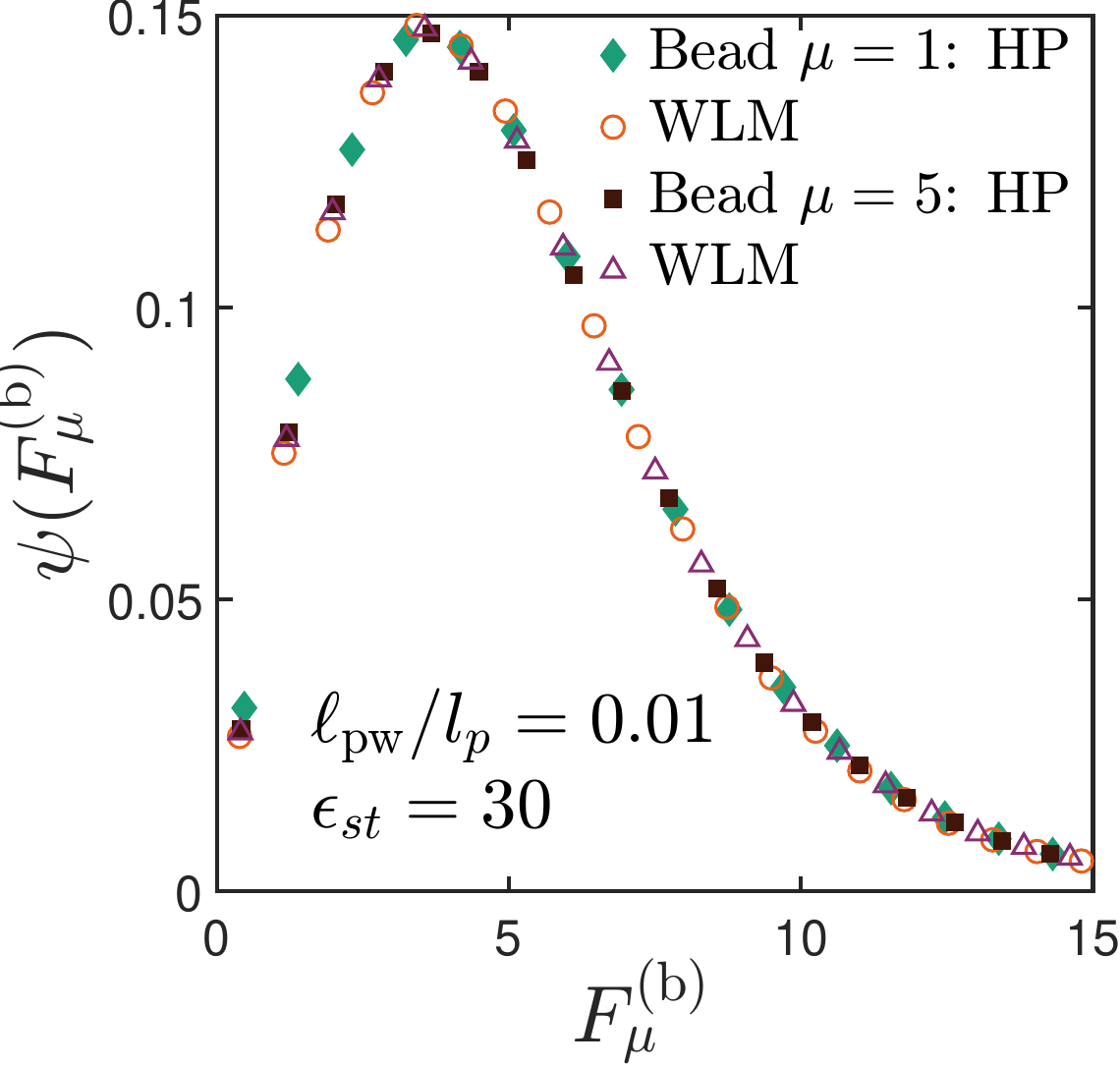} &
\includegraphics[width=5.4cm]{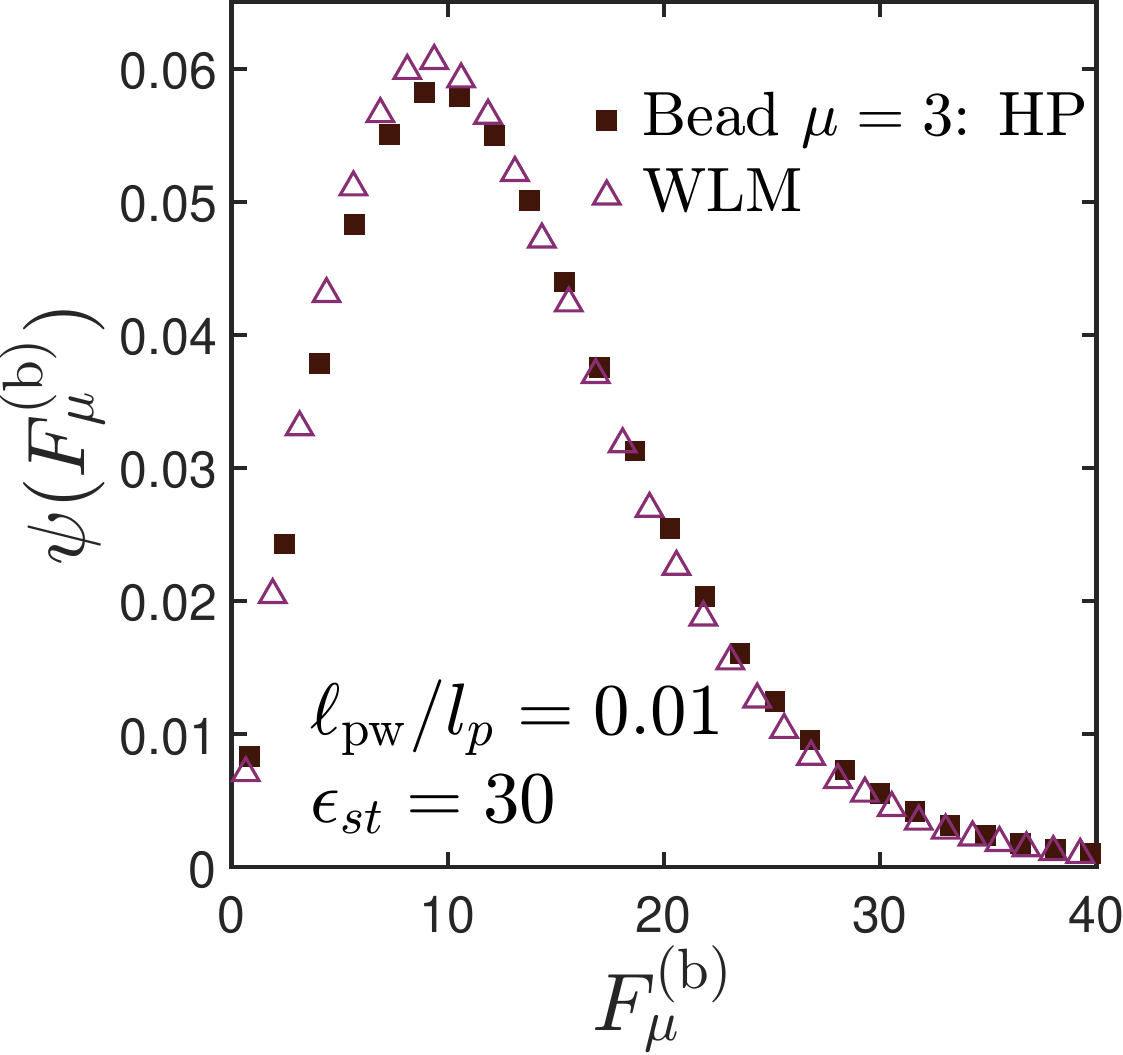} &
\includegraphics[width=5.3cm]{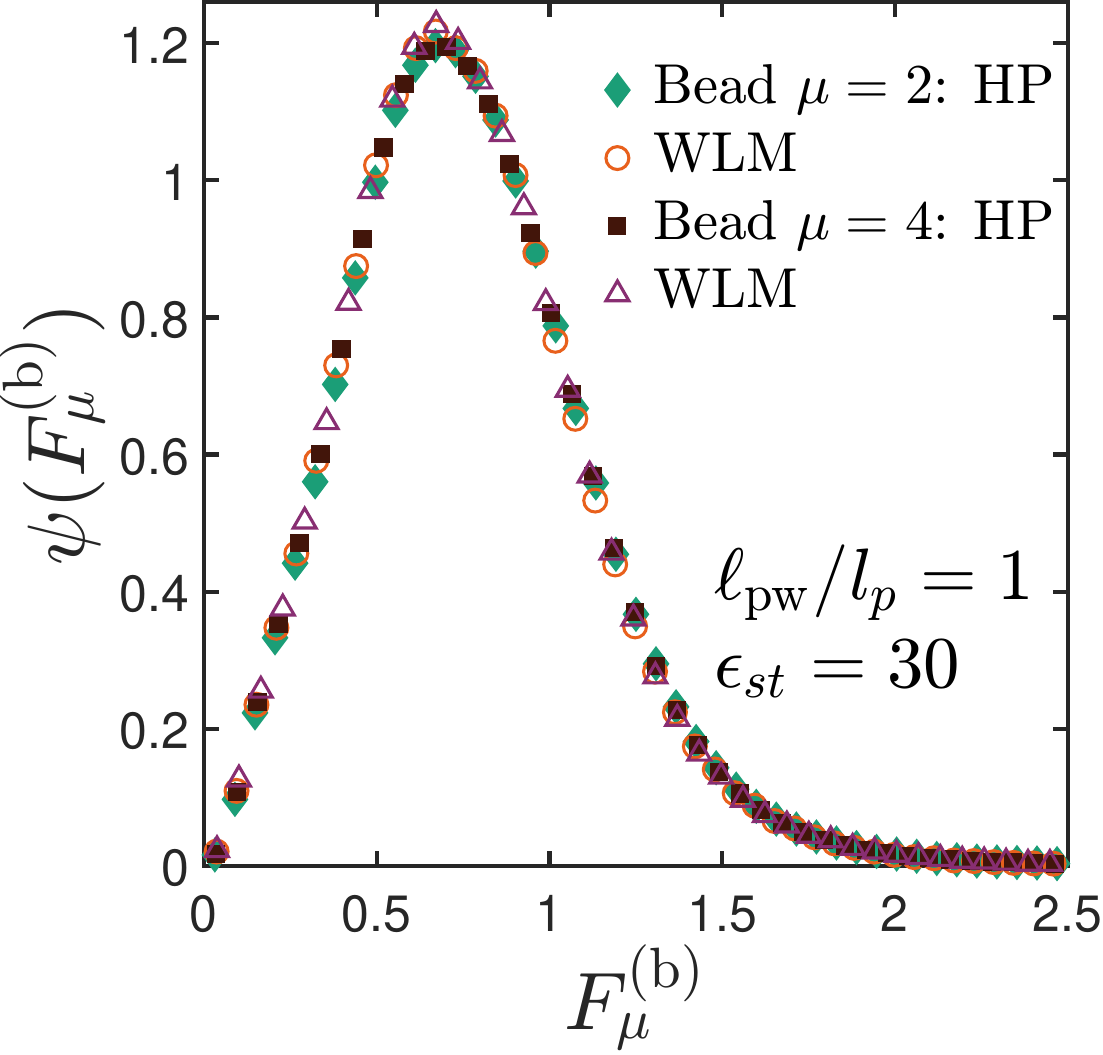} \\
(a)  & (b)  & (c) \\[10pt]
\includegraphics[width=5.4cm]{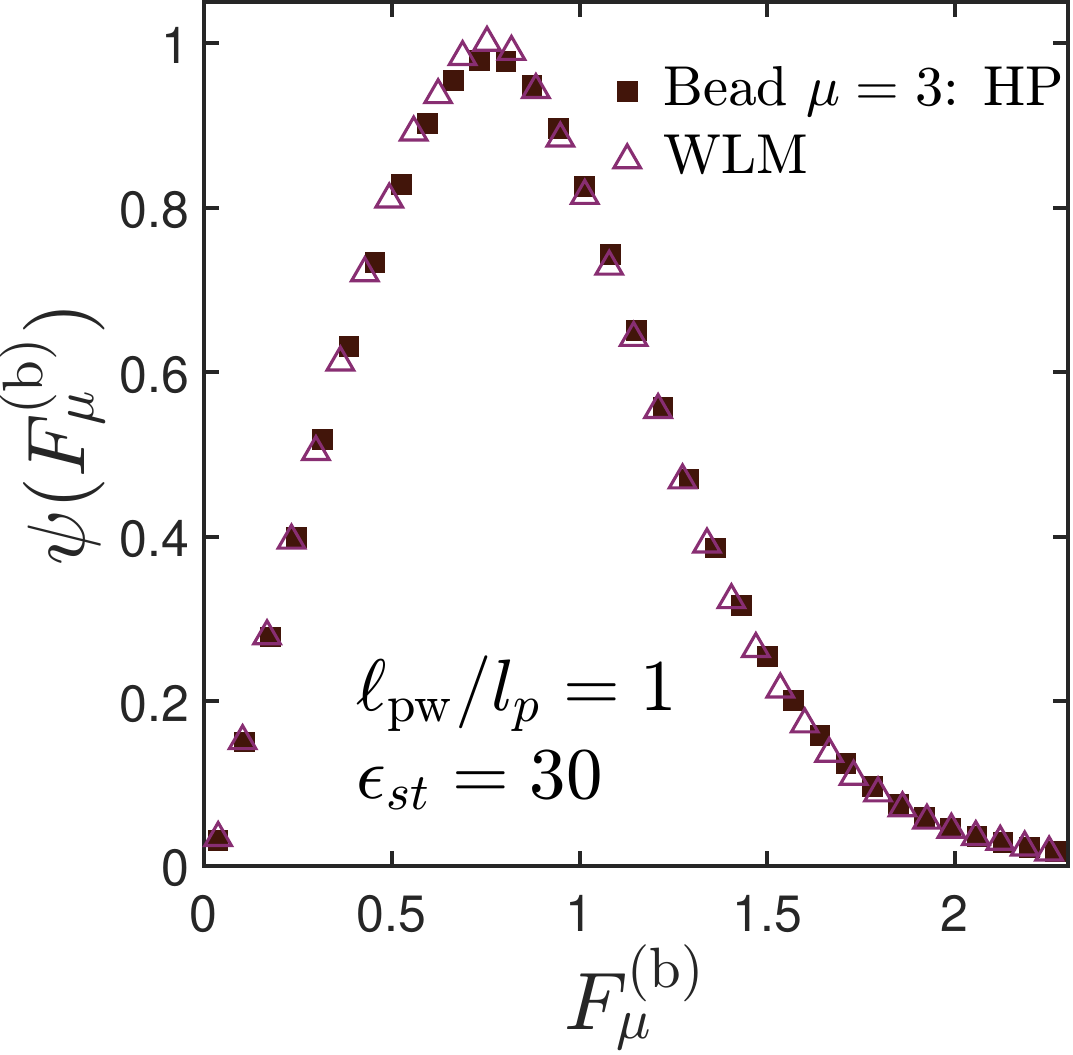} &
\includegraphics[width=5.6cm]{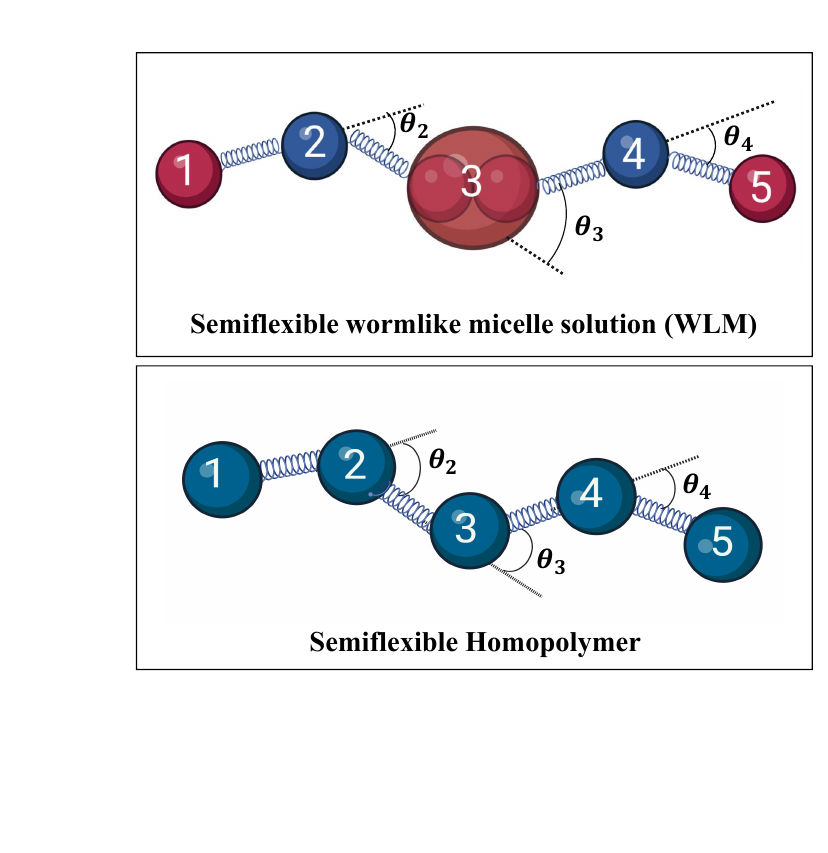} &
\includegraphics[width=5.4cm]{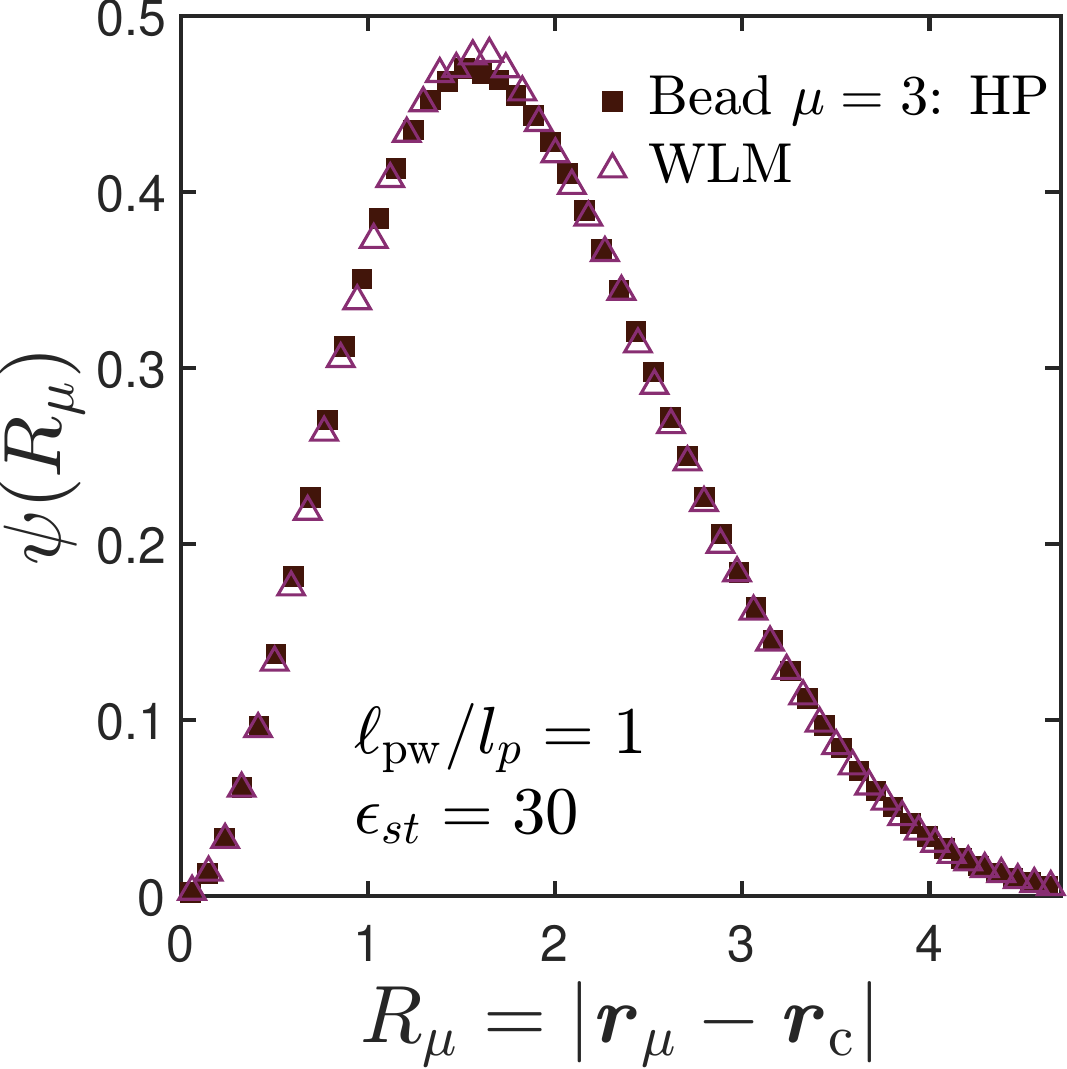} \\[10pt]
(d)  & (e)  & (f) \\[10pt]
\includegraphics[width=5.4cm]{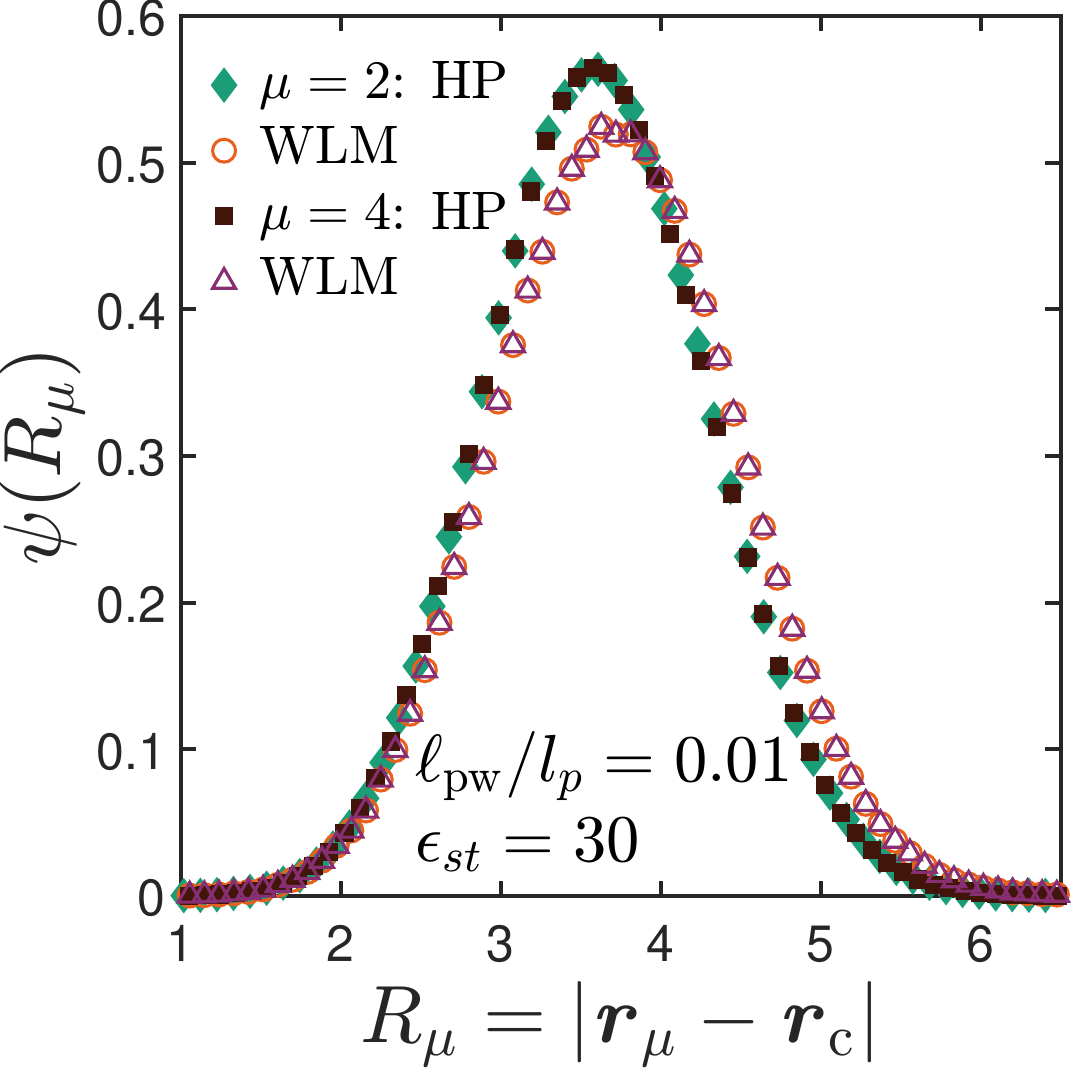} &
\includegraphics[width=5.4cm]{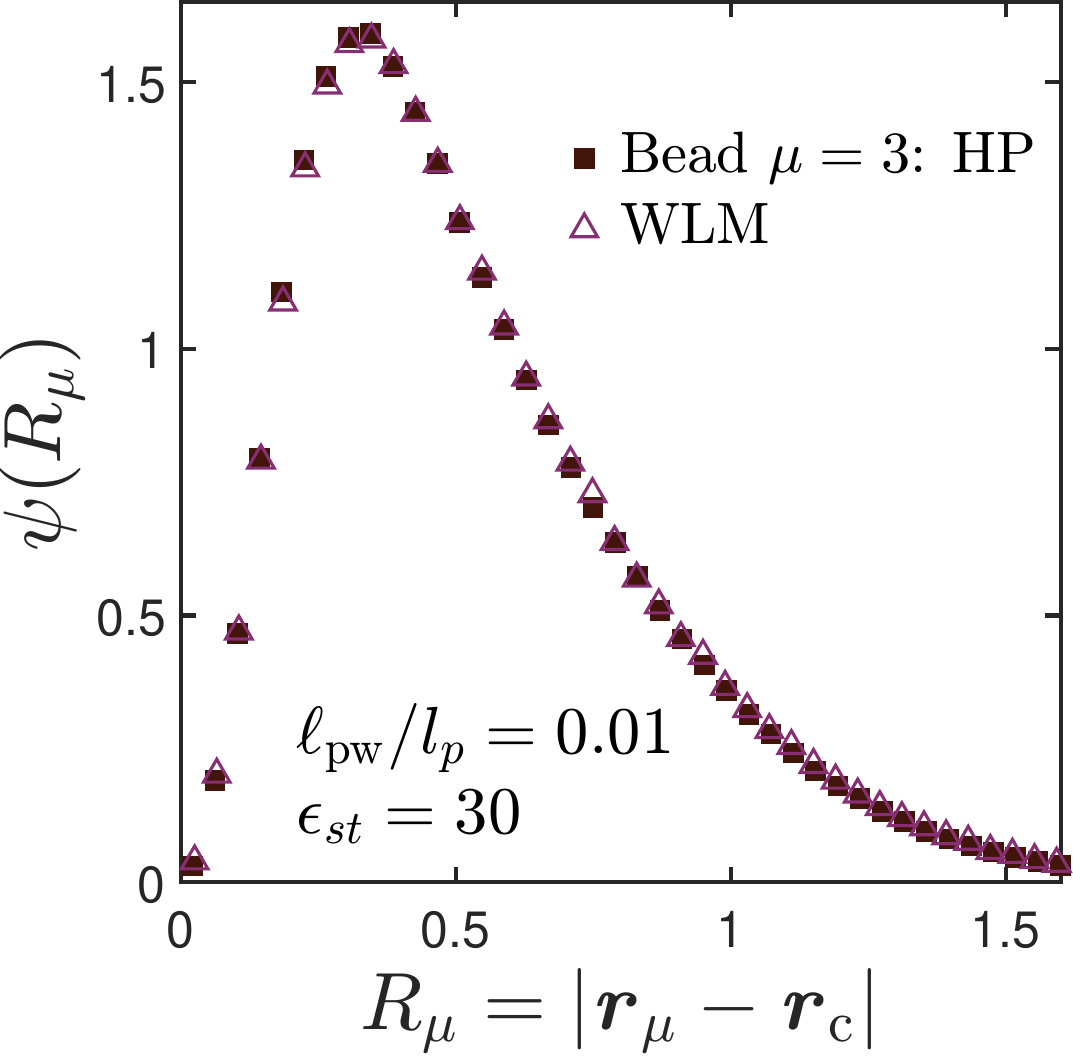} &
\includegraphics[width=5.6cm]{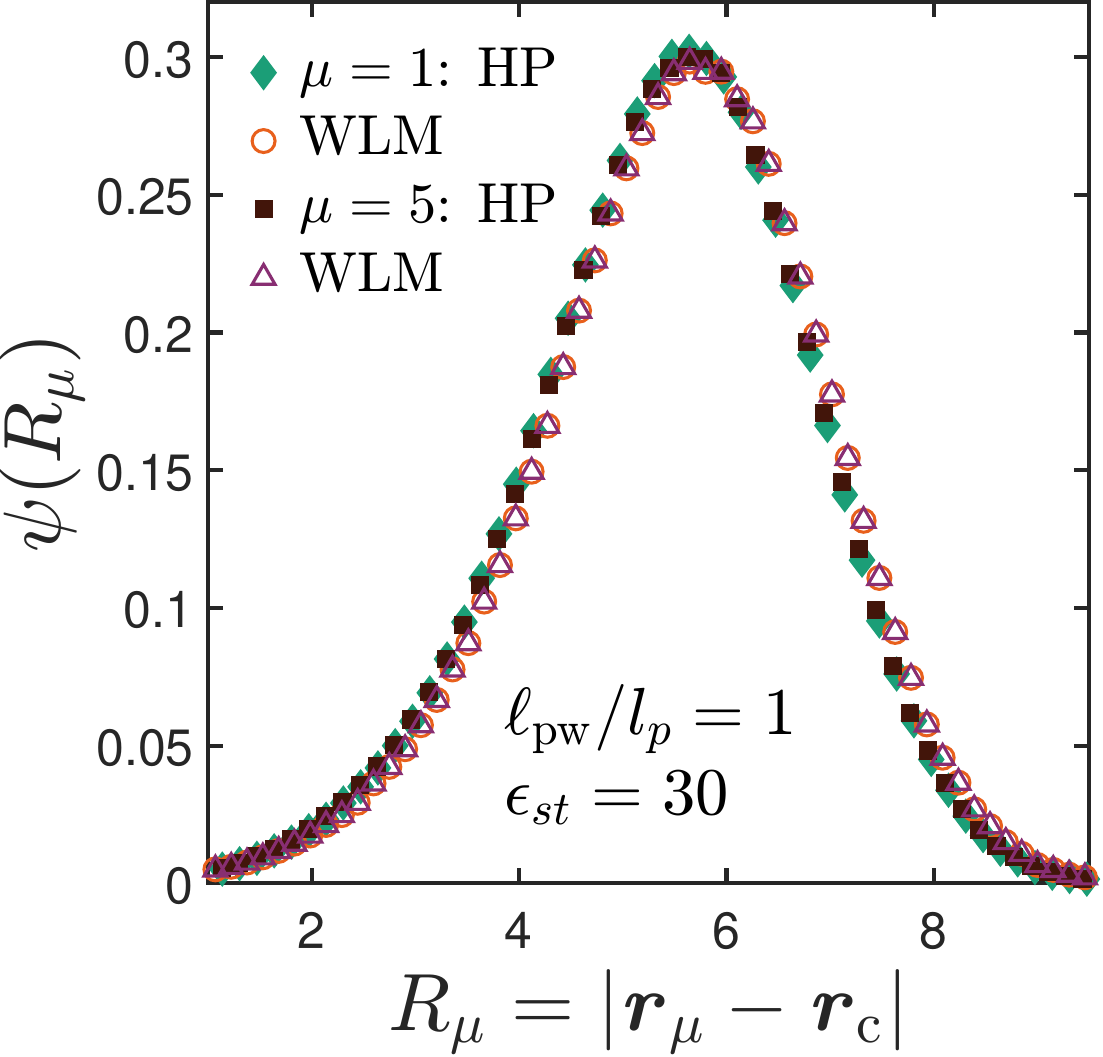} \\[10pt]
(g)  & (h)  & (i) 
\end{tabular}
    \caption {Validition of the implementation of the intra- and inter-persistent-worm bending potentials. A semiflexible linear wormlike micelle (WLM) composed of two persistent worms that are each trumbbells with one sticky end bead, is compared with a semiflexible homoplymer (HP) chain of five beads. In~(e), beads 1 to 5 in the wormlike micelle  correspond to beads with the same labels in the homopolymer chain. Bead ``3" in the wormlike micelle is positioned at the centre of mass of the positions of the stickers at the ends of the two persistent worms. In~(a) to~(d), the distribution of the magnitude of the bending force acting on each bead is compared across the two systems, for two different values of the bending stiffness $\ell_{\text{pw}}/l_p$. In~(f) to~(i), the distribution of the distance of each bead from the center of mass of the respective chains, is compared.}
    \label{fig6}
    \end{figure*}

The validity of the implementation of the intra- and inter-persistent-worm bending potentials described in~\sref{sec:interbp} is demonstrated here by comparing the behaviour of a semiflexible linear wormlike micelle with that of a semiflexible homoplymer chain. This is done by simulating two persistent worms in a simulation box, each of which is a trumbbell for which only one end bead is a sticker, with a high sticker energy of $\epsilon_{st} = 30$. This ensures that the persistent worms do not form a closed loop and become a ring, and once the stickers have fused to form a linear wormlike micelle at equilibrium, the high sticker energy prevents them from coming apart in the duration of a simulation. The backbone-backbone interaction strength is assumed to be purely repulsive, i.e., $\epsilon_{bb} =0$. Since the two fused stickers count as a single effective monomer, the linear wormlike micelle (composed of two persistent worms) has a total of $N_{\text{L,lin}}^{\text{eff}} = 5$ effective monomers. Consequently, it is compared to the behaviour of a semiflexible homopolymer chain with five beads, which is simulated in the dilute limit by imposing the same bending potential between the springs and with all the beads interacting with a purely repulsive potential, $\epsilon_{bb} =0$. As illustrated schematically in \fref{fig6}~(e), beads ``1", ``2", ``4", and ``5" in the linear wormlike micelle  correspond to beads with the same labels in the semiflexible homopolymer chain. Bead ``3" in the linear wormlike micelle is positioned at the centre of mass of the positions of the stickers at the ends of the two persistent worms, and corresponds to bead ``3" on the homopolymer chain. 

Two properties are compared across the two systems. The first property is the distribution $\psi(F^{\text{(b)}}_\mu)$ of the magnitude of the bending force $F^{\text{(b)}}_\mu = \vert \bm{F}^{\text{(b)}}_\mu \vert $ acting on each of the individual beads within the linear wormlike micelle, and the corresponding beads on the semi-flexible homopolymer chain. As displayed in~\fref{fig6}~(a) to~(d), for two different values of the bending stiffness represented by the ratios $\ell_{\text{pw}}/l_p =0.01$ and $\ell_{\text{pw}}/l_p = 1$, the force distributions on the respective beads are nearly identical in both the systems. It is worth noting that for each system, the force distributions for beads 1 and 5, and beads 2 and 4 lie on top of each other, as expected from symmetry considerations. All the remaining bead pairs in the two systems, which are not displayed for the sake of brevity, behave identically for both values of the bending stiffness.

The second property that is compared is the distribution $\psi(R_\mu)$ of the distance of each bead from the center of mass of the respective chains, $R_\mu = \vert \bm{r}_\mu - \bm{r}_c \vert \, ; \, \mu = 1 \ldots 5$. As displayed in~\fref{fig6}~(f) to~(i), for nearly all the beads, the distributions are fairly identical in the two systems for both values of the bending stiffness. Notably, however, for $\ell_{\text{pw}}/l_p =0.01$ (\fref{fig6}~(g)), while the distributions of distances for beads 2 and 4 lie on top of each other for the linear wormlike micelle and the homopolymer chains, there is an observable difference in the distribution of distances in the two different systems. A similar difference (not shown here) is observed between the distribution of distances in linear wormlike micelles and homopolymer chains for beads 1 and 5. This difference is, however, not observed when $\ell_{\text{pw}}/l_p = 1$ (\fref{fig6}~(i)), which represents a less rigid chain than the chain with $\ell_{\text{pw}}/l_p = 0.01$. Indeed, this is seen to be true for all larger values of $\ell_{\text{pw}}/l_p$ that have been examined here. It appears that in the case of highly rigid, nearly rod like systems, the location of beads relative to the centre of mass is more sensitive to the assumption that the linear wormlike micelle is a hypothetical chain of effective monomers, rather than the true situation, which is that it is composed of persistent worms that are bound together at the stickers. On the other hand, as can be seen from~\fref{fig6}~(a) and~(c), the distribution of bending forces is not as sensitive to the rigidity of the linear wormlike micelle.

In the rest of this work (the primary aim of which is to introduce a new mesoscopic model), preliminary results are presented for the case where all wormlike micelles are assumed to be fully flexible, and no bending potential is applied between springs. However, the results of this section establish that the methodology introduced here can be used to examine the rheological behaviour of wormlike micelles with an arbitrary degree of bending stiffness along their backbones. 

\subsection{\label{sec:cstar} The overlap concentration}

\begin{figure}[t]
\centering
\resizebox{0.90\columnwidth}{!}{\includegraphics{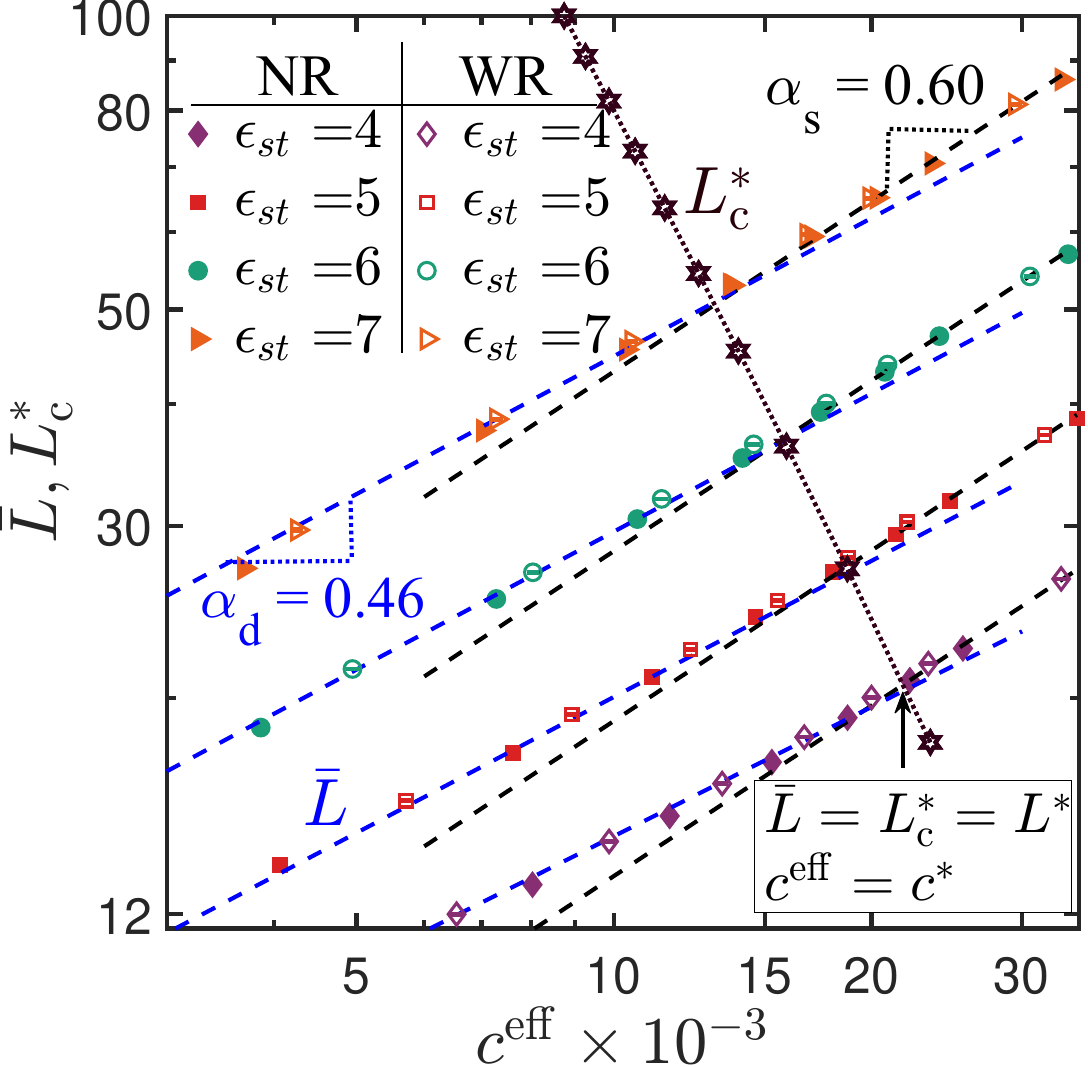}}\\
(a) \\[10pt]
\resizebox{0.90\columnwidth}{!}{\includegraphics{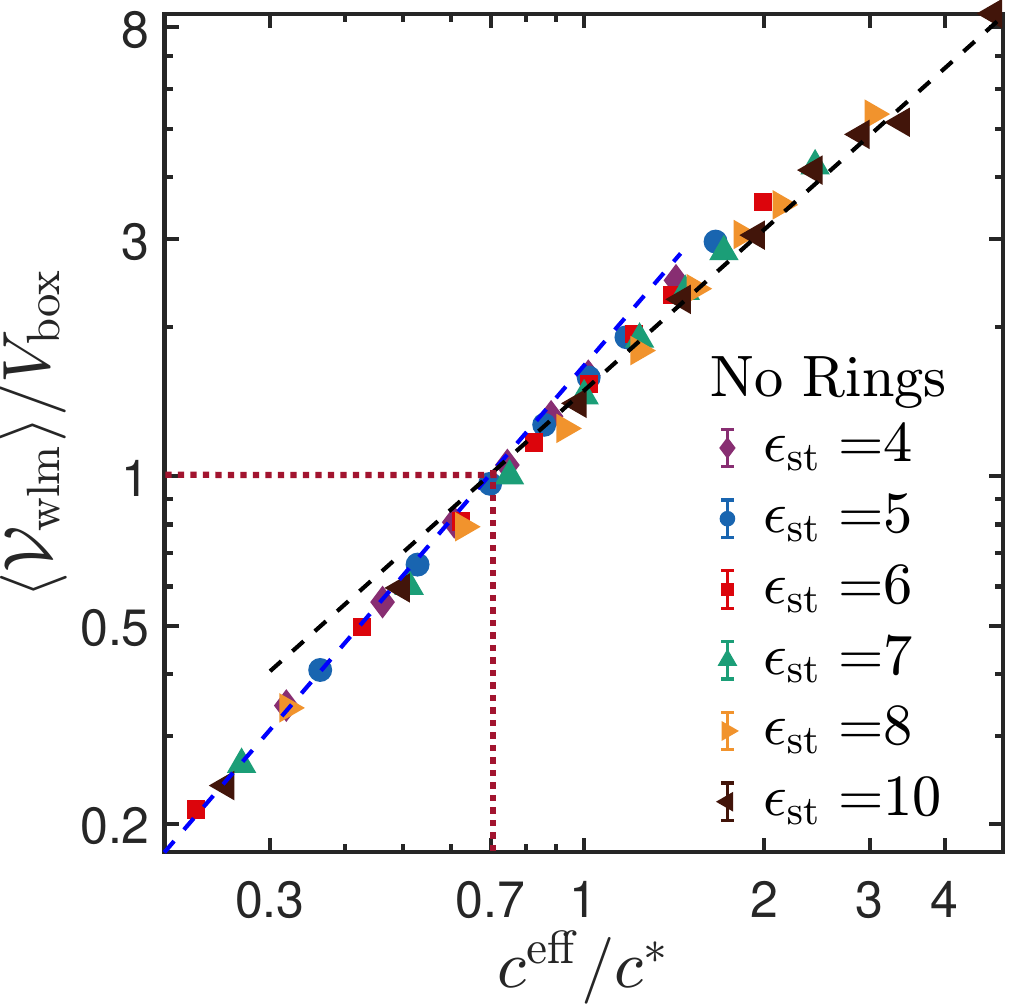}}\\
(b)
\vspace{-5pt}
\caption{(a) Determination of the overlap concentration in a wormlike micelle solution for various values of the sticker energy $\epsilon_{st}$. The intersection point of the two functions $L^*_{\text{c}} (c^{\text{eff}})$ and $\bar{L}(c^{\text{eff}})$, for any sticker energy determines the overlap length $L^*$ and the overlap concentration $c^*$, for that value of $\epsilon_{st}$. (b) Comparison of the two definitions of $c^*$ discussed in \sref{sec:overlap_conc}, where $\langle \mathcal{V}_{\text{wlm}}\rangle$ is the total volume occupied by all the linear wormlike micelles in the solution and $V_{\text{box}}$ is the volume of the simulation box. \label{fig7}}
\end{figure}

Following the procedure outlined in \sref{sec:overlap_conc}, the dependence of $L^*_{\text{c}}$ and $\bar{L}$ on $c^{\text{eff}}$, for various values of sticker energy $\epsilon_{st}$, was computed by varying $c/c^*_{\text{pw}}$ for systems with and without rings, and plotted together, as shown in \fref{fig7}~(a). As discussed in \sref{sec:overlap_conc}, the intersection point of the two functions $L^*_{\text{c}} (c^{\text{eff}})$ and $\bar{L}(c^{\text{eff}})$, for any sticker energy, determines the pair $(L^*, c^*)$ for that value of $\epsilon_{st}$. While, as expected, $L^*_{\text{c}}$ is a monotonically decreasing function of $c^{\text{eff}}$, the dependence of $\bar{L}$ on $c^{\text{eff}}$ clearly displays two distinct power law regimes that lie on either side of the intersection point, which corresponds to the overlap concentration, with slopes represented by $\alpha_{\text{d}} = 0.46$ and $\alpha_{\text{s}} = 0.6$. A more thorough discussion of this interesting observation is postponed to \sref{sec:lengths}, where it interpreted in the light of analytical mean-field theories. Notably, in systems in which both linear chains and rings are present, the presence of rings makes no difference to the estimation of $c^*$. 

\begin{figure}[t]
\centering
\resizebox{0.90\columnwidth}{!}{\includegraphics{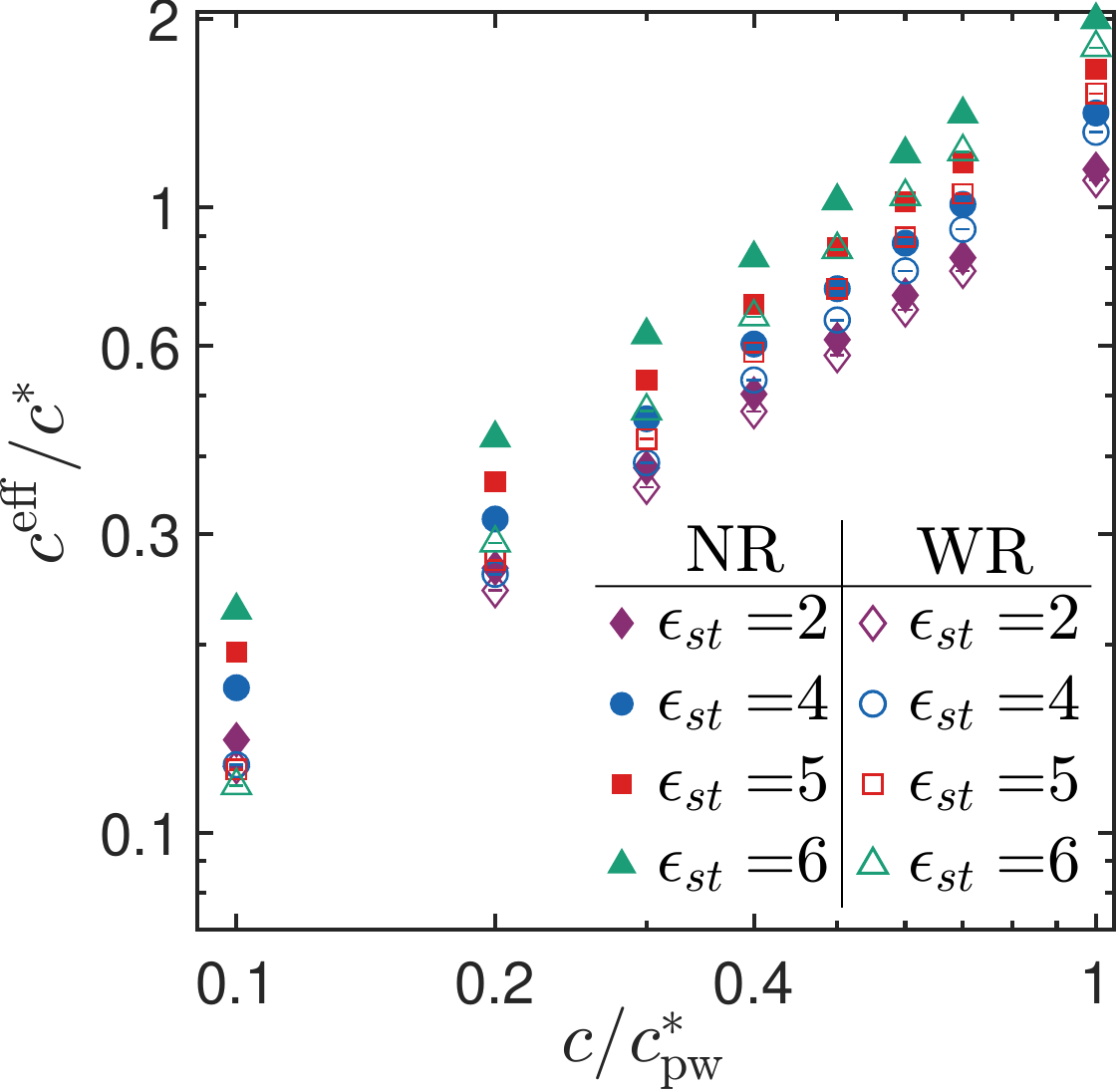}}\\
(a) \\[10pt]
\resizebox{0.90\columnwidth}{!}{\includegraphics{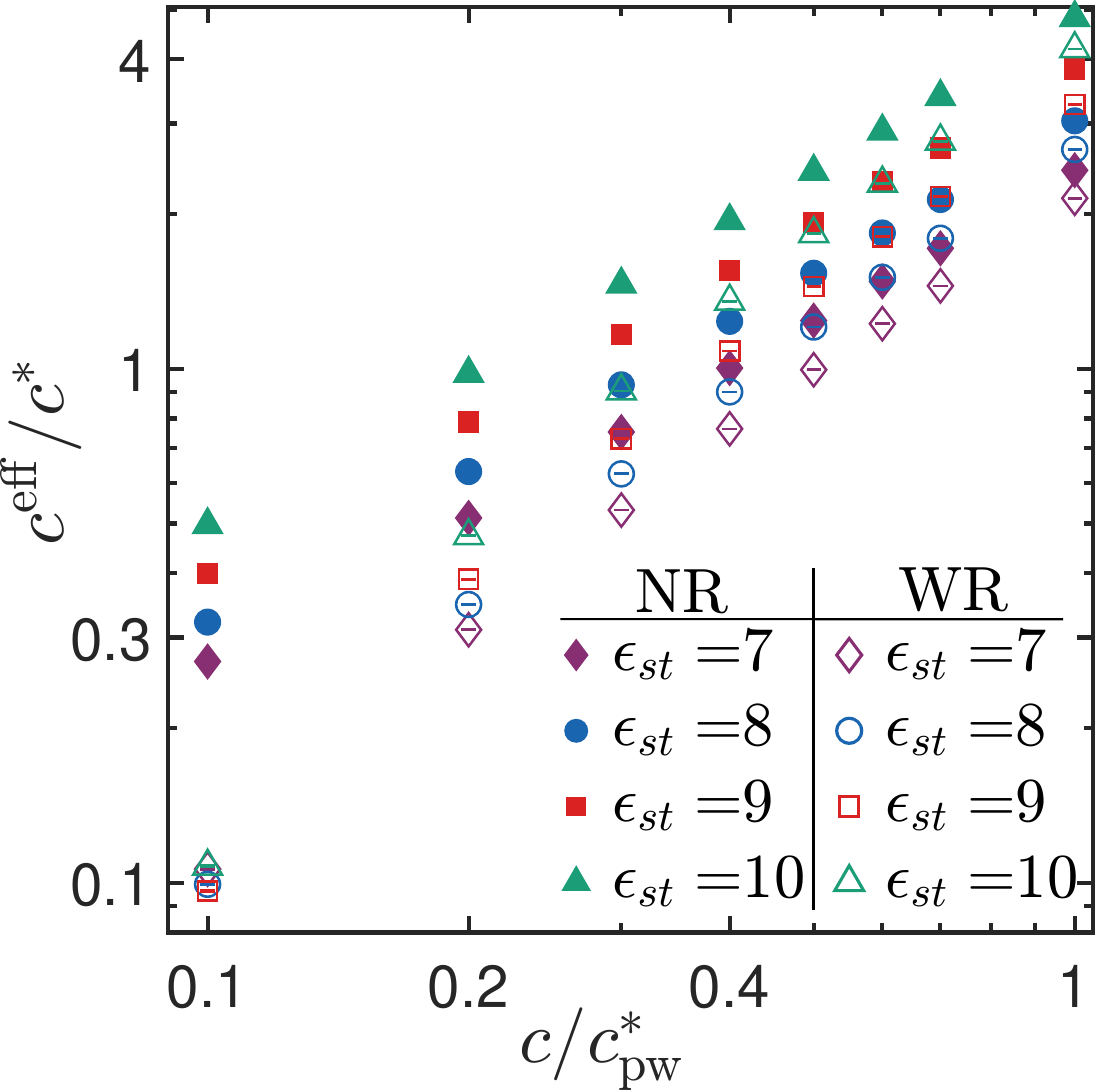}}\\
(b)
\vspace{-5pt}
\caption{Plots of $c^{\mathrm{eff}}/c^*$ versus $c/c_{\mathrm{pw}}^*$ for various sticker energies comparing systems without rings (NR, filled symbols) and with rings (WR, open symbols).  (a) $\epsilon_{st} \leq 6$, (b) $\epsilon_{st} > 6$.} {\label{fig8}}
\end{figure}

The scaled concentration $c^{\text{eff}}/c^*$, determined in this manner, for various values of $c/c^*_{\text{pw}}$ and $\epsilon_{st}$, is displayed in~\frefs{fig8}~(a) and~(b), for systems with and without rings. The plots show that $c^{\mathrm{eff}}/c^*$ is significantly higher for wormlike micellar solutions containing only linear chains (No Rings) compared to those with blend of linear chains and rings (With Rings), and this difference increases with the sticker energy $\epsilon_{st}$.

Figure~\ref{fig7}~(b) is a plot comparing the alternative definition of $c^*$ discussed in \sref{sec:overlap_conc}, with the value determined as discussed above. The $x$-axis in \fref{fig7}~(b) is the scaled concentration $c^{\text{eff}}/c^*$, with the value of $c^*$ determined from \fref{fig7}~(a), while the $y$-axis is the ratio $\langle \mathcal{V}_{\text{wlm}}\rangle/V_{\text{box}}$, with $\mathcal{V}_{\text{wlm}}$ defined in \eref{eq:vwlm}. Clearly, if the two definitions were identical, $\langle \mathcal{V}_{\text{wlm}}\rangle /V_{\text{box}}$ would be equal to one, when  $c^{\text{eff}}/c^* = 1$. While this is not the case, since \fref{fig7}~(b) indicates that $c^{\text{eff}}/c^* \approx 0.7$ when $\langle \mathcal{V}_{\text{wlm}}\rangle /V_{\text{box}} =1$, it suggests that the two definitions do not lead to values of $c^*$ that are very far apart. Remarkably, the data collapses onto a single curve, independent of sticker energy $\epsilon_{st}$, and a distinct change in slope is observed at $\langle \mathcal{V}_{\text{wlm}}\rangle /V_{\text{box}} =1$, suggesting a crossover from one regime to another. 

In this work, the procedure described in \sref{sec:overlap_conc} is used to determine the overlap concentration (displayed in~\fref{fig7}~(a) and~\frefs{fig8}). As will be demonstrated in subsequent sections, the use of this value $c^*$ to separate the dilute and semidilute concentration regimes leads to results that are completely consistent with the predictions by analytical mean-field theories.


\subsection{\label{sec:staticprop} Static properties of wormlike micellar solutions}

Computational results for a number of static properties of wormlike micellar solutions are presented in this section, along with a comparison with the predictions of scaling and mean-field theories. A brief summary of the results of analytical theories is given in \sref{sec:anal} below, for ease of reference. 

\begin{figure*}[t]
\begin{center}
\begin{tabular}{ccc}
\includegraphics[width=5.5cm]{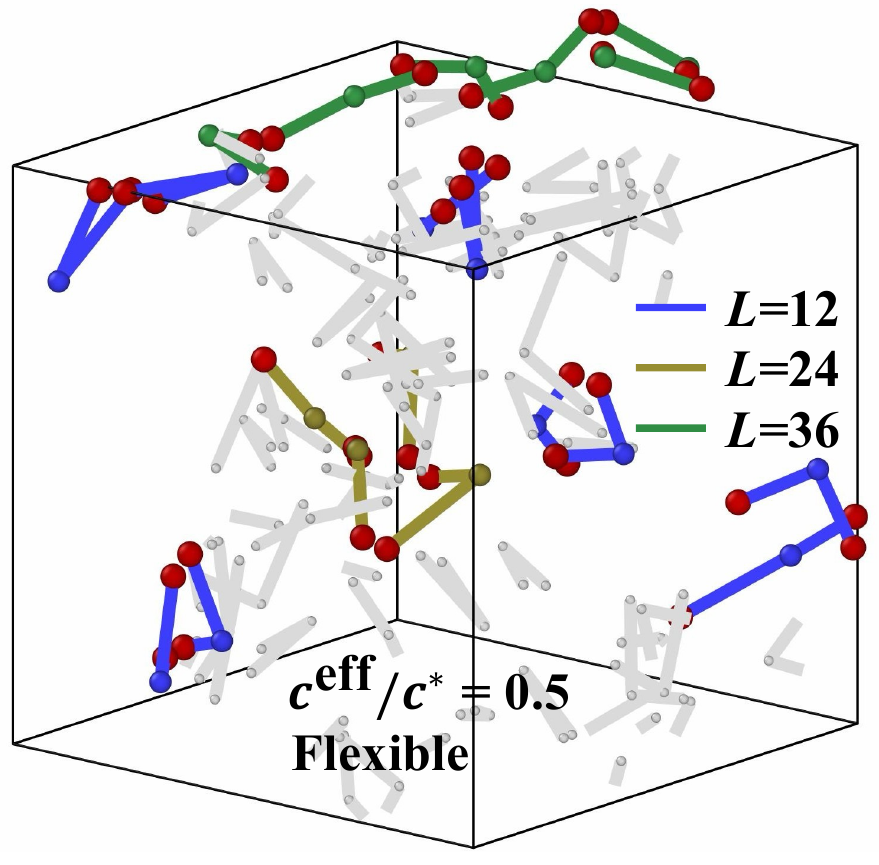} &
\includegraphics[width=5.5cm]{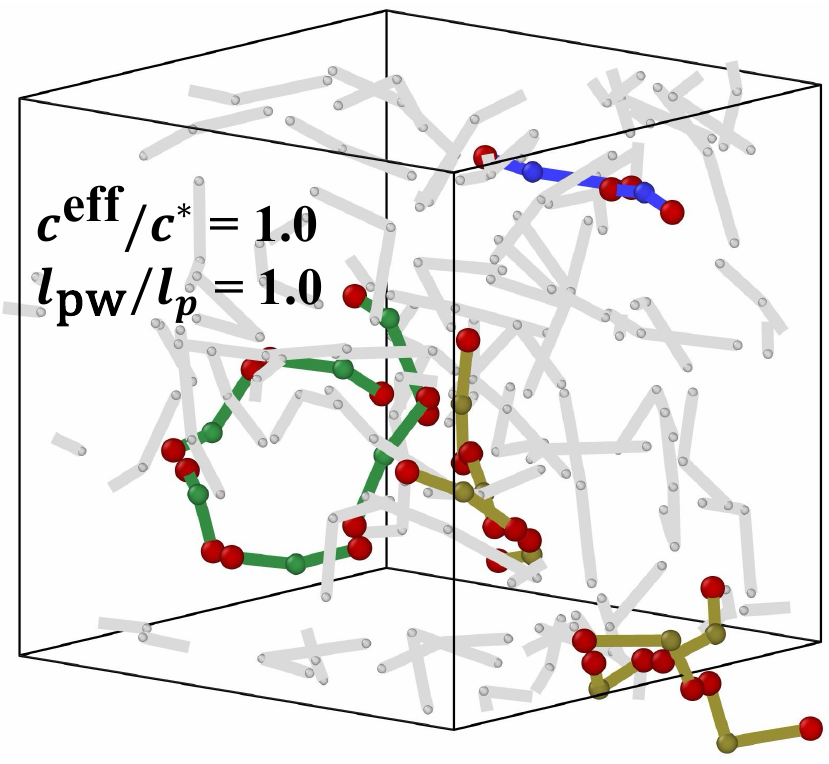} &
\includegraphics[width=6.0cm]{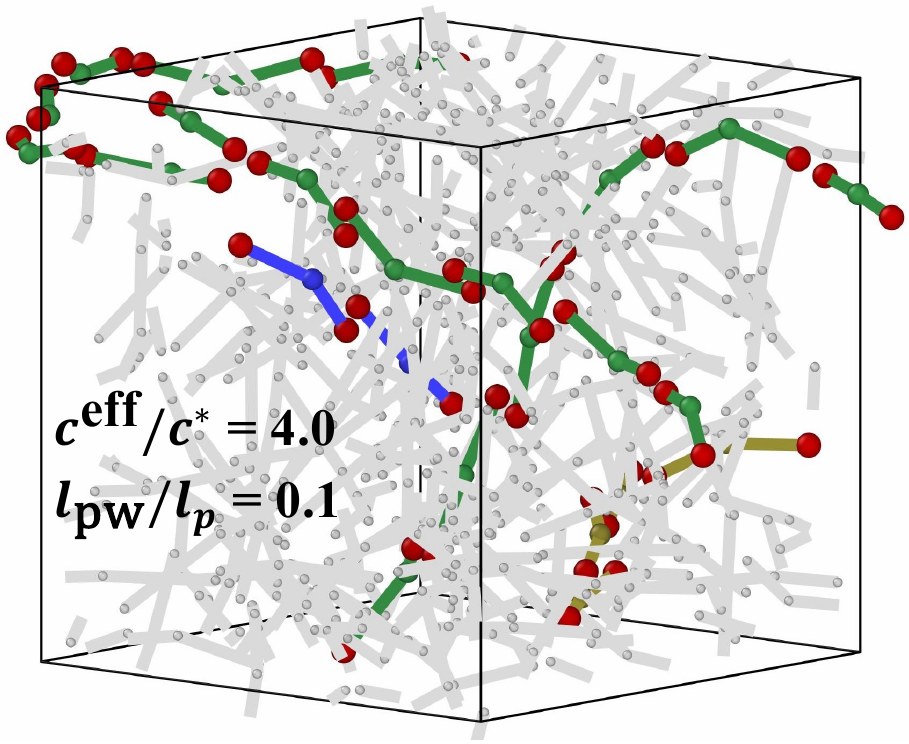} \\
(a) Dilute & (b) Overlap & (c) Semidilute
\end{tabular}
\end{center}
\vspace{-10pt}
\caption{Simulation snapshots of wormlike micelle conformations at different concentrations and bending stiffnesses. Periodic images of chains have not been depicted while chains leaving a simulation box are shown. Three different lengths, $L =12$, $24$ and $36$, have been highlighted, with the respective colours indicated in the legend to (a). Note that $L = m^{\text{L}}_{\text{pw}} (N_{\text{pw}} -1) b$, where $m_{\text{L}}^{\text{pw}}$ is the number of persistent worms in a micelle. Persistent worms are trumbbells in these simulations.  In all figures, $\epsilon_{st} = 7, b = 3, N_{\text{pw}} = 3$. (a) Dilute concentration regime with $c^{\text{eff}}/c^* = 0.5$,  $\ell_{\text{pw}}/l_p \to \infty$, $n_{12}^{\text{lin}} = 5$, $n_{24}^{\text{lin}} = 1$ and $n_{32}^{\text{lin}} = 1$. (b) Overlap concentration with $c^{\text{eff}}/c^* = 1.0$,  $\ell_{\text{pw}}/l_p = 1.0$, $n_{12}^{\text{lin}} = 1$, $n_{24}^{\text{lin}} = 2$ and $n_{32}^{\text{lin}} = 1$. (c) Semidilute concentration regime with $c^{\text{eff}}/c^* = 4.0$,  $\ell_{\text{pw}}/l_p = 0.1$, $n_{12}^{\text{lin}} = 1$, $n_{24}^{\text{lin}} = 1$ and $n_{32}^{\text{lin}} = 3$.  \label{fig9}}
\end{figure*}

To give a sense of the typical conformations of wormlike micelles that are captured in the simulations (before getting into the details of various specific property predictions),  \fref{fig9} displays snapshots of simulation boxes for various concentrations and wormlike micelle bending stiffnesses. In particular, a snapshot in the dilute regime is depicted in \fref{fig9}~(a), with $c^{\text{eff}}/c^* = 0.5$, with the simulation box containing fully flexible wormlike micelles ($\ell_{\text{pw}}/l_p \to \infty$), \fref{fig9}~(b) is a snapshot at the overlap concentration depicting semiflexible wormlike micelles with the ratio $\ell_{\text{pw}}/l_p = 1.0$, and \fref{fig9}~(c) is a snapshot in the semidilute concentration regime, with $c^{\text{eff}}/c^* = 4.0$, and with the simulation box containing significantly more rigid wormlike micelles with $\ell_{\text{pw}}/l_p = 0.1$. Note that  the overlap concentration $c^*$ has been determined as described in \fref{fig5}. Three different lengths of micelles ($L = 12$, $24$, and $36$) have been highlighted (with the remaining wormlike micelles greyed out for clarity), and the figure caption indicates the number of micelles of each highlighted length,  $n_{12}^{\text{lin}}, n_{24}^{\text{lin}}$ and $n_{32}^{\text{lin}}$, respectively, that are present in a simulation box. 

Equilibrium simulations have been carried out to compare results with the predictions of static properties by analytical theories, which are summarised in the section below. 

\subsubsection{\label{sec:anal} Analytical predictions of static properties}

According to mean-field theory~\cite{Wittmer1998,Huang2006}, the probability distribution of scaled lengths of linear wormlike micelles, $p(x)$, is given by a Schultz-Zimm distribution for dilute wormlike micelle solutions and a purely exponential distribution in the semidilute regime,
    \begin{equation}\label{lendist}
    p(x) =
    \begin{cases}
      \dfrac{\gamma^{\gamma}}{\Gamma(\gamma)} \, x^{\gamma-1} \, e^{-\gamma x} \, ;  \quad \bar{L} \ll L^* \quad(\text{dilute}) \\[10pt]
      e^{-x} \, ; \quad \bar{L} \gg L^* \quad (\text{semidilute}) 
    \end{cases}  
    \end{equation}
where $\gamma$ is the self-avoiding walk susceptibility exponent, for which the most accurate value to date, calculated by \citet{Clisby2017} using high precision Monte Carlo, is $\gamma = 1.156 953 00(95)$. Earlier studies  of linear wormlike micelles in dilute solutions, carried out with a variety of different numerical methods~\cite{Wittmer1998,Wittmer2000,PaddingPRE2004,Huang2006}, have used values of $\gamma$ ranging from 1.158 to 1.165 to fit their predictions of the distribution of lengths with the Schultz-Zimm  distribution.  

In mean-field theory, the mean length of a chain is a function of monomer volume fraction $\phi$ and sticker energy $\epsilon_{st}$, and is predicted to be given by,
\begin{equation}{\label{meanL}}
  \bar{L} =  L^* \left( \frac{\phi}{\phi^*} \right)^\alpha =  A \, \phi^\alpha \exp \left( \delta \, \epsilon_{st} \right)
\end{equation}
where $\phi^*$ is the overlap volume fraction. The amplitude $A$ and exponents $\alpha$ and $\delta$ assume different values in the dilute and semidilute regime, as discussed further below. The quantities $\phi^*$ and $L^*$ are expected to be exponential functions of the sticker energy,
\begin{align}\label{cstarLstar}
    \phi^* & = P \, e^{- {\epsilon_{st}}/ {\varphi} } \nonumber \\[10pt] 
    L^* & = Q \, e^{ {\epsilon_{st}}/ {\kappa} }
\end{align}
with the exponents $\varphi$ and $\kappa$ in \eref{cstarLstar} related to the exponents $\alpha$ and $\delta$ in \eref{meanL}, 
\begin{align}\label{cstarlstarexp}
    	\varphi & = \frac{\alpha_s - \alpha_d}{\delta_s - \delta_d}  \nonumber \\[10pt] 
	\kappa & = \left( \nu d - 1 \right) \varphi
\end{align}
where $(\alpha_d, \delta_d)$ and $(\alpha_s, \delta_s)$ denote the pair of exponents $(\alpha, \delta)$ in the dilute and semidilute regime, respectively, $\nu$ is the Flory exponent and $d$ is the dimension of space. The amplitudes $P$ and $Q$ are related to the prefactors $A_d$ and $A_s$ in the expression for the mean-chain length (\eref{meanL}) defined in the dilute and semidilute density regimes, respectively, and given by, 
\begin{align}\label{amplitudePQ}
    P & = \left( \dfrac{A_d}{A_s} \right)^{{1}/({\alpha_s - \alpha_d})} \nonumber \\[10pt] 
    Q & = A_s P^{\alpha_s}
\end{align}
Finally, the exponents $\alpha$ and $\delta$ in each regime can be shown analytically to obey,
\begin{align}\label{exponents}
  \alpha_d & = \dfrac{1}{1+\gamma} = 0.46\,; \quad \delta_d = 0.46 \nonumber \\[10pt] 
   \alpha_s & = \dfrac{1}{2} \left\lbrack 1+ \dfrac{\gamma - 1}{ \nu d - 1} \right\rbrack = 0.6 \,; \quad \delta_s =0.5 
\end{align}
With these values of the exponents $\alpha$ and $\delta$ in the two regimes, it follows from \eref{cstarlstarexp} that  $\varphi \approx 3.5$ and $\kappa \approx 2.8$ (using $\nu = 0.6$). The non-universal amplitudes $A_d$, $A_s$, $P$  and $Q$ are system dependent, but must obey the relationships given in~\eref{amplitudePQ}. 

Mean-field theories have also been developed for wormlike micellar solutions that contain rings as well as linear chains~\cite{Milchev2000,Wittmer2000,PaddingPRE2004}. Of the many static and dynamic properties that have been discussed, attention is restricted here to the probability $\psi_{\text{R}} (L) \, dL$ of finding ring wormlike micelles of length between $L$ and $L + \Delta L$, with the probability density predicted to have the form,
\begin{equation}\label{ringdist}
\psi_{\text{R}} (L) = \lambda_0 \, L^{- (1 + \nu d)} \, e^{-\mu L} \, H \left( L - L^R_{\text{min}} \right)
\end{equation}
where $\lambda_0$ is a constant, $\mu = \gamma_{\text{eff}}  / \bar{L} $, with $\bar{L}\, (c^{\text{eff}}/c^*, \epsilon_{st})$ being the mean length of linear wormlike micelles at the given concentration and  persistent worm sticker energy, and the Heaviside function $H\left( L - L^R_{\text{min}} \right)$ enforces a minimum ring size of length $L^R_{\text{min}}$. In the present model, since the stickers at the two ends of a single persistent worm can adhere to form a ring, $ L^R_{\text{min}} = \left( N_{\text{pw}} - 1 \right) b$. In contrast to the exponential length dependence of linear wormlike micelles, the distribution of ring lengths is expected to obey a power law distribution, with an exponential damping factor that depends on the mean length of linear wormlike micelles.

Since the computations in this work have been carried out in terms of the effective monomer concentration $c^{\text{eff}}$, results are reported here, and comparisons with analytical expressions are carried out, with the concentration ratio $c^{\text{eff}}/c^*$ in place of the volume fraction ratio $\phi/\phi^*$.

\begin{figure*}[t]
\begin{center}
\begin{tabular}{cc}
\includegraphics[width=8.2cm]{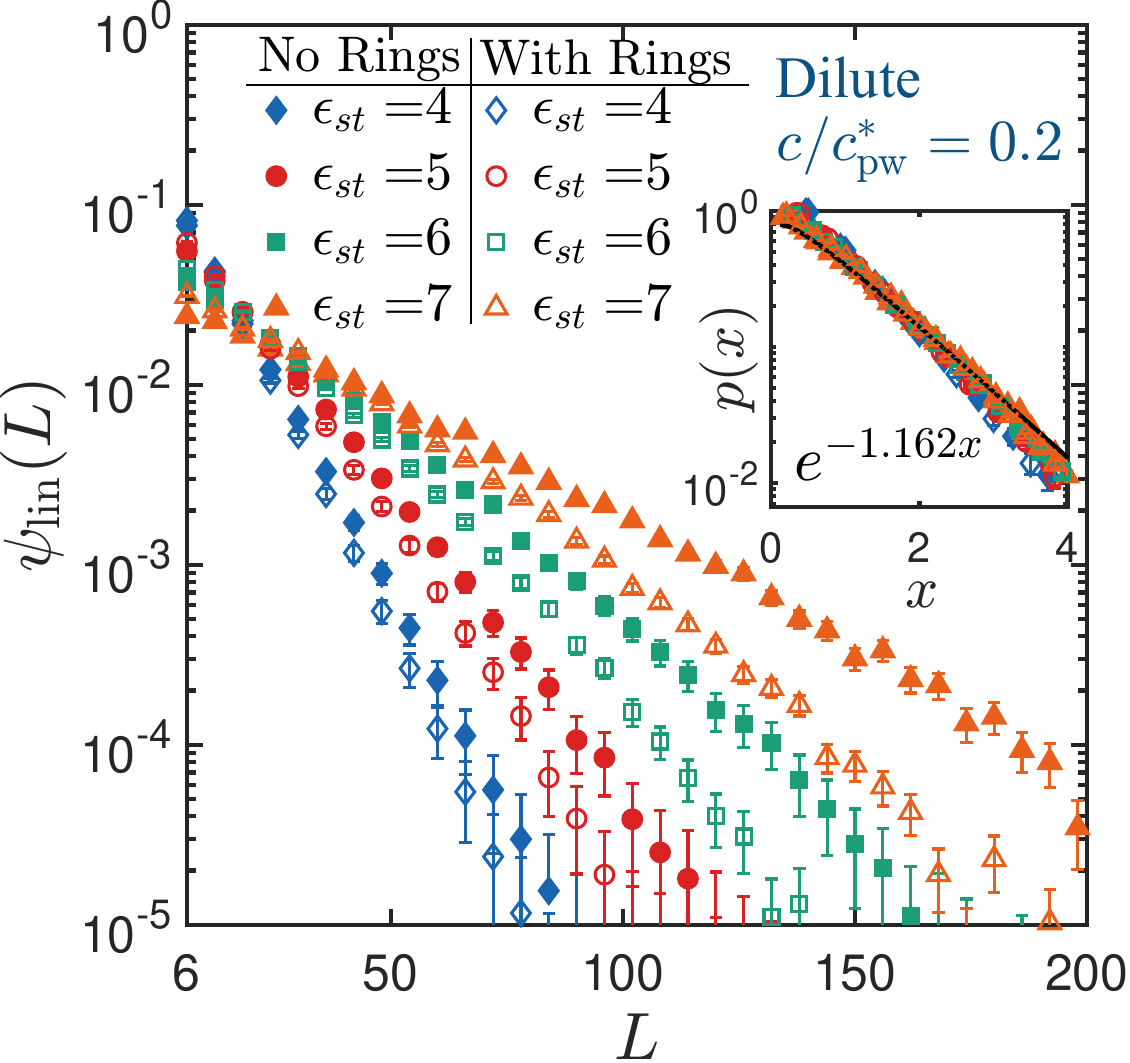} &
\includegraphics[width=8.2cm]{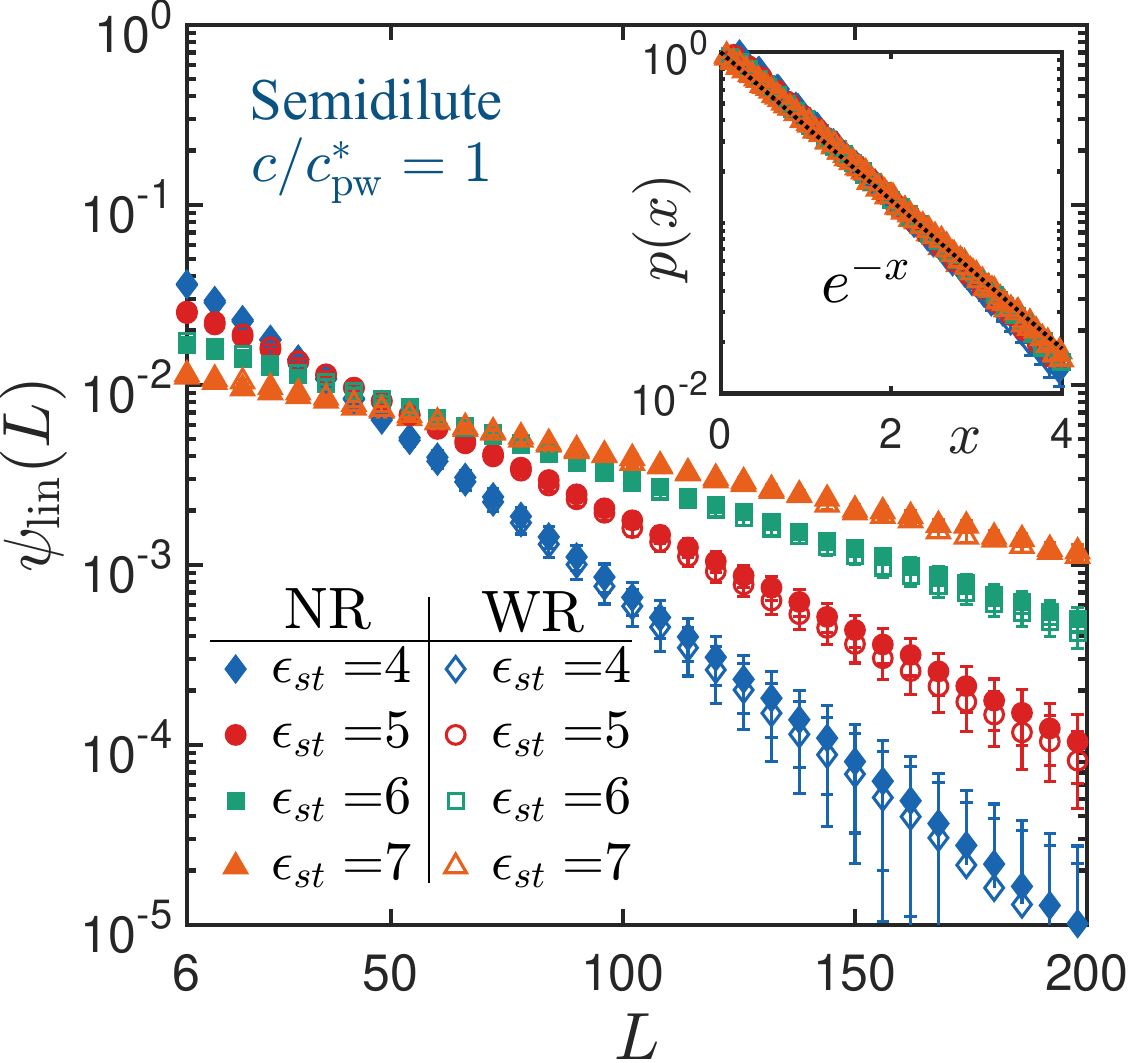} \\
(a)  & (b) 
\end{tabular}
\end{center}
\vspace{-10pt}
\caption{Probability distribution of lengths of linear wormlike micelles for systems with (WR) and without (NR) rings, at various values of the sticker energy  $\epsilon_{st}$, in (a) dilute and (b) semidilute solutions. The concentration is reported in terms of $c/c^*_{\text{pw}}$, which is held constant in the two regimes. Each set of symbols at a fixed value of $\epsilon_{st}$ corresponds to a unique value of scaled concentration $c^{\text{eff}}/c^*$, which is related to $c/c^*_{\text{pw}}$ and $\epsilon_{st}$ as displayed in~\fref{fig8}. 
\label{fig10}}
\end{figure*}

\subsubsection{\label{sec:lengths} Length distributions and the mean length of linear wormlike micelles}

The distribution of lengths $\psi_{\text{lin}} (L)$ of linear wormlike micelles at various values of the sticker energy, for systems with and without rings, in the dilute and semidilute regimes is displayed in \fref{fig10}~(a) and~(b), respectively. When represented in terms of  $\psi_{\text{lin}} (L)$, a difference in the distributions is clearly visible when rings are present or absent in the system, though the difference is more pronounced at the lower concentration and at higher sticker energies. This is because, as will be demonstrated later, the fraction of rings is higher in these cases. However, as shown in the insets, when represented in terms of the distribution $p(x)$ of scaled lengths, $x = L/ \bar{L}$, the data collapses onto a unique curve, independent of sticker strength, whether there are rings present or absent. The dependence of the length distribution  $\psi_{\text{lin}} (L)$ on $c^{\text{eff}}/c^*$ and $\epsilon_{st}$ in each regime is absorbed into the mean length $\bar{L}$ when represented in terms of $p(x)$. In order to estimate $x$ and $p(x)$, it is necessary to determine the mean length of the linear wormlike micelles in the system. 

\begin{figure}[b]
    \centering
\resizebox{8.2 cm}{!}{\includegraphics*[width=12cm]{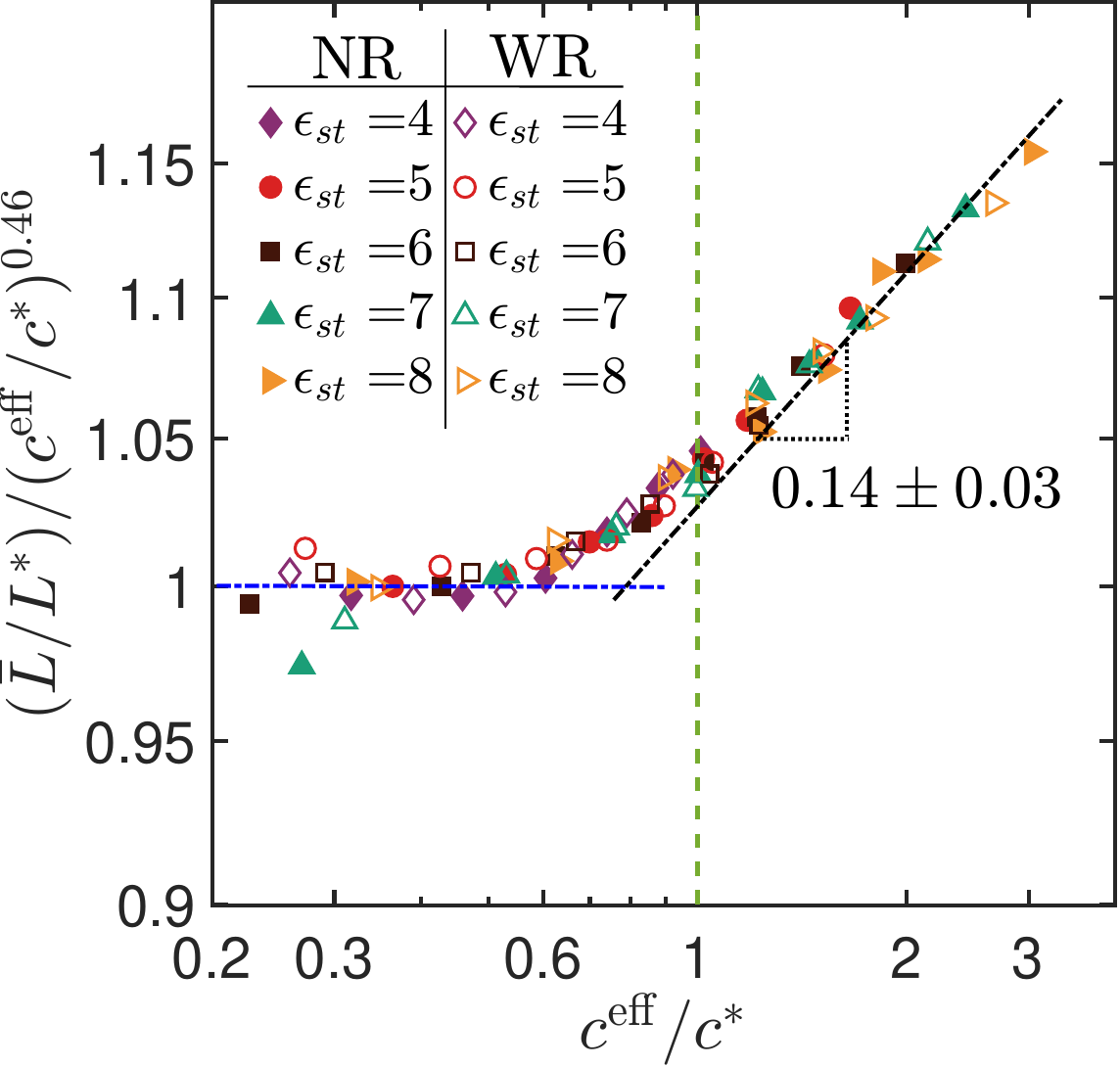}} 
    \vspace{-10pt}
    \caption {The dependence of the ratio of the mean length $\bar{L}$ to the overlap length $L^*$ on the scaled concentration $c^{\text{eff}}/c^*$ for systems with (WR) and without rings (NR), at various values of the sticker energy  $\epsilon_{st}$. The collapse of data, and the values of the power law exponents in the two regimes agree with the predictions of mean-field theory.}
    \label{fig11}
    \end{figure} 
    
According to mean-field theory, when the mean length $\bar{L}$ is scaled by the overlap length $L^*$ and plotted as a function of $c^{\text{eff}}/c^*$, data for different sticker energies is expected to collapse on to a master plot, as expressed in~\eref{meanL}, with the exponent $\alpha$ assuming values of $0.46$ and $0.6$ in the dilute and semidilute regimes, respectively. \fref{fig11} is a plot of simulation results in terms of these variables, where the data is plotted as $\left( \bar{L} / L^{*} \right) / \left[ ( c^{\mathrm{eff}} /c^{*} )^{0.46} \right]$ versus $c^{\mathrm{eff}} /c^{*}$ to better illustrate the crossover from the dilute to the semidilute regime. The collapse of data for different sticker energies is clearly visible, with the expected values of the exponents in the two concentration regimes. The gradual change in the effective exponent occurs in the concentration range $0.7 \lesssim c^{\mathrm{eff}}/c^* \lesssim 1$. The scatter in the data at low concentrations and sticker energies is because the mean length of micelles is too short to obey scaling theories. The collapse of data when represented in terms of scaled variables, and the scatter at low values of $c^{\text{eff}}/c^*$ and $\epsilon_{st}$, mirrors the identical observations made earlier in their Monte Carlo simulations by~\citet{Wittmer1998}.

\begin{figure*}[t]
\begin{center}
\begin{tabular}{cc}
\includegraphics[width=8cm]{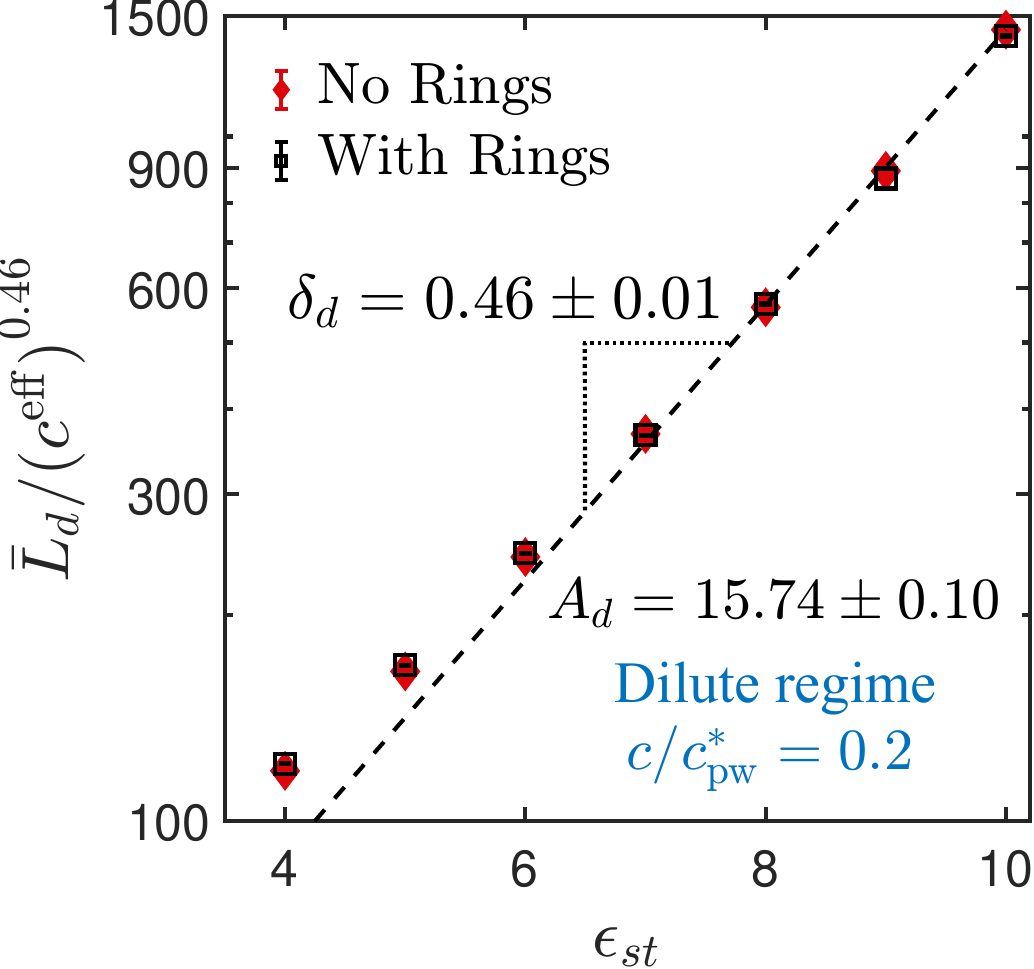} &
\includegraphics[width=7.95cm]{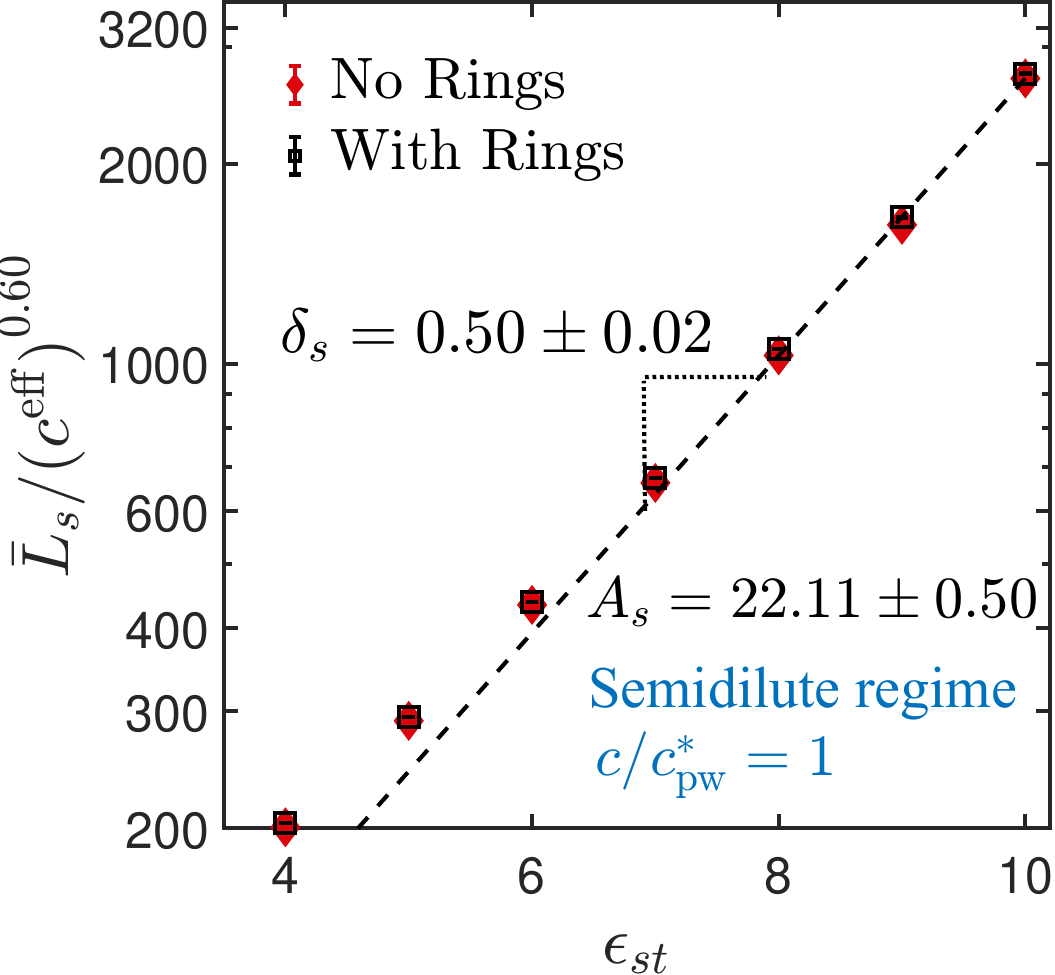} \\
(a)  & (b) \\[10pt]
\includegraphics[width=8cm]{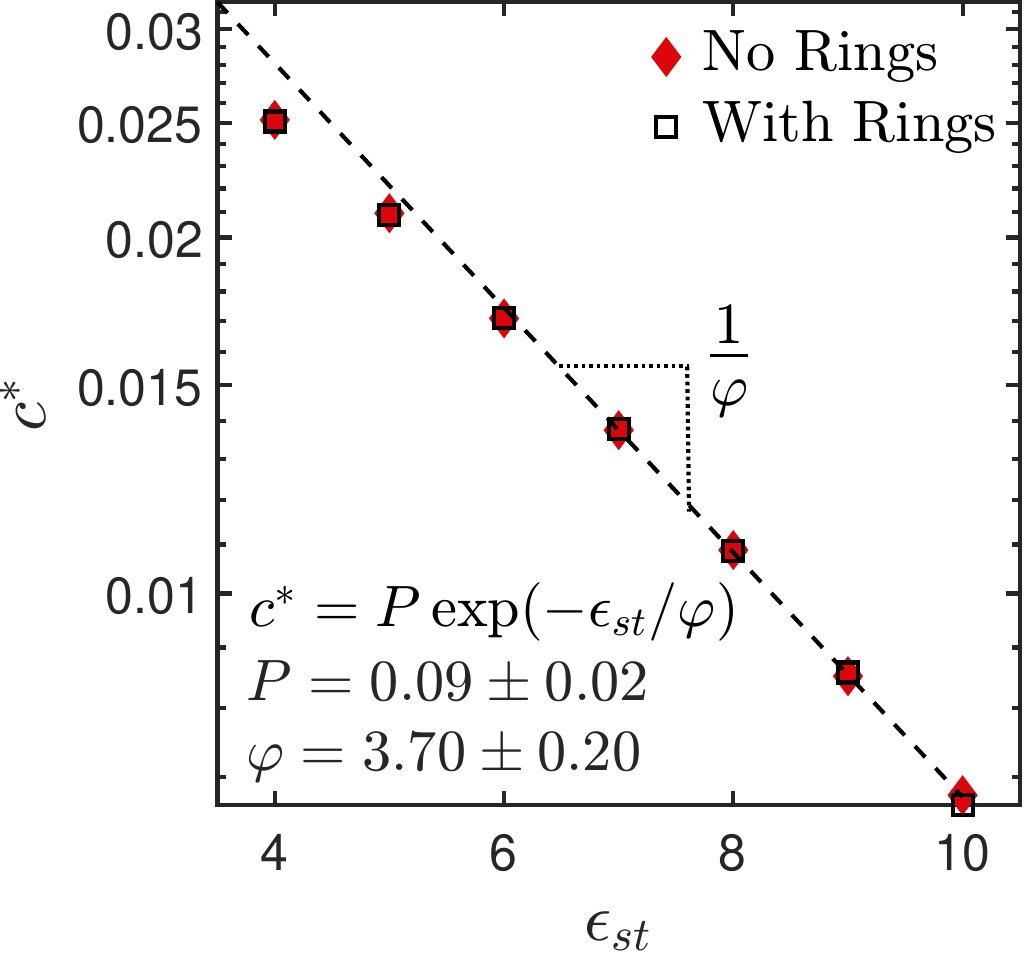} &
\includegraphics[width=7.65cm]{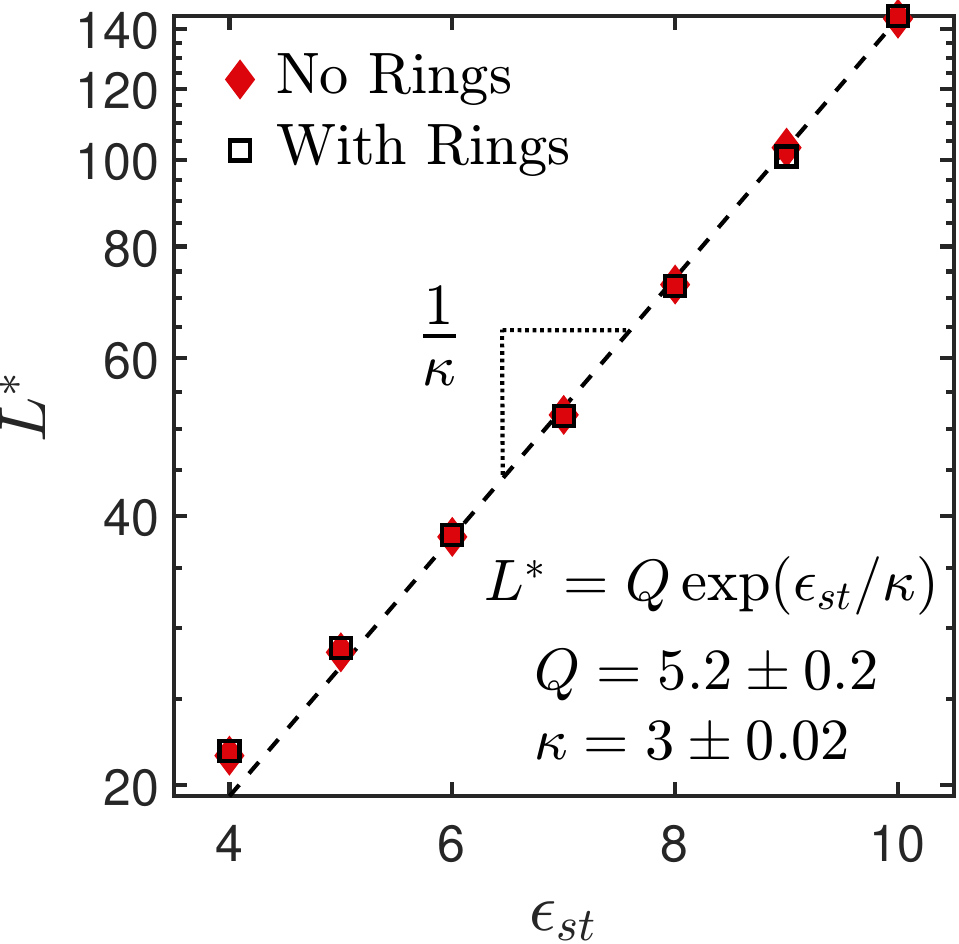} \\
(c)  & (d) 
\end{tabular}
\end{center}
    \vspace{-10pt}
\caption{Prefactors and exponents in the expression for the mean length (\eref{meanL}) in (a) dilute and (b) semidilute solutions, and amplitudes and exponents  in the expression for (c) the overlap concentration $c^*$ and (d) the overlap length $L^*$ (\eref{cstarLstar}), predicted by mean-field theory.   \label{fig12}}
\end{figure*}

The overlap concentration $c^*$ and the overlap length $L^*$  have been computed as described in~\sref{sec:overlap_conc}, with the results displayed earlier in~\fref{fig7} for various values of $\epsilon_{st}$. The power law exponents of $0.46$ and $0.6$ for the dependence of $\bar{L}$ on $c^{\text{eff}}$, seen in~\fref{fig7}~(a),  can now be understood as a validation of the prediction by mean-field theory (see~\eref{meanL}). The data collapse for systems with and without rings also indicates that the presence of rings makes no difference to the static properties of linear wormlike micelles, in line with a similar observation made earlier by~\citet{Wittmer2000}.

\begin{figure*}[t]
\begin{center}
\begin{tabular}{cc}
\includegraphics[width=8.1cm]{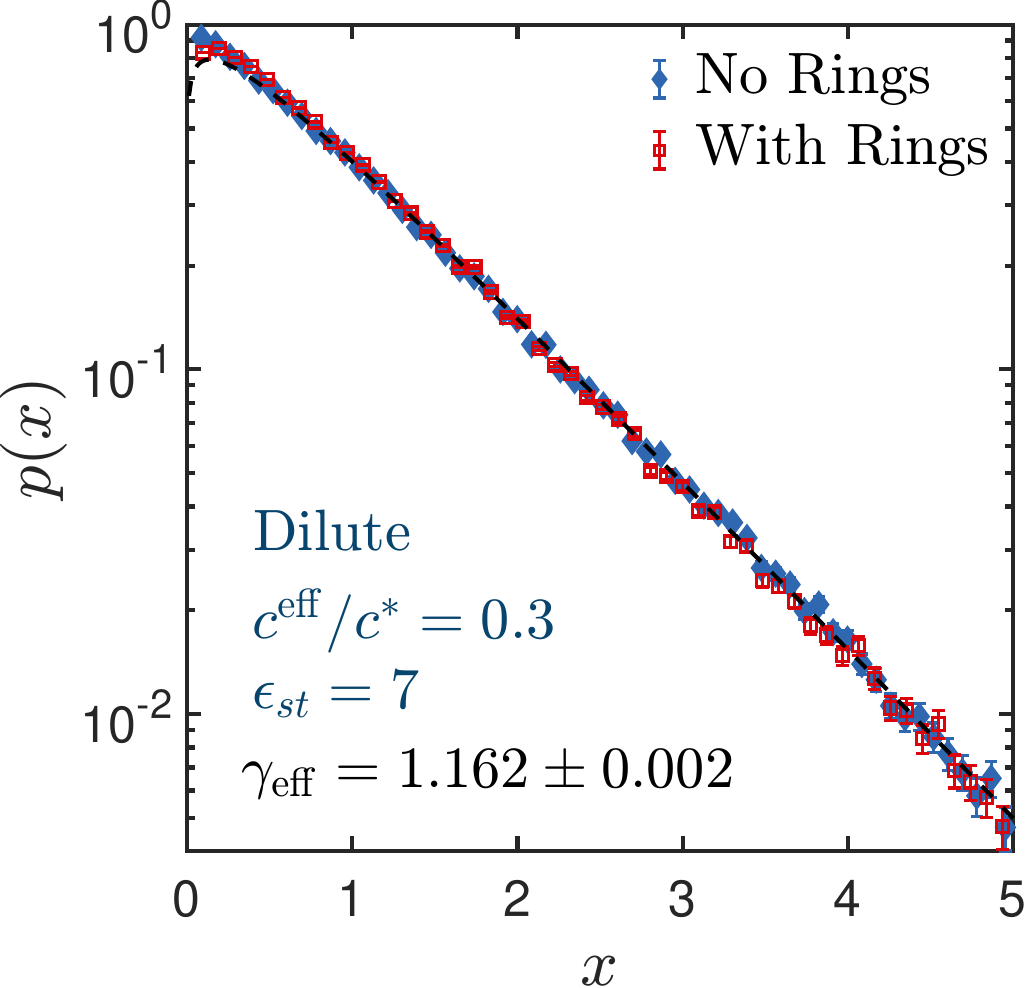} &
\includegraphics[width=8.1cm]{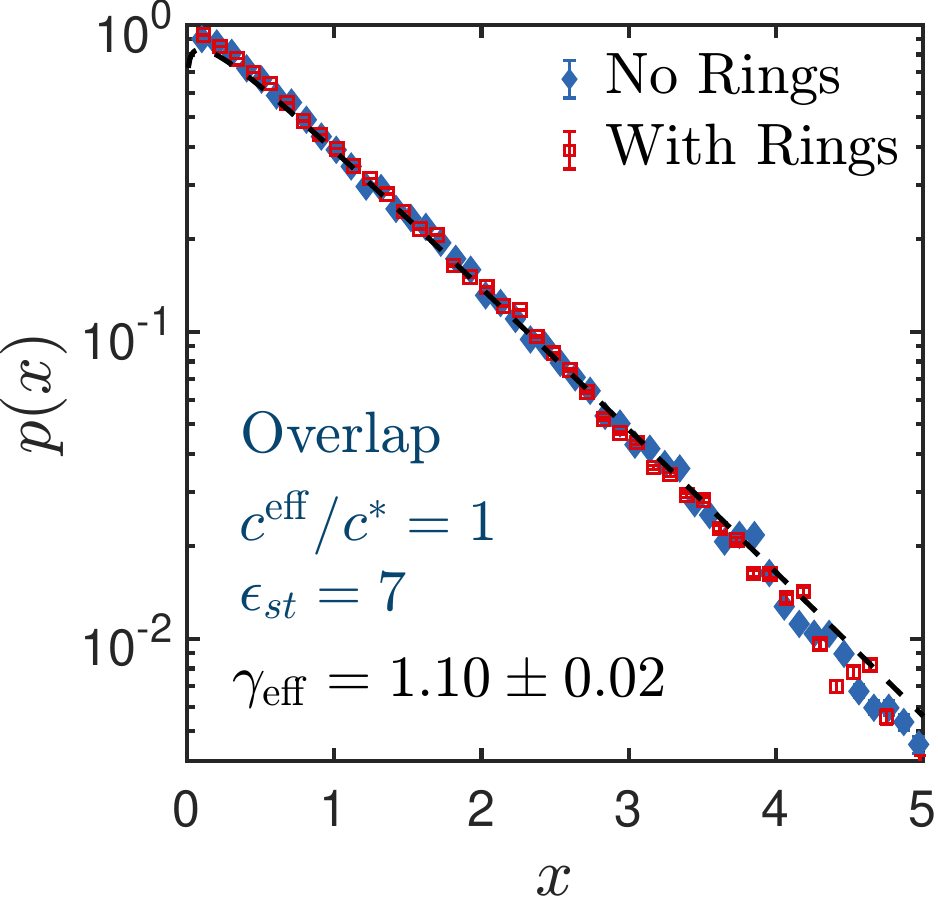} \\
(a)  & (b) \\[10pt]
\includegraphics[width=8.1cm]{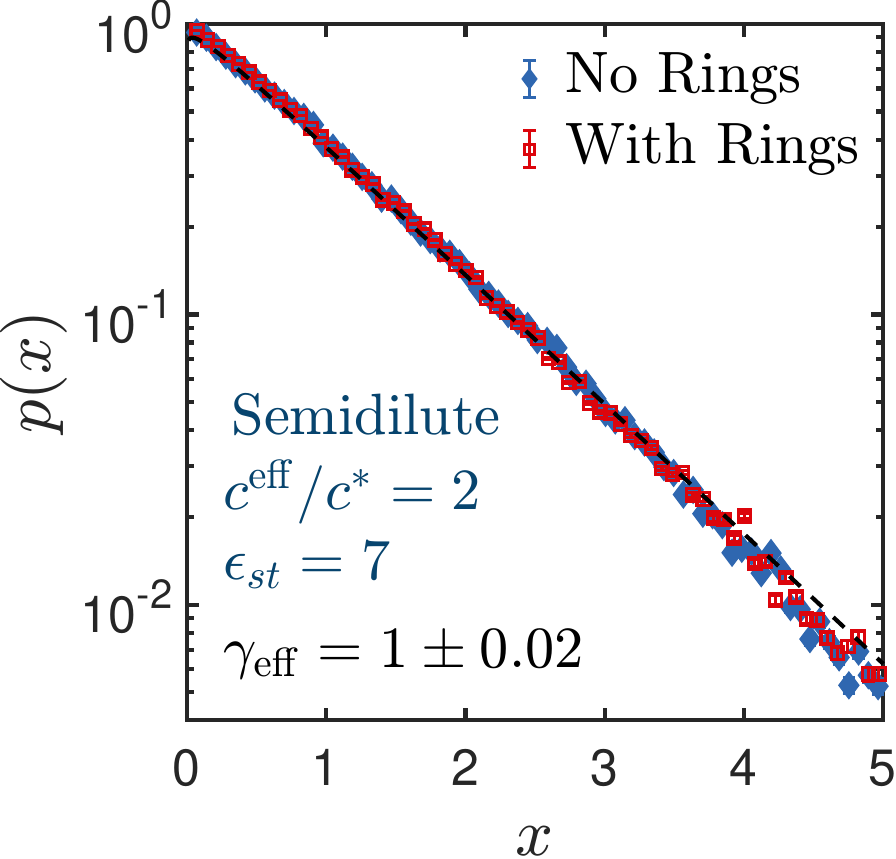} &
\includegraphics[width=8.1cm]{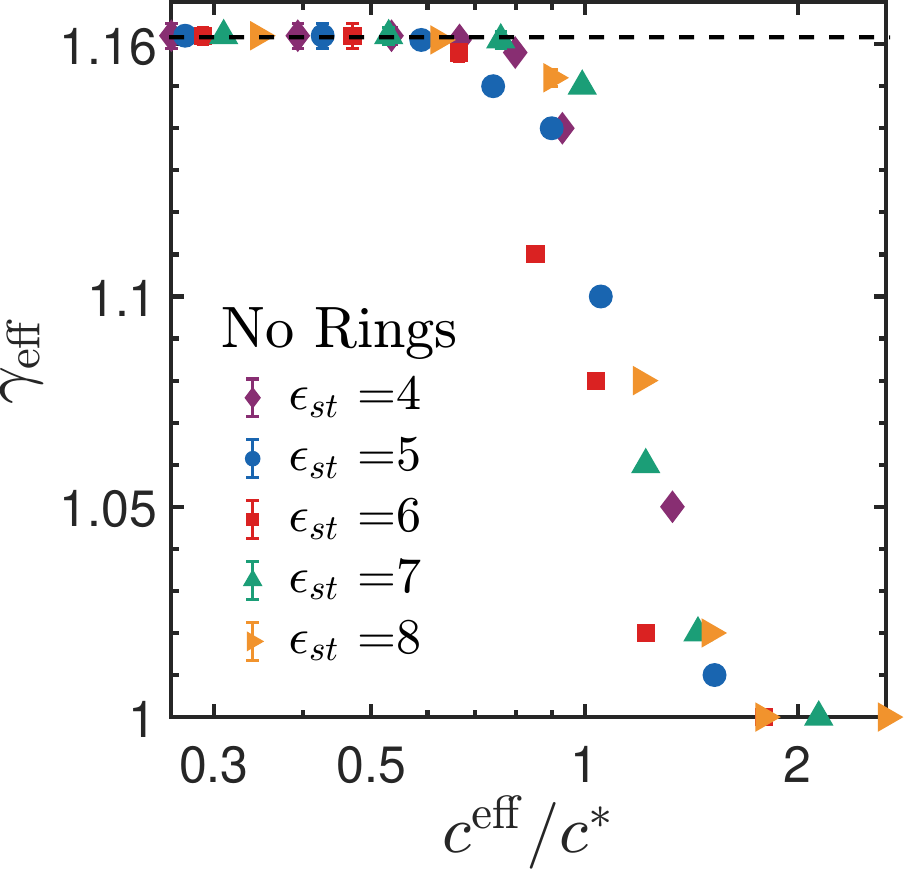} \\
(c)  & (d) 
\end{tabular}
\end{center}
    \vspace{-10pt}
\caption{The probability distribution $p(x)$ as a function of the scaled length $x$ in (a) the dilute regime, (b) at the overlap concentration,  and (c) in the semidilute regime. The sticker energy is held constant at $\epsilon_{st} = 7$. Similar behaviour is observed at other values of  $\epsilon_{st}$. The dependence of the effective exponent  $\gamma_{\text{eff}}$ on the scaled concentration, for various values of $\epsilon_{st}$, is displayed in (d).  \label{fig13}}
\end{figure*}

Since the value of the exponent $\alpha$ in the two concentration regimes has been determined, the amplitudes $A_d$ and $A_s$, and exponents $\delta_d$ and  $\delta_s$ in~\eref{meanL}, can be determined by plotting $\bar{L}/\left( c^{\text{eff}} \right)^\alpha$ versus $\epsilon_{st}$ in the two regimes, as displayed in~\fref{fig12}~(a) and~(b). The values of $\delta_d = 0.46$ and $\delta_s= 0.5$, determined from data at sufficiently high $\epsilon_{st}$ (where scaling laws are obeyed), are in complete agreement with the predictions of mean-field theory. The expectation that both $c^*$ and $L^*$ are exponential functions of $\epsilon_{st}$ is borne out by~\fref{fig12}~(c) and~(d). The exponents $\varphi$ and $\kappa$  in~\eref{cstarLstar}  and the  prefactors $P$ and $Q$, can be determined by fitting the data in these figures. The value of $\varphi$ is found to be $3.70 \pm 0.20$ and that of $\kappa$ is found to be $3.00 \pm 0.02$. These values are reasonably close to the values of $3.5$ and $2.8$ expected from mean-field theory. The prefactors $A_d = 15.74 \pm 0.10$ and $A_s = 22.11 \pm 0.50$, and the amplitudes $P= 0.09 \pm 0.02$ and $Q = 5.2 \pm 0.2$ can be seen to satisfy the relations given in~\eref{amplitudePQ}.

Once the mean length is determined, it is possible to plot the probability distribution $p(x)$ as a function of the scaled length $x$ at various concentrations. \fref{fig13}~(a) to~(c) display $p(x)$ as a function of $x$ in the dilute regime at $c^{\text{eff}}/c^* = 0.3$, at the overlap concentration, $c^{\text{eff}}/c^* = 1$, and in the semidilute regime at $c^{\text{eff}}/c^* = 2$, respectively. In all the cases, the sticker energy is held fixed at $\epsilon_{st} = 7$, and systems with and without rings have been computed. The simulation data is fitted with the expression for the Schulz-Zimm distribution, given in~\eref{lendist}. Note that when $\gamma = 1$, the Schulz-Zimm distribution reduces to a purely exponential distribution. It is possible, consequently to fit simulation data for various values of $c^{\text{eff}}/c^*$ by using $\gamma$ as a fitting parameter. The  value of $\gamma$ that fits the data at any concentration is denoted here as $\gamma_{\text{eff}}$. It is clear from \fref{fig13}~(a) that the Schulz-Zimm distribution is obeyed in the dilute solution, with the exponent $\gamma_{\text{eff}} = 1.162 \pm 0.002$, which is close to the value expected from mean-field theory. In a semidilute   solution at $c^{\text{eff}}/c^* = 2$, the exponent $\gamma_{\text{eff}} = 1.00 \pm 0.02$, suggesting that the probability distribution is a pure exponential function, as predicted by mean-field theory. At intermediate concentrations, the value of $\gamma_{\text{eff}}$ lies between $1.162$ and $1$, and decreases with increasing concentration.  \fref{fig13}~(b) indicates that $\gamma_{\text{eff}} = 1.10 \pm 0.02$ at $c^{\text{eff}}/c^* = 1$. Similar behaviour was observed for other values of the sticker energy, and \fref{fig13}~(d), which is a plot of $\gamma_{\text{eff}}$ versus $c^{\text{eff}}/c^*$, shows that the crossover from the dilute to the semidilute concentration regime occurs roughly between $c^{\text{eff}}/c^* = 1$ and $2$ in all cases.

\begin{figure}[t]
\centering
\resizebox{0.96\columnwidth}{!}{\includegraphics{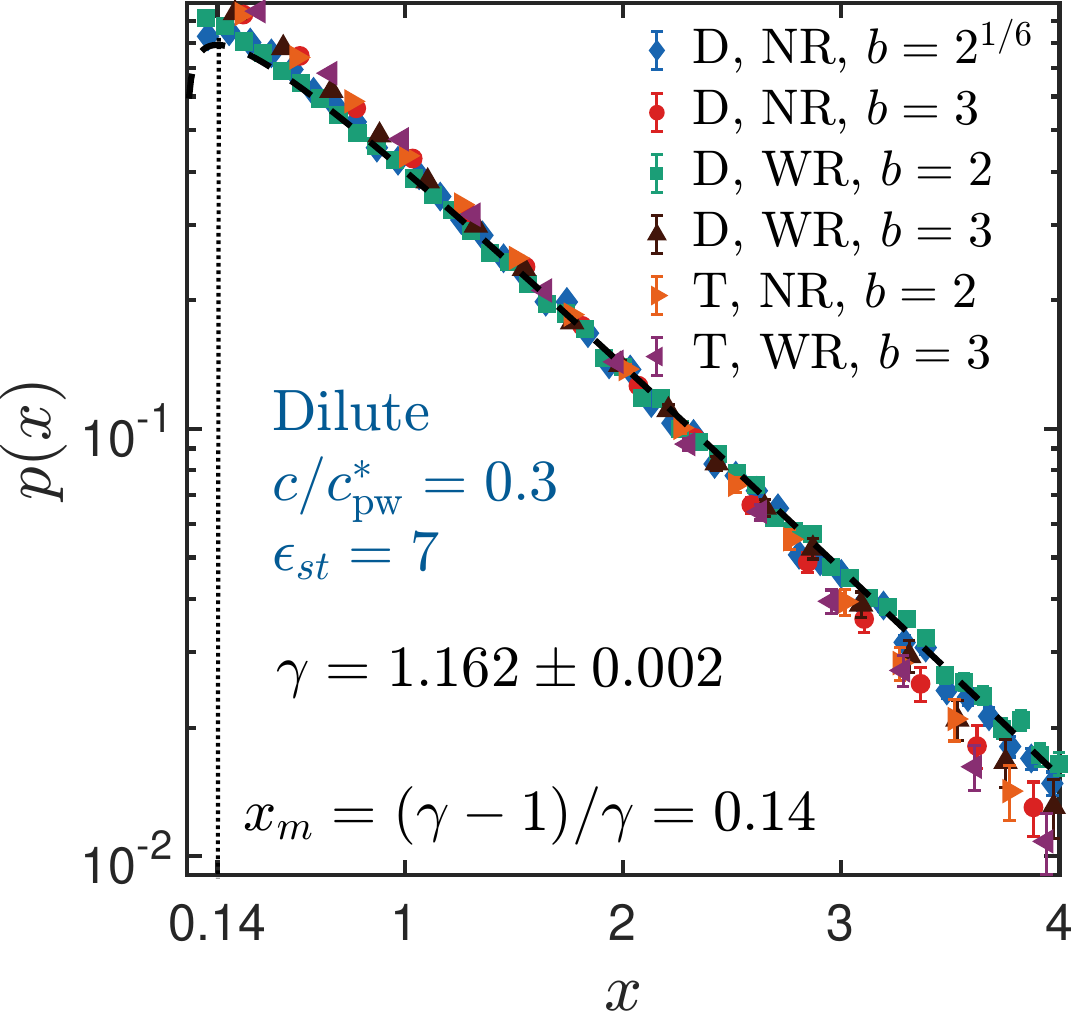}}\\
(a) \\[10pt]
\resizebox{0.96\columnwidth}{!}{\includegraphics{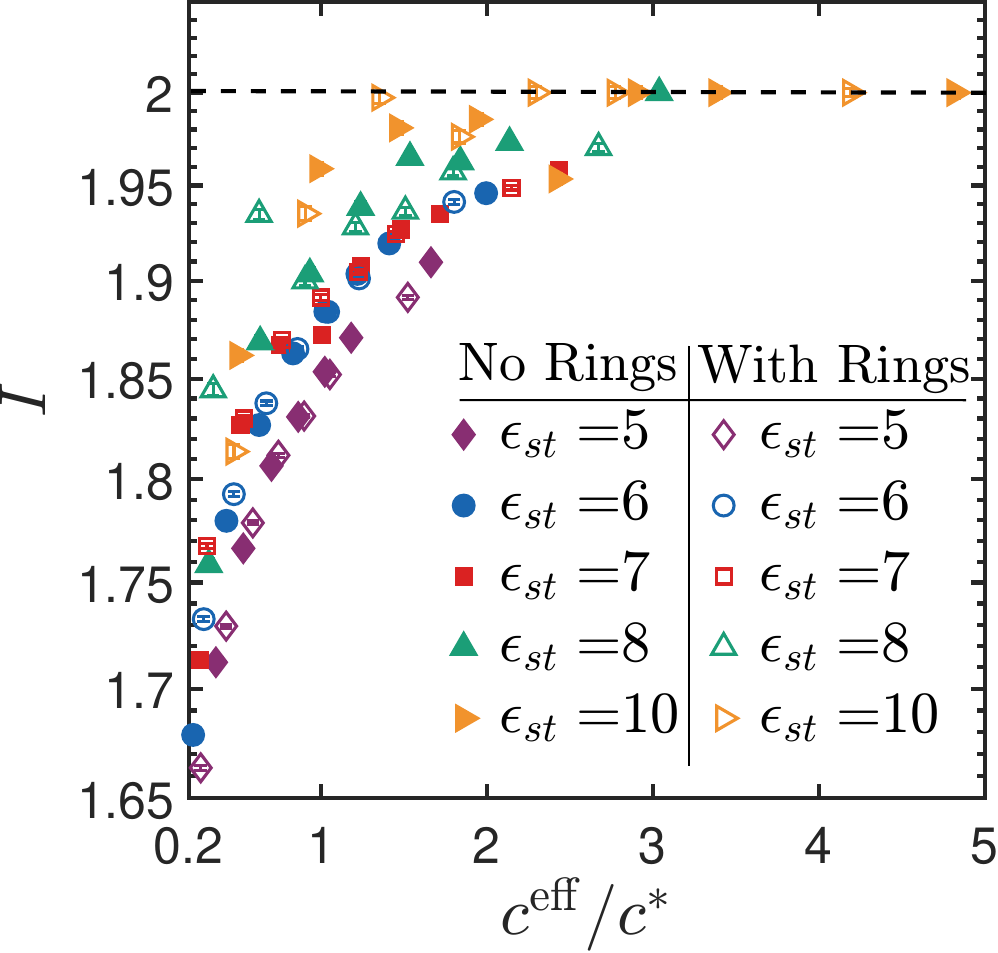}}\\
(b)
\vspace{-7pt}
\caption{(a) Independence of the probability distribution of scaled lengths $p(x)$ from the choice of the model for the persistent worm, whether it is a trumbbell (T) with spring rest lengths $b =2$ or $3$, or a dumbbell (D) with $b=2^{1/6}$, $2$ or $3$. (b) Polydispersity index $I$ as a function of scaled concentration for systems with and without rings, at various values of the sticker energy $\epsilon_{st}$. Data is shown only for cases where ${\text{c}}^{\text{eff}}/\text{c}^* > 0.2$ and $\bar{L} > 20$, where the scaling laws are obeyed.}
\label{fig14}
\end{figure}

The forms of the distribution of scaled lengths $p(x)$ predicted by mean-field theory in the different concentration regimes (\eref{lendist}) are universal functions, and do not depend on the particular model chosen to represent a wormlike micelle. This has been demonstrated previously by the validation of the mean-field predictions by the many earlier studies of wormlike micelle solutions, that have been based on different models~\cite{Wittmer1998,Wittmer2000,Kroger2004,PaddingPRE2004,Huang2006}. In the present work, results have so far been presented for the choice of trumbbells as persistent worms. This is the smallest bead spring chain (with 3 beads) that enables the application of a bending potential, and was used here to validate the implementation of the intra- and inter-persistent worm bending potentials. However, when the wormlike micelles are fully flexible, it is possible to use simple dumbbells as persistent worms, as has been done in some earlier studies~\cite{Vasquez2007,Zhou2014,Adams2018,Cook2020}. Here, the universal nature of the prediction of $p(x)$ by the current mesoscopic model is demonstrated in~\fref{fig14}~(a). Regardless of the model used (dumbbells with spring rest lengths $b=2^{1/6}$, $2$, or $3$, or trumbbells with $b=2$ or $3$) to represent a persistent worm, the simulation results show that the scaled length distribution is accurately captured by a Schultz-Zimm distribution  with a value of $\gamma_{\text{eff}} = 1.162 \pm 0.002$ used to fit the data, for a dilute wormlike micelle solution at $c/c^*_{\text{pw}} = 0.3$, where the sticker energy has the value  $\epsilon_{st} = 7$. A similar observation of model independence was made at other values of $c^{\text{eff}}/c^*$ and $\epsilon_{st}$, with appropriate values of $\gamma_{\text{eff}}$.

It should be noted that the parameter $\gamma$ in the Schultz-Zimm distribution can be determined by fitting either the tail or the peak of the distribution, which occurs at $x_m = (\gamma - 1)/\gamma$. However, since wormlike micelle lengths change in discrete units of persistent worm length $\ell_{\text{pw}}$, it is challenging to locate $x_m$ at small values of $x$ when trumbbells are used as persistent worms, due to insufficient resolution. The values of $\gamma_{\text{eff}}$ displayed in \fref{fig13}~(a) to~(c) was consequently determined by fitting the tail of the distributions. On the other hand, it is easier to locate $x_m$ for dumbbells with $b=2^{1/6}$ (the shortest possible spring rest length), as shown in~\fref{fig14}~(a), where its location is indicated with the vertical line at $x_m = 0.14$. A Schultz-Zimm distribution with $\gamma_{\text{eff}} = 1.162 \pm 0.002$ fits the data well and matches the location of $x_m$.

\begin{figure*}[t]
\begin{center}
\begin{tabular}{cc}
\includegraphics[width=8.5cm]{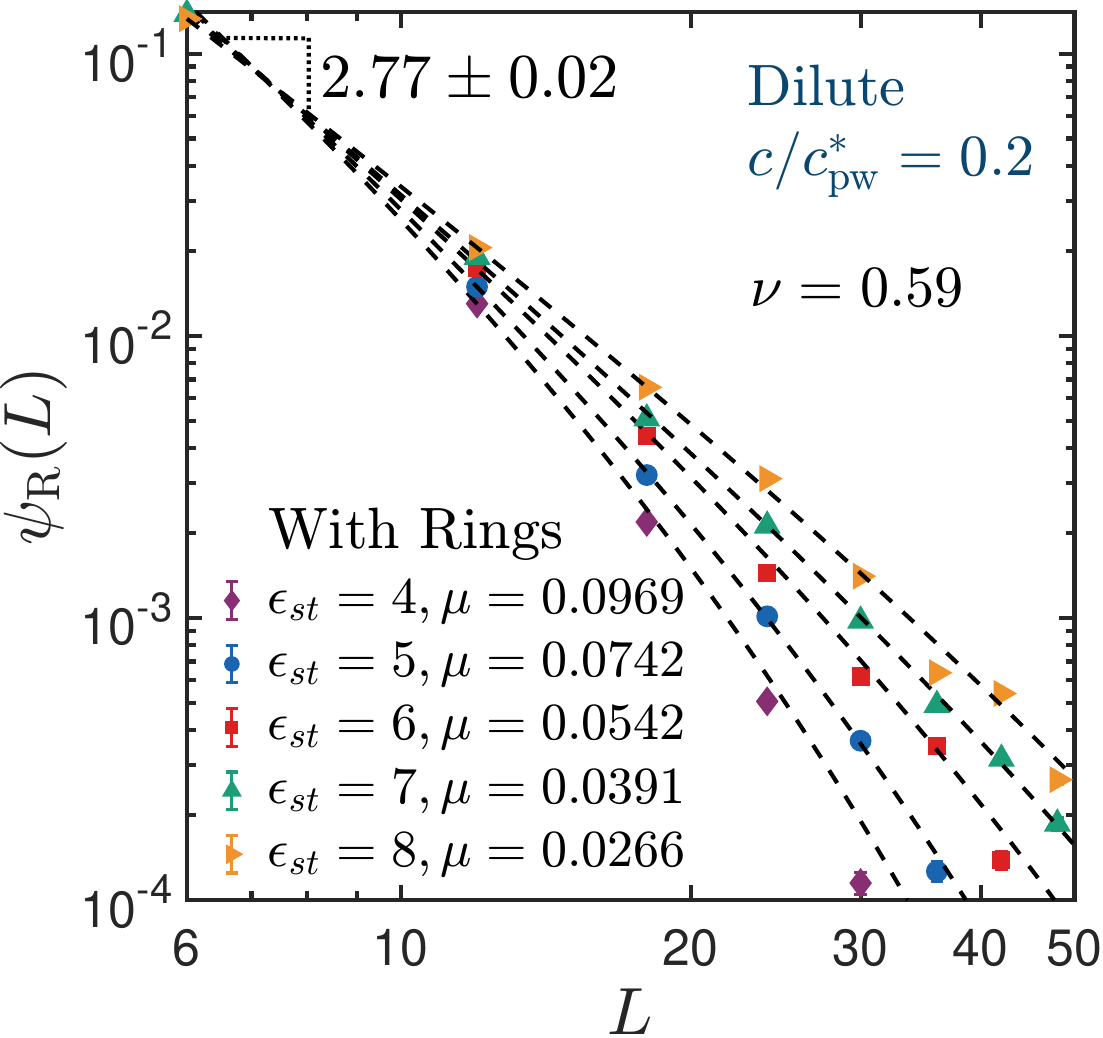} &
\includegraphics[width=8.5cm]{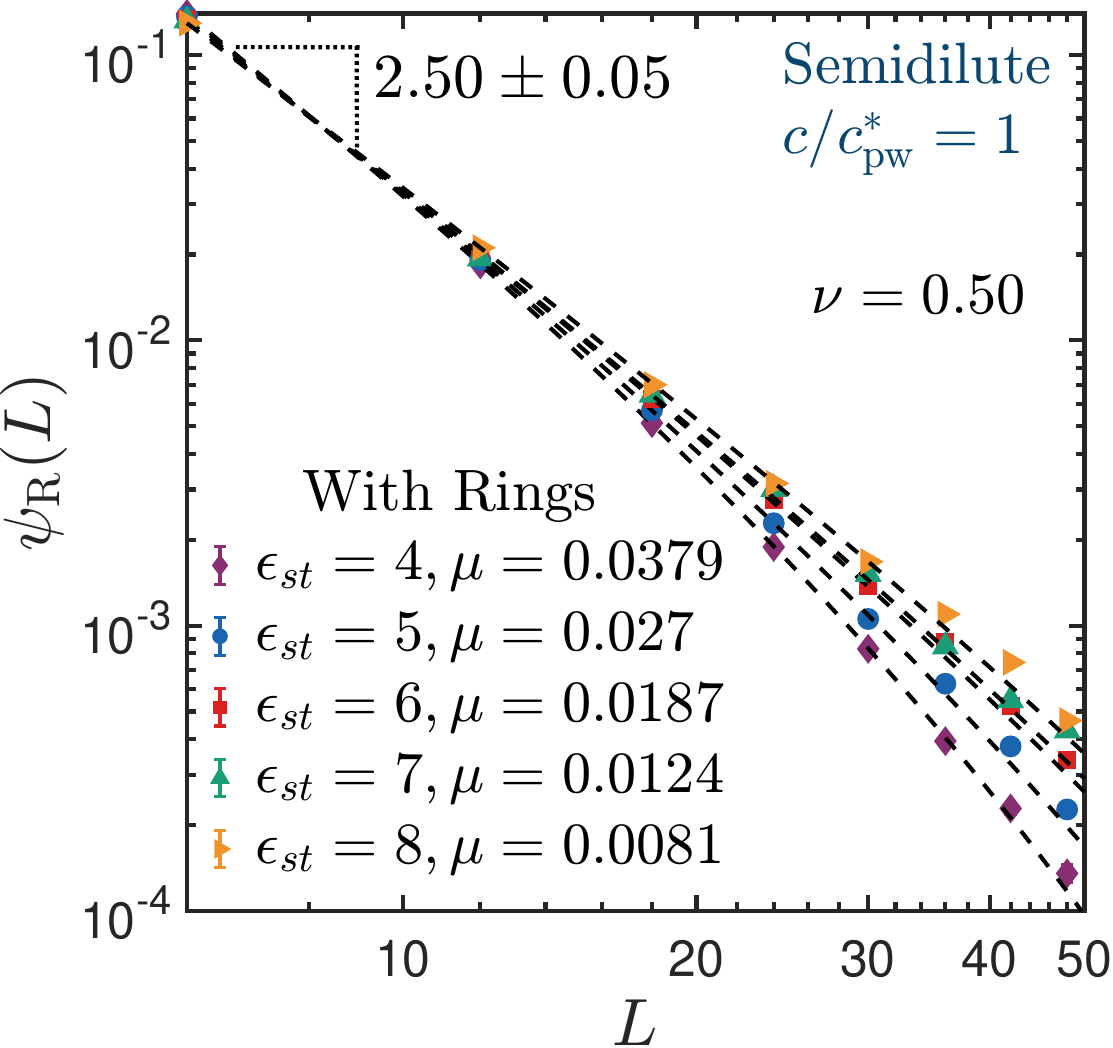} \\
(a)  & (b) 
\end{tabular}
\end{center}
    \vspace{-10pt}
\caption{Probability distribution of lengths of ring wormlike micelles, at various values of the sticker energy  $\epsilon_{st}$, in (a) dilute and (b) semidilute solutions. The concentration is reported in terms of $c/c^*_{\text{pw}}$, which is held constant in the two regimes. Each set of symbols at a fixed value of $\epsilon_{st}$ corresponds to a unique value of scaled concentration $c^{\text{eff}}/c^*$, which is related to $c/c^*_{\text{pw}}$ and $\epsilon_{st}$ as displayed in~\fref{fig8}.
\label{fig15}}
\end{figure*}

\fref{fig14}~(b) displays the polydispersity index $I$ for wormlike micelle solutions as a function of concentration, where $I$ is defined by~\cite{Wittmer1998}, 
\[ I = \dfrac{\overline{L^2}}{ \left( \bar{L} \right)^2} = 1 + \dfrac{1}{\gamma_{\text{eff}}} \]
and represents the ratio of weight-average to number-average molecular weight. The index approaches a value of  2 with increasing concentration (for ${\text{c}}^{\text{eff}}/\text{c}^* > 2$). While in principle, the value of $I$ in the limit of low concentration can be used to estimate the exponent $\gamma_{\text{eff}}$, as pointed out by \citet{Wittmer1998} (who estimated $I$ similarly from their Monte Carlo simulations), this is difficult in practice because of the difficulty of determining the distribution accurately for small values of $x$, leading to a wide range of values for $\gamma_{\text{eff}}$.

As observed in the case of other static properties presented earlier, the presence of rings makes no difference to the length distribution of linear chains, and in all the cases considered above, curves for various values of  $\epsilon_{st}$ collapse on to a master curve when plotted in terms of the appropriate variables. Essentially, the presence of rings simply reduces the population of linear chains in the solution, but does not influence the properties in any significant way.

\subsubsection{\label{sec:rings} Length distributions and the fraction of ring wormlike micelles}

The mean-field prediction for the distribution of lengths of ring wormlike micelles is given in~\eref{ringdist}. Since the simulations have been carried out with trumbbells that have a spring rest length $b=3$, in the present study $L^R_{\text{min}} = 6$. Figures~\ref{fig15}~(a) and~(b) display the distributions at various values of $\epsilon_{st}$ for dilute and semidilute solutions, respectively. The initial power law decay with an exponent $(3\nu +1)$ is clearly visible, with the respective values of the Flory exponent $\nu$  in each regime, followed by an exponential damping that is dependent on the mean length $\bar{L}$ of linear wormlike micelles, which is a function of $c^{\text{eff}}/c^*$ and $\epsilon_{st}$. 

The constant $\lambda_0$ and the variable $\mu$ in~\eref{ringdist} are used as fitting parameters to draw the curves through the simulation data as displayed in \fref{fig15}~(a) and~(b). Since $\mu =  \gamma_{\text{eff}}  / \bar{L} $, where the mean length is given by the expression $\bar{L} =  A \,  \left({c^{\text{eff}}} \right)^\alpha \exp \left( \delta \, \epsilon_{st} \right)$ (with the appropriate values of $A$, $\alpha$ and $\delta$ in each regime), it is possible to use the fitted values of $\mu$ to estimate $ \gamma_{\text{eff}}^* = \mu \bar{L}$ for each pair of values $(c^{\text{eff}}/c^*, \epsilon_{st})$. Values obtained in this manner are displayed in \tref{table:ringfit}, and compared with values of $ \gamma_{\text{eff}}$ obtained by fitting the Schulz-Zimm distribution to the probability distribution of scaled lengths $p(x)$. It can be seen that both estimates are close to each other.

\begin{table*}[t]
\vskip-10pt
\centering
\caption{Values of parameters in~\eref{ringdist} determined by fitting simulation data and the quantity $\gamma_{\text{eff}}^*$, determined from $\gamma_{\text{eff}}^* = \mu \bar{L}$. The values of $\gamma_{\text{eff}}^*$ match the values obtained from fitting the scaled length distribution of linear chains. In the dilute regime ($c/c_{\text{pw}}^* = 0.2$), $\gamma_{\text{eff}}^* \approx \gamma_{\text{eff}} = 1.162$, and in the semidilute regime ($c/c_{\text{pw}}^* = 1$), $\gamma_{\text{eff}}^* \approx \gamma_{\text{eff}} = 1$.}
\label{table:ringfit}
\vskip5pt
\bgroup
\setlength{\tabcolsep}{1.1em}
{\def\arraystretch{1.1}
\begin{tabular}{|c|c|c|c|c|c|c|c|c|}
\hline  
\multirow{2}{*}{$\epsilon_{st}$} & \multicolumn{4}{c|}{$c/c^*_{\text{pw}} =0.2$ }
& \multicolumn{4}{c|}{$c/c^*_{\text{pw}} =1$ } \\
\cline{2-9} 
& {$\lambda_0$} & {$\mu$}  & {$\bar{L}$}  & {$\gamma_{\text{eff}}^*$}  & {$\lambda_0$} & {$\mu$}  & {$\bar{L}$}  & {$\gamma_{\text{eff}}^*$}     \\
\hline
\multicolumn{1}{|l|}{4} & 37.53 $\pm$ 0.30 & 0.0969 $\pm$ 0.0004 & 11.99 & 1.162 & 15.98 $\pm$ 0.10 & 0.0379 $\pm$ 0.0002 & 26.47 &  1  \\ \hline
\multicolumn{1}{|l|}{5} & 32.81 $\pm$ 0.40 & 0.0742 $\pm$ 0.0003 & 15.68 & 1.163 & 14.68 $\pm$ 0.15  & 0.027 $\pm$ 0.002 & 37.17 &  1  \\ \hline
\multicolumn{1}{|l|}{6} & 28.13 $\pm$ 0.10 & 0.0542 $\pm$ 0.0002 & 21.40 & 1.160 & 13.62 $\pm$ 0.20 & 0.0187 $\pm$ 0.0002 & 54.07 &  1  \\ \hline
\multicolumn{1}{|l|}{7} & 23.98 $\pm$ 0.05 & 0.0391 $\pm$ 0.0002 & 29.71 & 1.162 & 13.02 $\pm$ 0.02  & 0.0124 $\pm$ 0.0003 & 81.15 &  1  \\ \hline
\multicolumn{1}{|l|}{8} & 22.30 $\pm$ 0.15 & 0.0266 $\pm$ 0.0001 & 43.73 & 1.163 & 11.92 $\pm$ 0.10 & 0.0081 $\pm$ 0.0002 & 125.64 &  1  \\ \hline     
\end{tabular}
}
\egroup
\end{table*}

\begin{figure*}[t]
\begin{center}
\begin{tabular}{cc}
\includegraphics[width=8.5cm]{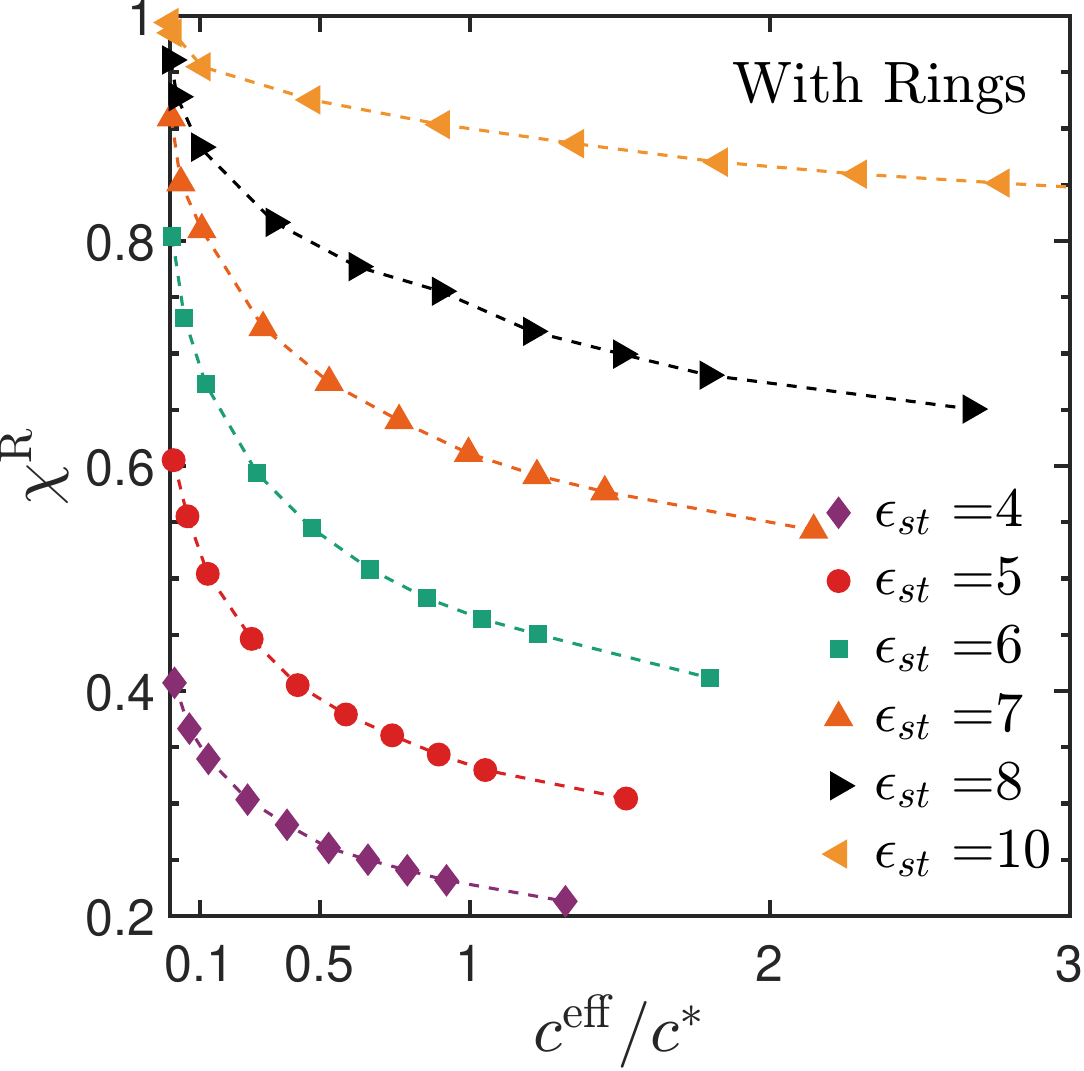} &
\includegraphics[width=8.5cm]{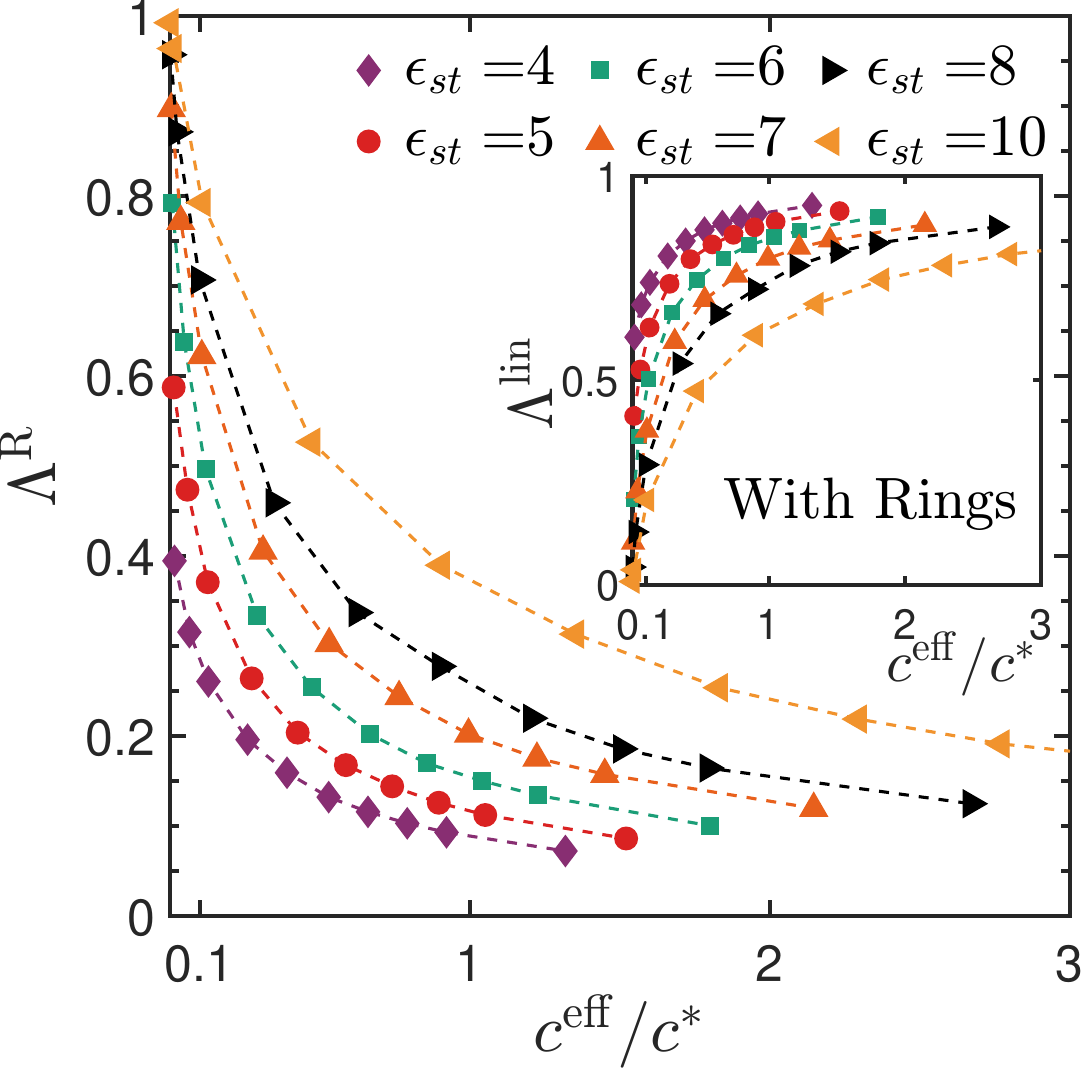} \\
(a)  & (b) \\[10pt] 
\end{tabular}
\end{center}
\vspace{-20pt}
\caption{(a) Fraction of rings $\chi^{\text{R}}$ in a wormlike micelle solution, and (b) fraction of stickers incorporated into rings, $\Lambda^{\text{R}}$, with the fraction belonging to linear wormlike micelles $\Lambda^{\text{lin}}$ shown in the inset, as functions of the scaled concentration.   
\label{fig16}}
\end{figure*}

It was seen earlier in \fref{fig10} that the distribution of lengths of linear wormlike micelles was affected by the existence of rings, with the change being more pronounced at lower concentrations and higher sticker energies. The formation of rings clearly depends on several factors such as the bending rigidity, the effective wormlike micelle concentration, and the sticker energy. The presence of rings in the solution is tracked here in two different ways. In the first, the fraction of all wormlike micelles in the solution that are rings, $\chi^{\text{R}}$, is calculated from,    
\[
\chi^{\text{R}} = \langle \, \mathcal{N_{\text{wlm}}^{\text{R}}} \, \rangle/\langle \,  \mathcal{N}_{\text{wlm}} \, \rangle \]
where $\mathcal{N_{\text{wlm}}^{\text{R}}}$ and $ \mathcal{N}_{\text{wlm}}$ have been defined previously in~\eref{eq:Nwlm}. In the second, the fraction of all stickers in the solution that have been incorporated into rings, $\Lambda^{\text{R}}$, is obtained from,
\begin{align*}
\Lambda^{\text{R}} &= \frac{\langle N_{\text{st,R}}^{\text{bound}} \rangle}{N_{\text{st}}^{\text{T}}} \,, \\
\text{where} \quad 
N_{\text{st,R}}^{\text{bound}} &= \sum_{L_{\text{min}}}^{L_{\text{max}}} 2\, m_{\text{pw,R}}^{\text{L}}\, n^{\text{R}}_{\text{L}}(L) \,,\:  \:  \text{and} 
\:  \:   N_{\text{st}}^{\text{T}} = 2 \, n_{\text{pw}}^{\text{T}}
\end{align*}
The factors of two account for the presence of two sticky end beads per persistent worm. Clearly, it follows that, $\Lambda^{\text{lin}} = 1- \Lambda^{\text{R}}$ is the fraction of stickers that belong to linear wormlike micelles.

Figures~\ref{fig16}~(a) and~(b) present the fractions $\chi^{\text{R}}$ and $\Lambda^{\text{R}}$ as functions of the scaled effective concentrations for a range of different sticker energies. In both figures it is clear that the fraction of rings, and of stickers incorporated in rings, is larger at low concentrations and at high sticker energies. At low concentrations, it is much easier for a sticker on a linear wormlike micelle to find another sticker at the end of the same chain than to find a sticker on another chain, and when the sticker energy is high, the stickers remain bound for longer. With increasing concentration, the probability of ring formation decreases since stickers at the ends of linear wormlike micelles are more likely to find stickers on other chains with which they can bind, and the wormlike micelles increase in length. This is also reflected in the inset to \fref{fig16}~(b), where the fraction of stickers belonging to linear chains, $\Lambda^{\text{lin}}$, increases monotonically with increasing concentration. 

A closer examination (not shown here) of the surprisingly large fraction of rings that are present, regardless of concentration ($\chi^{\text{R}}$ greater than 60\% for sticker energies $\epsilon_{st}$ greater than $7 \, k_{\text{B}} T$), reveals that a majority of the rings are just single persistent worms that have fused at the ends to become rings. Nevertheless, because the overwhelming number of rings are small in size, the fraction of all the available stickers that are incorporated in rings, $\Lambda^{\text{R}}$, does not remain large with increasing concentration, but rather decreases rapidly, as can be seen in \fref{fig16}~(b), since the linear wormlike micelles that are increasing in number and length with increasing concentration, contain most of the stickers. 

\begin{figure}[t]
    \centering
\resizebox{8.5 cm}{!}{\includegraphics*[width=10cm]{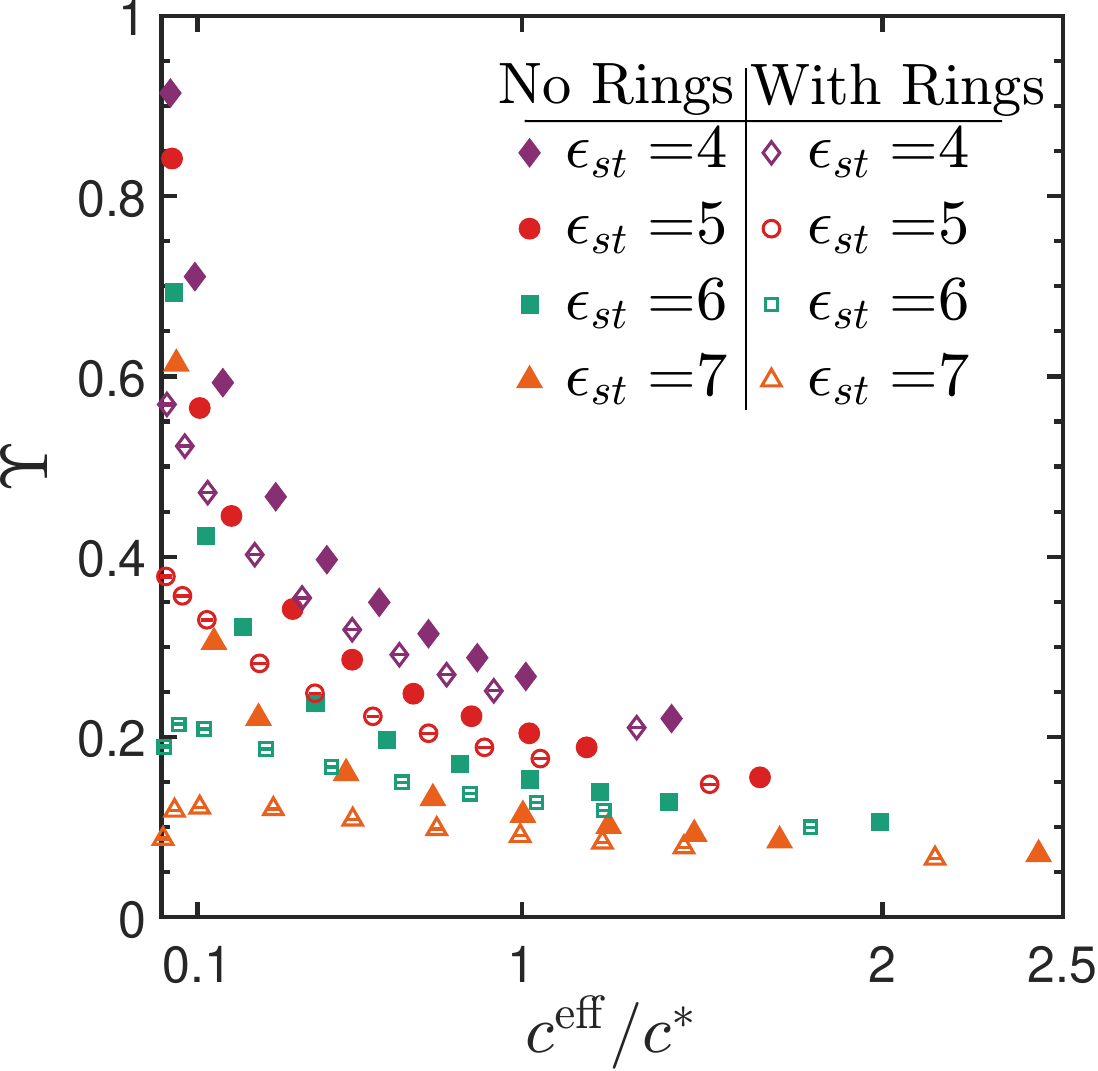}} 
    \vspace{-10pt}
    \caption {Fraction of free stickers ($\Upsilon$) in wormlike micelles as a function of scaled concentration. Systems with and without rings have been displayed. 
    \label{fig17}}
    \end{figure} 
    
An alternative way to examine the influence of rings is to calculate the number of \textit{free} stickers in solution, $\Upsilon$, in the presence and absence of rings, which can be calculated from,
\[ 
\Upsilon = 1- \frac{ \left \langle N_{\text{st}}^{\text{bound}} \right \rangle } { N_{\text{st}}^{\text{T}} } 
\]
where, $ N_{\text{st}}^{\text{bound}}=N_{\text{st,lin}}^{\text{bound}} + N_{\text{st,R}}^{\text{bound}}  \, , \:    \text{with}  \:  \ N_{\text{st,lin}}^{\text{bound}} = \sum_{L_{\text{min}}}^{ L_{\text{max}}} \, \left ( 2 \, m_{\text{pw,lin}}^{\text{L}}-2 \right ) n^{\text{lin}}_{\text{L}} (L) $. The subtraction of $2$ in the definition of $N_{\text{st,lin}}^{\text{bound}}$ accounts for the fact that the end beads in linear chains remain unbound. Clearly, when the possibility of ring formation is permitted in the model, there are far fewer free stickers in solution, as the stickers at the end of linear chains combine and lead to the formation of rings. At high sticker energies, when the ends of even single persistent worms combine to form rings, stickers remain bound as the concentration increases. In both cases of rings and no rings, as the fraction of linear wormlike micelles grows with increasing concentration, the fraction of free stickers that are at ends of the chains decreases monotonically (\fref{fig17}).

\subsubsection{\label{sec:radius_of_gyration} The radius of gyration for linear wormlike micelles}

\begin{figure*}[t]
    \centering
\resizebox{17cm}{!}{\includegraphics*[width=17cm]{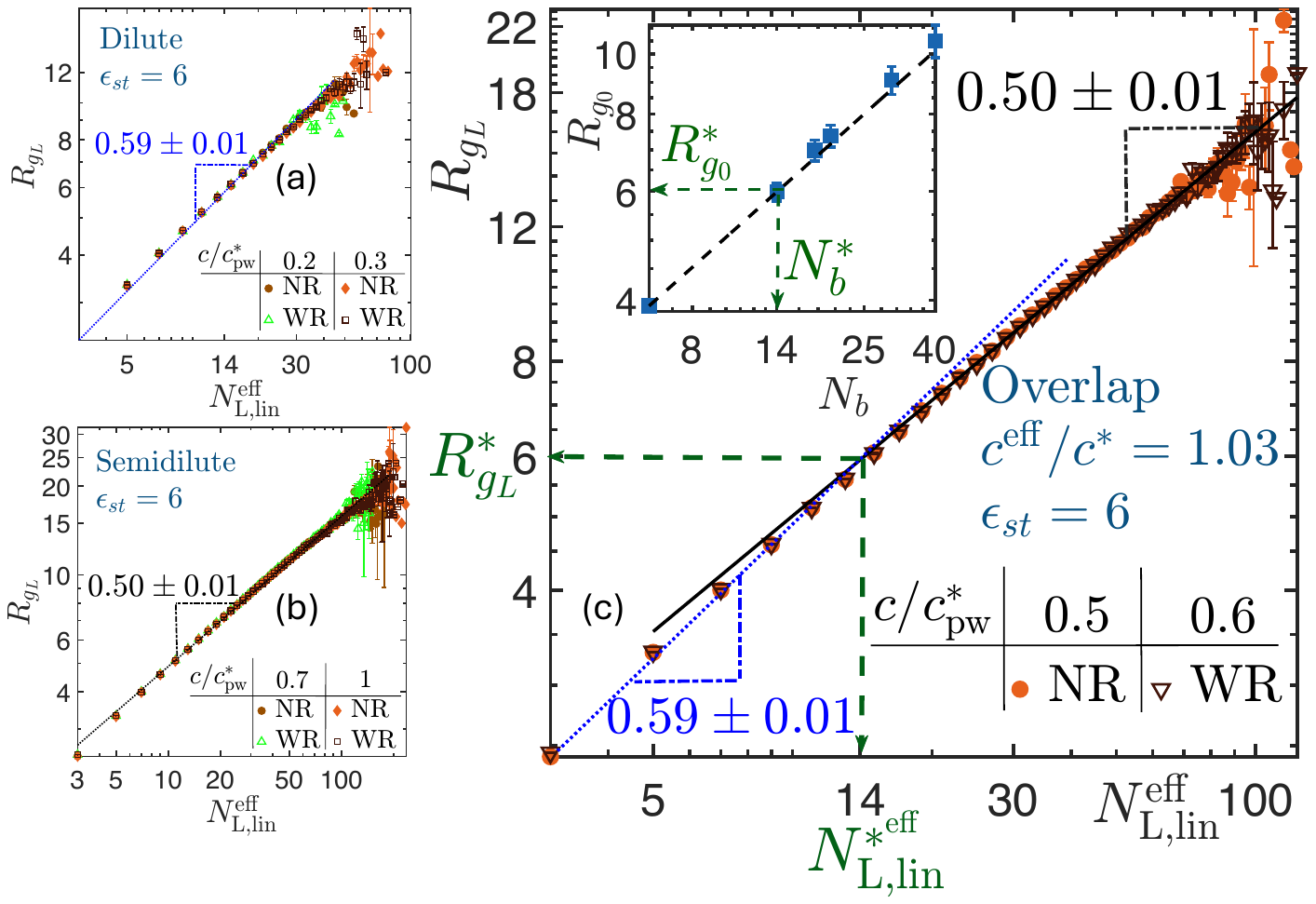}} 
    \vspace{-10pt}
\caption{The scaling of the radius of gyration $R_{g_L}$ with $N_{{\text{L,lin}}}^{\text{eff}}$, which is the effective number of monomers in a linear wormlike micelle of length $L$, in (a) the dilute concentration regime, (b) the semidilute concentration regime, and (c) at $c^{\text{eff}}/c^* = 1.03$, which is close to the overlap concentration. Systems with and without rings have been considered at a fixed sticker energy of $\epsilon_{st} = 6$. The inset to (c) shows the scaling with the number of beads $N_b$ of the radius of gyration $R_{g0}$ of a homopolymer chain in a dilute solution under athermal solvent conditions.
\label{fig18}}
\end{figure*}

The radius of gyration $R_{g_L}$ of linear wormlike micelles of length $L$, as a function of the number of effective monomers $N_{\mathrm{L,lin}}^{\mathrm{eff}}$ in the chain, at various concentrations, is displayed in \frefs{fig18}~(a) to~(c). In \fref{fig18}~(a), data in the dilute regime at two different values of $c/c^*_{\text{pw}}$, is presented for systems with and without rings, at $\epsilon_{st} = 6$. The data collapses on to a single curve, with $R_{g_L} \sim \left( N_{\mathrm{L,lin}}^{\mathrm{eff}} \right)^{0.59}$, as expected for chains obeying self-avoiding walk statistics. The significant fluctuations in $R_{g_L}$ for high values of $N_{\mathrm{L,lin}}^{\mathrm{eff}}$ is due to poor statistics at large chain lengths. Figure~\ref{fig18}~(b) displays data at the same value of $\epsilon_{st}$ for two different values of $c/c^*_{\text{pw}}$ in the semidilute regime. In this case, the data collapses on to a curve with $R_{g_L} \sim \left( N_{\mathrm{L,lin}}^{\mathrm{eff}} \right)^{0.5}$, indicating that Flory screening is present and leads to the chains obeying random walk statistics. Figure~\ref{fig18}~(c) examines the dependence of $R_{g_L}$ on $N_{\mathrm{L,lin}}^{\mathrm{eff}}$ at an effective concentration that is very close to the overlap concentration ($c^{\text{eff}}/c^* = 1.03$). It is clear that $R_{g_L} \sim \left( N_{\mathrm{L,lin}}^{\mathrm{eff}} \right)^{0.59}$ until a critical value of the number of effective monomers ${N_{\mathrm{L,lin}}^{\mathrm{eff^*}}}$, after which it scales as $R_{g_L} \sim \left( N_{\mathrm{L,lin}}^{\mathrm{eff}} \right)^{0.5}$. This is consistent with the blob scaling picture, with chains obeying self-avoiding walk statistics within a correlation blob, and random walk statistics on length scales larger than a blob. It follows that the crossover value of the radius of gyration, $R^*_{g_L}$, is the size of a correlation blob in a wormlike micellar solution at the overlap concentration. Interestingly, it turns out that $R^*_{g_L}$, which is a property of linear wormlike micelles formed by the fusion of persistent worms in a dilute polydisperse solution, is identical in magnitude to the size of a homopolymer chain of the same length obeying self-avoiding walk statistics in a dilute solution, as indicated in the inset to~\fref{fig18}~(c), where $N_b^* = N_{\mathrm{L,lin}}^{\mathrm{eff^*}}$ leads to $R_{g0}^* = R_{g_L}^*$. Note that the homopolymer data in the inset was obtained by simulating bead-spring chains in the dilute limit, with $\epsilon_{bb} = 0$.

It may be recalled that the overlap concentration $c^*$, at a fixed sticker energy $\epsilon_{st}$, was defined as the effective concentration at which the overlap length $L^*$ is equal to the mean length $\bar{L}$ (see \fref{fig5}). One can calculate the number of effective monomers $N^*$ that correspond to the overlap length $L^*$ from $N^* = (L^*/b) + 1$. When estimated in this manner for $\epsilon_{st} = 6$, it turns out that  $N^* = N_{\mathrm{L,lin}}^{\mathrm{eff^*}}$, as determined in \fref{fig18}~(c) above. Thus, at the overlap concentration, the correlation blob is just equal in size to the size of the swollen coil conformation adopted by a  linear wormlike micelle of overlap length $L^*$, reinforcing the validity of determining the overlap concentration by the procedure described in \sref{sec:overlap_conc}. By fitting data at various values of $N^*$ determined by varying $\epsilon_{st}$ at $c^{\mathrm{eff}}/c^*=1$, it is found that $R_{g_L}^* (\epsilon_{st}) = B  \left( N^* \right)^{0.59}$, where $B \approx 1.41$ is a fitting parameter. 

\begin{figure}[b]
\vspace{-15pt}
\centering
\resizebox{0.90\columnwidth}{!}{\includegraphics{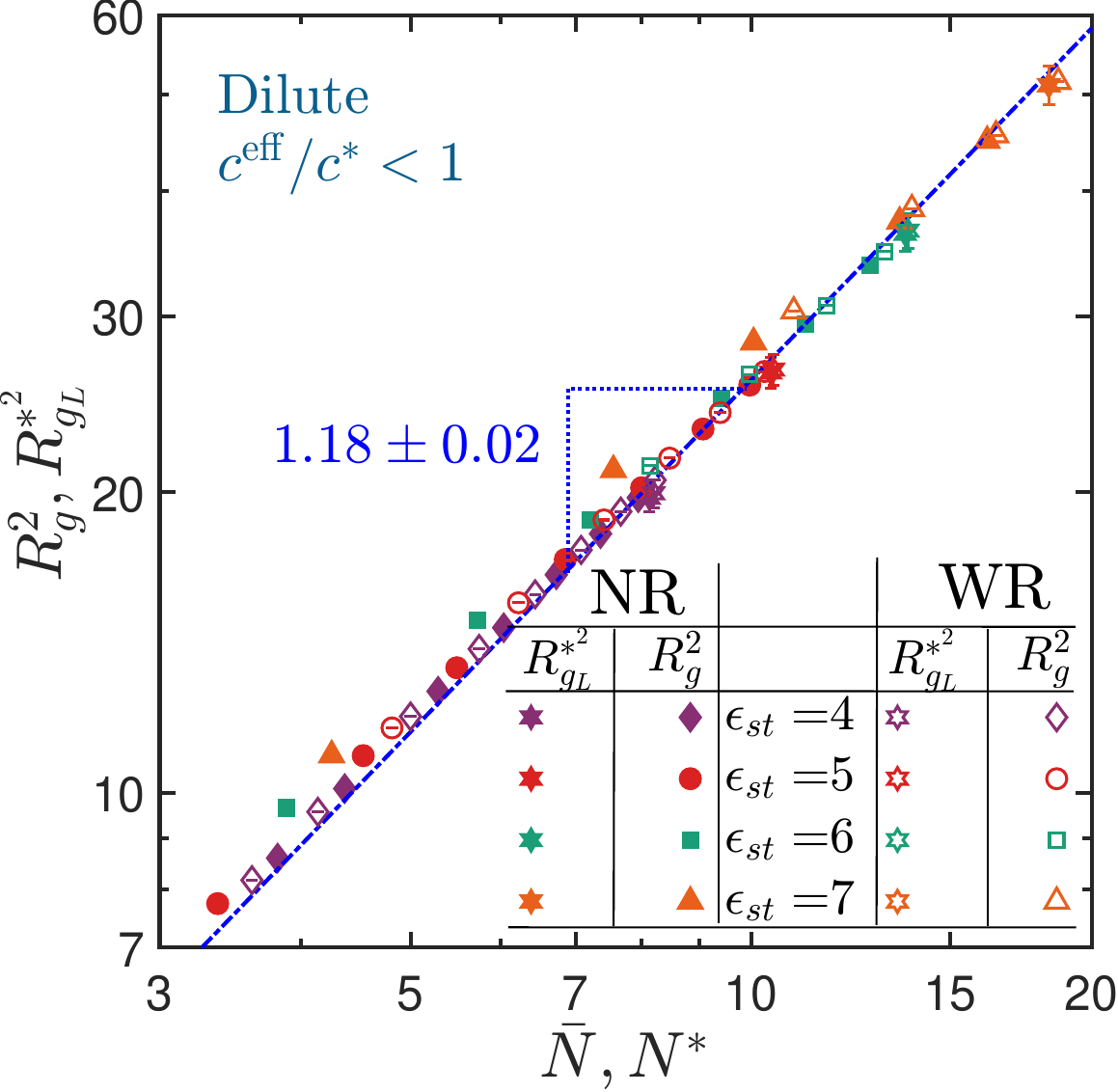}}\\
(a) \\[10pt]
\resizebox{0.93\columnwidth}{!}{\includegraphics{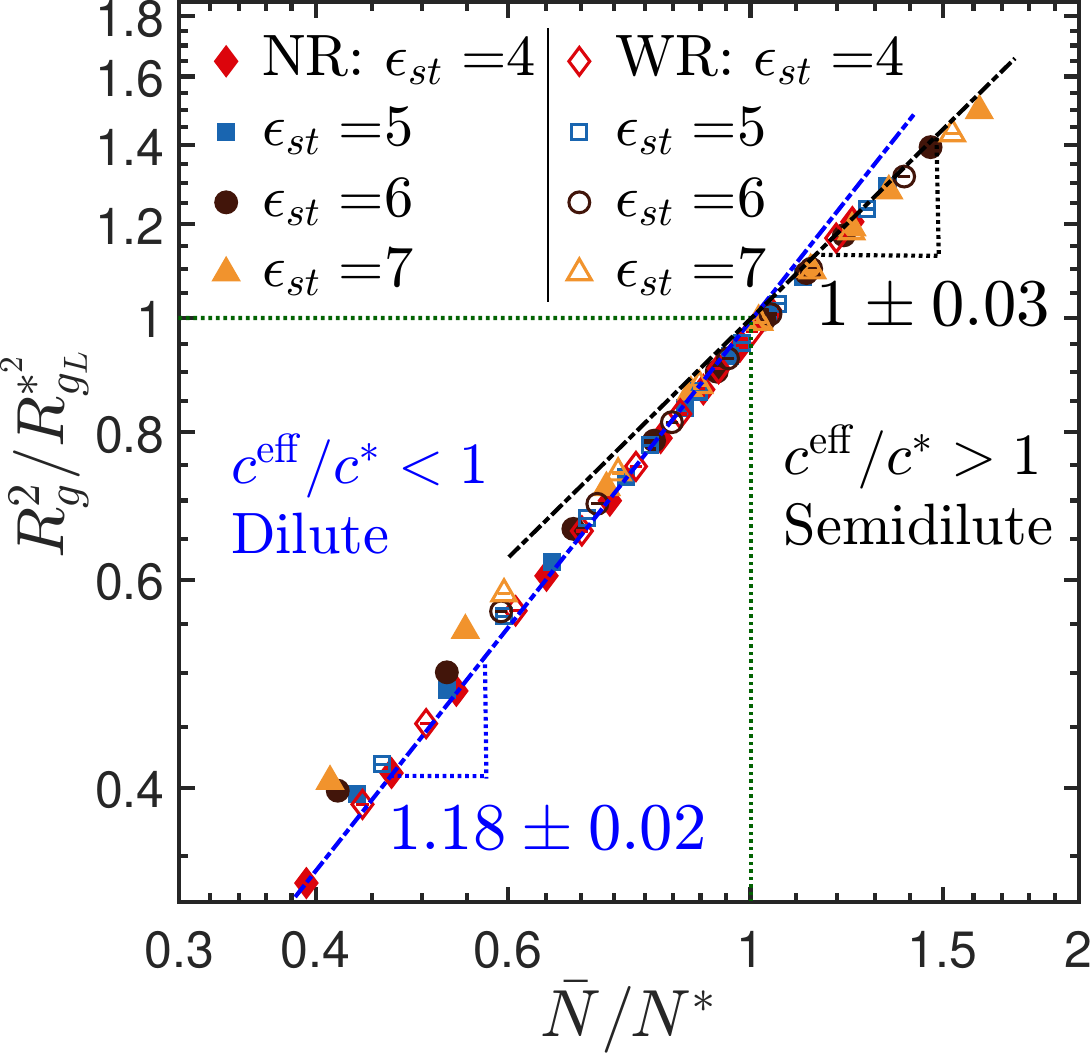}}\\
(b)
\vspace{-6pt}
\caption{(a) Scaling in the dilute regime of the square of the radius of gyration $R_g^2$ with $\bar{N}$, which is the number of beads in a linear wormlike micelle with length equal to the mean length $\bar{L}$ at the respective values of $c^{\text{eff}}/c^*$ and $\epsilon_{st}$, and of the square of the crossover radius of gyration $R_{g_L}^{*^2}$ with $N^*$, which is the number of beads in a linear wormlike micelle with length equal to the overlap length $L^*$ at the respective value of $\epsilon_{st}$. (b) Scaling of the ratio $R_g^2/R_{g_L}^{*^2}$ in the dilute and semidilute regimes, with the number of correlation blobs $(\bar{N}/N^*)$, at different sticker energies. Systems with (WR) and without rings (NR) have been considered. \label{fig19}}
\end{figure}

As defined earlier in~\eref{meanradius}, the alternative definition of the radius of gyration used here is $R_{g}$, which is the mean size of all linear wormlike micelles in the solution, regardless of their length. The square of this radius of gyration is displayed in Figure~\ref{fig19}~(a) in the dilute regime, as a function of $\bar{N}$, which is the number of beads in a linear wormlike micelle with length equal to the mean length $\bar{L}$. Each of the data points in the plot correspond to simulation results obtained at values of $c/c^*_{\text{pw}}$ and $\epsilon_{st}$ in~\fref{fig8} that lead to values of $c^{\text{eff}}/c^* < 1$, with $\bar{L}$ evaluated at these particular values of concentration and sticker energy. The number of beads  is calculated from $\bar{N} = (\bar{L}/b)+1$. Remarkably, the dependence of $R_{g}^2$ on $\bar{N}$ appears to be identical to that of ${R^*_{g_L}}^{\!\!\!2}$ on $N^*$, i.e., $R_{g} = B \left( \bar{N} \right)^{0.59}$, implying one can write $R_g/R^*_{g_L} = \left( \bar{N}/N^*\right )^{0.59}$, regardless of the value of $\epsilon_{st}$. Indeed, when the ratio $(R_g^2/{R^*_{g_L}}^{\!\!\!2})$ is plotted as a function of the ratio $(\bar{N}/N^*)$ in both the dilute and semidilute regimes, as displayed in~\fref{fig19}~(b), data for various values of sticker energy $\epsilon_{st}$ collapse on to a master plot, that exhibits a sharp transition at $\bar{N}/N^* = 1$ from one power law regime with slope $1.18 \pm 0.02$ to another power law regime with slope $1.0 \pm 0.03$. This suggests that one could write $R_g = R^*_{g_L} \left( \bar{N}/N^*\right )^\nu$, where $\nu$ is the Flory exponent with value 0.59 in the dilute regime and 0.5 in the semidilute regime. Such a representation is consistent with the picture that at any sticker energy $\epsilon_{st}$ in the semidilute regime, the mean size of all the wormlike micelles in a simulation box is equivalent to that of a linear wormlike micelle of mean length $\bar{L}$, whose conformation is a random walk of correlation blobs, where $R^*_{g_L}$ is the size of the correlation blob at the overlap concentration, and $(\bar{N}/N^*)$ is the number of such correlation blobs in the chain, while in the dilute regime where  $\bar{N}$ is less than $N^*$, the mean size of all the wormlike micelles is equivalent to that of a linear wormlike micelle of mean length $\bar{L}$, whose conformation is a self-avoiding walk. 

It is worth noting that for any value of $\epsilon_{st}$, the overlap length $L^*$ can be evaluated since the parameters $Q$ and $\kappa$ are known (see~\eref{cstarLstar}) and hence $N^*$ can be determined. Similarly, at any values of $c^{\text{eff}}/c^*$ and $\epsilon_{st}$, the number of beads $\bar{N}$ can be determined from the mean length $\bar{L}$, which is given by~\eref{meanL}, where the parameters $A$, $\alpha$ and $\delta$ are known in both concentration regimes. Finally, since the constant $B$ is known, the crossover radius of gyration $R^*_{g_L}$ can be calculated for any value of $N^*$. It follows that at any value of $c^{\text{eff}}/c^*$ and $\epsilon_{st}$, the value of the radius of gyration $R_g$ can be found rapidly using the expression $R_g = R^*_{g_L} \left( \bar{N}/N^*\right )^\nu$.

\begin{figure}[t]
\begin{center}
\includegraphics[width=8.5cm]{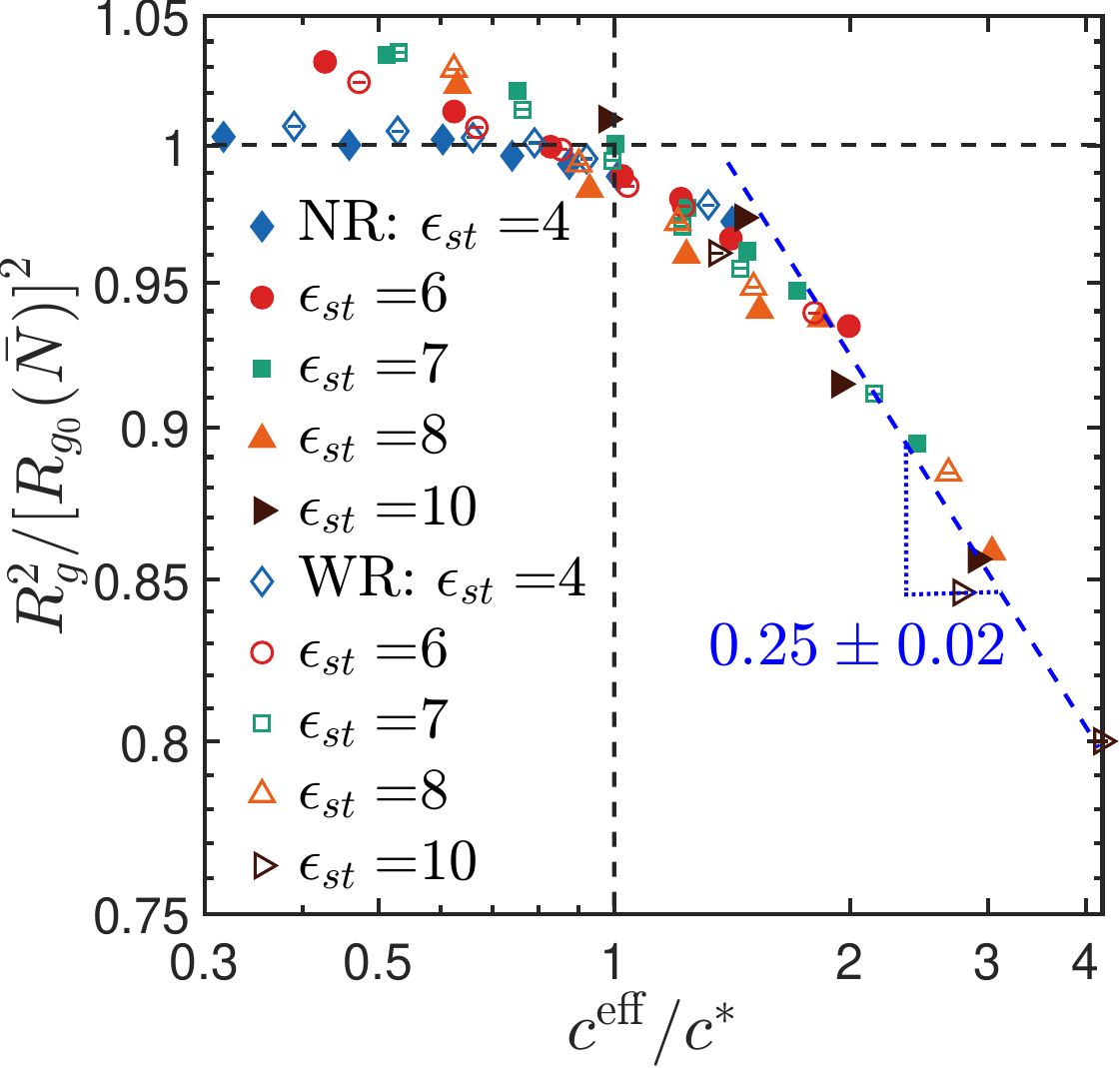}
\end{center}
\vspace{-10pt}
\caption{Dependence of the ratio of the squared radius of gyration of linear wormlike micelles $R_g^2$ to the squared radius of gyration of a homopolymer chain in the dilute limit $R_{g0}^2$, as a function of the scaled concentration $c^{\text{eff}}/c^*$. Note that at any value of $c^{\text{eff}}/c^*$ and $\epsilon_{st}$, $R_{g0}^2$ is evaluated for $ N_b = \bar{N}$, which is the number of beads in a linear wormlike micelle with length $\bar{L}$. Systems with (WR) and without rings (NR) have been considered at various values of sticker energy $\epsilon_{st}$.
\label{fig20}}
\end{figure}

It was seen above that in the dilute regime, the radius of gyration $R_g$ scales identically with chain length $\bar{N}$, as the crossover radius of gyration $R^*_{g_L}$ does with $N^*$ (see~\fref{fig19}~(a)), which in turn scales identically as the radius of gyration $R_{g0}$ with chain length $N_b$, of a homopolymer chain in an athermal solvent (see inset to~\fref{fig18}~(c)). We expect therefore that in the dilute regime, at any value of $c^{\text{eff}}/c^*$ and $\epsilon_{st}$, $R_g = R_{g0}(N_b = \bar{N})$, where $\bar{N}$ is the number of beads in a linear wormlike micelle with length equal to the mean length $\bar{L}$, at $c^{\text{eff}}/c^*$ and $\epsilon_{st}$. In~\fref{fig20}, the ratio $R_g/R_{g_0}(\bar{N})$ is plotted as a function of $c^{\text{eff}}/c^*$, at various values of $\epsilon_{st}$, for systems with and without rings. The data collapses on to a master plot, with a value of the ratio close to one in the dilute regime, as expected, and a shrinkage in size with increasing concentration due to Flory screening. With increasing concentration, the slope of the power law region becomes equal to 0.25, which is consistent with well known scaling law for homopolymer chains under athermal solvent conditions in the semidilute regime, $R_g^2/{R_{g0}^2} \sim \left ( c/c^* \right)^{(2\nu-1)/(1-3\nu)}$. 

\begin{figure*}[t]
\begin{center}
\begin{tabular}{cc}
\includegraphics[width=8.5cm]{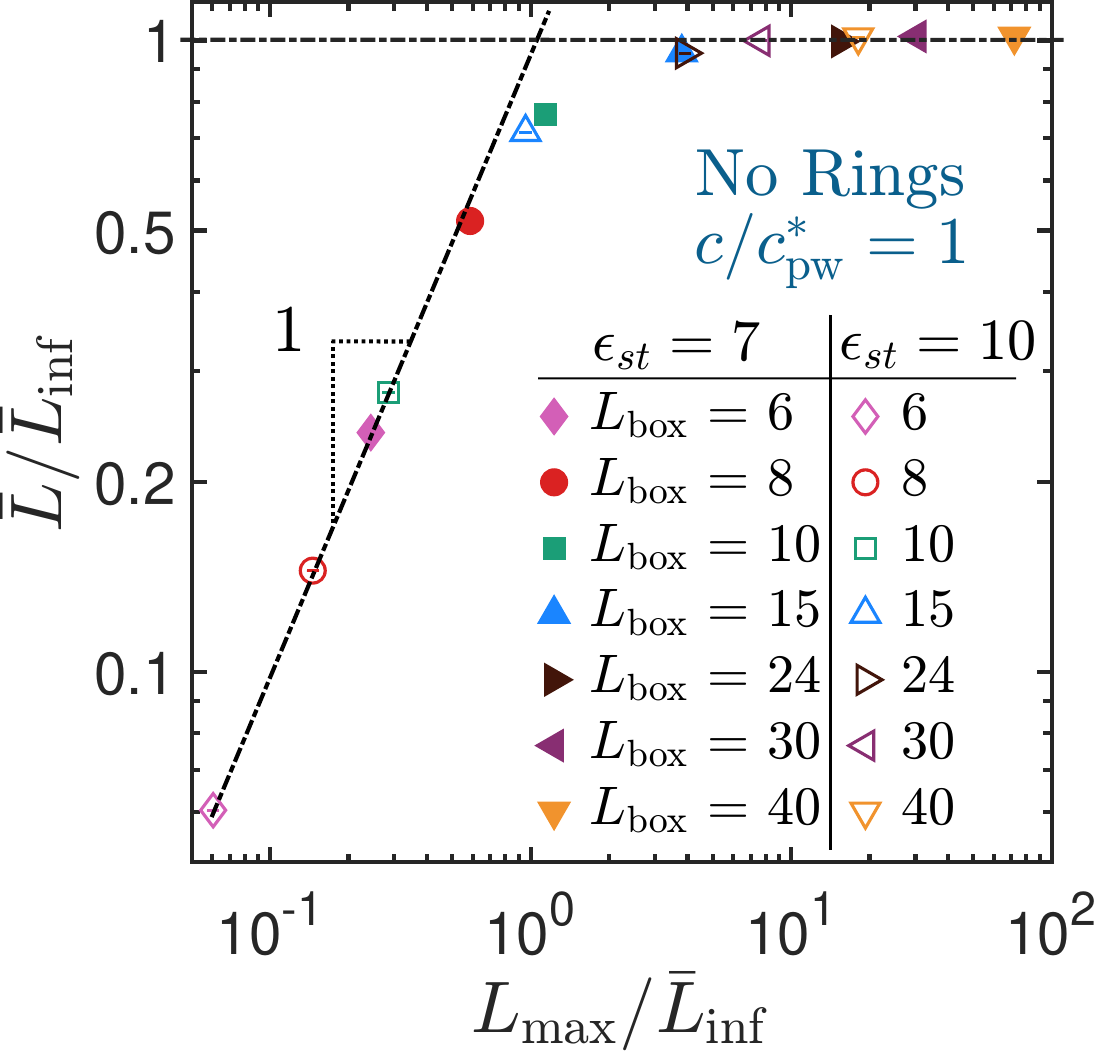} &
\includegraphics[width=8.5cm]{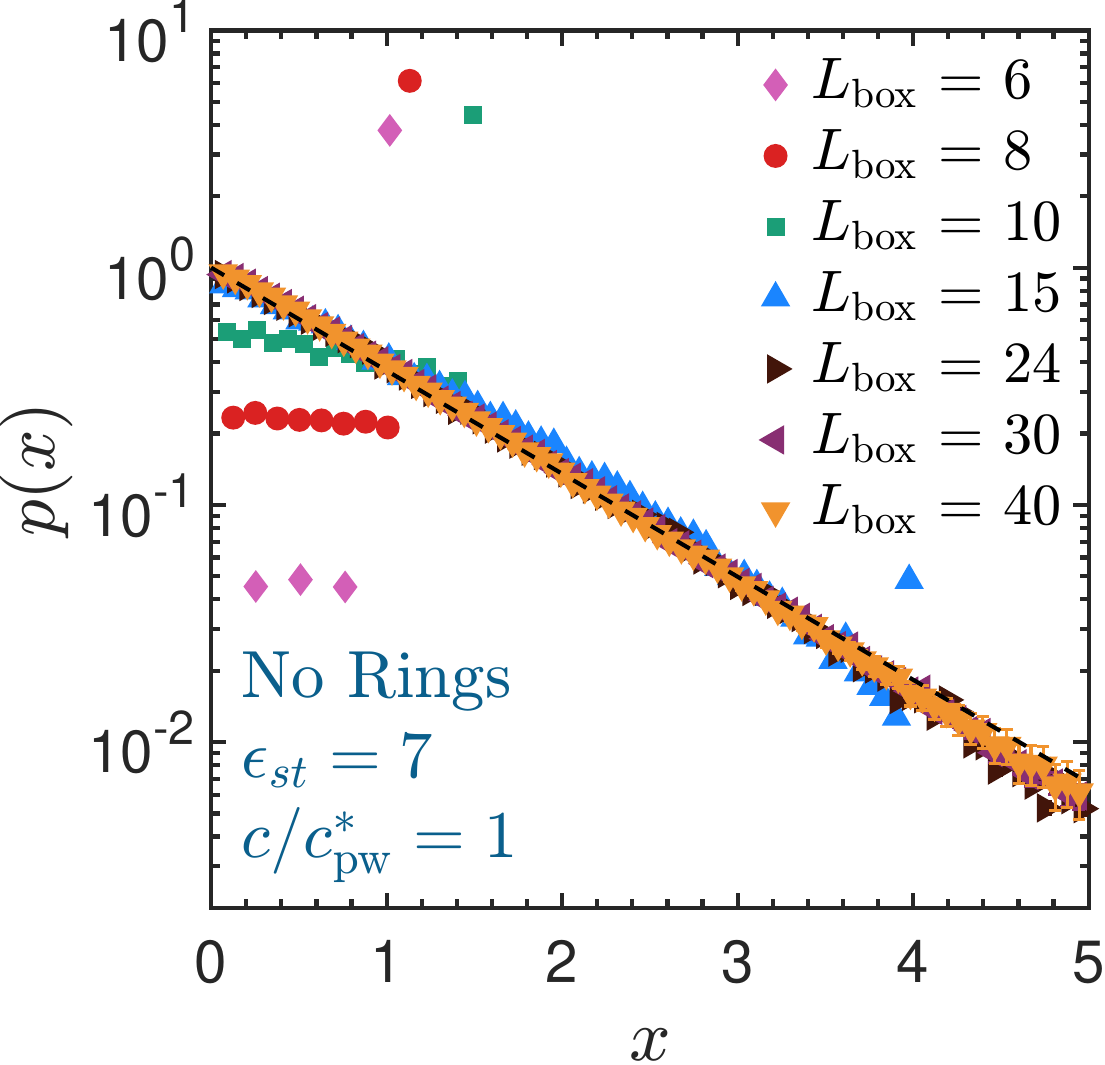} \\
(a)  & (b)
\end{tabular}
\end{center}
\vspace{-10pt}
\caption{Dependence of mean length and scaled length distribution on simulation box size for a semidilute solution with $c/c_{\text{pw}}^*=1$. (a) Reduced mean length of linear wormlike micelles $\bar{L}/\bar{L}_{\text{inf}}$ versus the ratio $L_{\text{max}}/\bar{L}$ for various simulation box sizes $L_{\text{box}}$, at two values of the sticker strength $\epsilon_{st} = 7$ and $10$. (b) Scaled length distribution $p(x)$ versus scaled length $x= L/\bar{L}$, for various simulation box sizes at sticker strength $\epsilon_{st} = 7$. \label{fig21}}
\end{figure*}

It should be noted that many of the observations reported here with regard to the scaling of the radius of gyration for linear wormlike micelles have been reported previously in the pioneering paper by~\citet{Wittmer1998}. Here, however, systems with and without rings have been considered and some of the observations, such as the results in~\fref{fig19}, are discussed for the first time.

\subsubsection{\label{sec:Fse} Finite-size effects}

Finite-size effects in simulations may be considered to be responsible for: (1) a failure to observe predictions of power law scaling due to the exhibition of non-universal behaviour, (2) the existence of too few chains in a system to determine averages sufficiently accurately to observe behaviour predicted for long chains, and (3) chains in a simulation box interacting with their own periodic images. The first of these signs was observed here in~\fref{fig11}, where the predicted scaling behaviour for the dependence of the ratio $\bar{L}/L^*$ on the scaled concentration $c^{\text{eff}}/c^*$ was not observed at low concentrations and sticker energies. This was essentially due to the mean length of linear wormlike micelles being too short, typically with $\bar{L} < 20$, with the finite-size effect disappearing with increasing concentrations and sticker energies, both of which favour longer chains. The second sign of a finite-size effect was observed in~\fref{fig18}, where there are sizeable fluctuations in the predicted radius of gyration for chains with large values of the number of effective monomers, $N_{\mathrm{L,lin}}^{\mathrm{eff}}$. The poor statistics are a consequence of an insufficient number of long chains in the simulation box. The influence of finite-size effects and the minimum box size necessary to avoid these effects, is examined here systematically following the procedure introduced previously by~\citet{Wittmer1998}. 

The problem of chains being too short is only an issue when it comes to the predictions of universal scaling behaviour, since the latter is valid only for sufficiently long chains, and does not indicate a concern with the simulations themselves in terms of the choice of the simulation box size. However, in the case of points (2) and (3) above, the problem can be remedied by choosing a sufficiently large box size. 

\begin{table*}[t]
\vskip-10pt
\centering
\caption{Demonstration of finite-size effects. The simulation box size $L_{\text{box}}$ is varied at a fixed semidilute concentration of $c/c_{\text{pw}}^*=1$, for two different sticker energies $\epsilon_{st}$. The maximum possible length  of a linear wormlike micelle $L_{\text{max}} =  n_{\text{pw}}^T \ell_{\text{pw}}$, where $n_{\text{pw}}^T$ is the number of persistent worms in a simulation box, and $\ell_{\text{pw}} =6$ is the rest length for the trumbbells used here as persistent worms. $\bar{L}_{\text{inf}}$ represents the scaling prediction for the mean length of a wormlike micelle (see \eref{meanL}. The mean length of linear wormlike micelles, $\bar{L}$, the effective concentration  $c^{\text{eff}}$, and the end-to-end vector $\langle R_e\rangle$, are obtained directly from simulations. $c_{\text{inf}}^*$ is the box size independent value of the overlap concentration for the given values of  $\epsilon_{st}$. }
\label{table:finite_size_effect}
\vskip5pt
\bgroup
\setlength{\tabcolsep}{1.8em}
{\def\arraystretch{1.3}
\begin{tabular}{ | c | c | c | c | c | c | c | c | c | c | }
\hline 
 \hline
$\epsilon_{st}$ & {$L_{\text{box}}$}  & {$n_{\text{pw}}^T$} & {$L_{\text{max}}$}  & {$c^{\text{eff}}/c_{\text{inf}}^*$} & {$\bar{L}_{\text{inf}}$}  & {$\bar{L}$} & {$\langle\mathcal{N}_{\text{wlm}}^{\text{lin}}\rangle$}  & {$\langle R_e\rangle$} \\
\hline
\multirow{8}{*}{7} & 6 & 4 & 24 & 3.08 & 98.504 & 23.57 & 1.02 & 8.09 \\ 
 \cline{2-9}
& 8 & 9 & 54 & 2.75 & 92.162 & 47.71 & 1.13 & 13.15 \\
 \cline{2-9}
& 10 & 17 & 102 & 2.61 & 89.36 & 68.25 & 1.49 & 16.90 \\ 
 \cline{2-9}
& 15 & 55 & 330 & 2.46 & 86.73 & 82.97 & 3.97 & 17.35  \\ 
 \cline{2-9}
& 24 & 223 & 1338 & 2.43 & 86.14 & 85.98 & 15.58 & 17.57 \\ 
 \cline{2-9}
& 30 & 436 & 2616 & 2.43 & 86.17 & 86.42 & 29.92 & 17.61 \\ 
 \cline{2-9}
& 40 & 1032 & 6192 & 2.43 & 86.50 & 87.33 & 70.89 & 17.65 \\ 
 \hline
  \hline
\multirow{8}{*}{10} & 6 & 4 & 24 & 6.91 & 396.87 & 23.97 & 1.00 & 7.84 \\ 
\cline{2-9}
& 8 & 9 & 54 & 6.15 & 370.29 & 53.57 & 1.01 & 12.86 \\ 
\cline{2-9}
& 10 & 17 & 102 & 5.81 & 357.61 & 98.94 & 1.03 & 17.89 \\ 
\cline{2-9}
& 15 & 55 & 330 & 5.47 & 344.96 & 246.88 & 1.34 & 29.55  \\ 
\cline{2-9}
& 24 & 223 & 1338 & 5.39 & 342.50 & 326.67 & 4.09 & 33.22 \\ 
\cline{2-9}
& 30 & 436 & 2616 & 5.39 & 342.38 & 341.90 & 7.65 & 34.54 \\ 
\cline{2-9}
& 40 & 1032 & 6192 & 5.39 & 342.07 & 342.15 & 17.88 & 34.61 \\  
\cline{2-9}
 \hline
  \hline
\end{tabular}
}
\egroup
\vspace{-10pt}
\end{table*}

When the sticker energy is high and there are relatively few persistent worms in a simulation box, it is highly likely that all the persistent worms combine to form a single long wormlike micelle, leading to $\bar{L} \approx L_{\text{max}}$, and the number of wormlike micelles in a box $\langle\mathcal{N}_{\text{wlm}}^{\text{lin}}\rangle \approx   L_{\text{max}}/\bar{L} \approx \mathcal{O} (1)$. In this limit, statistics can be expected to be poor, and estimating length distributions for long wormlike micelle lengths becomes problematic. Clearly, the number of persistent worms $n_{\text{pw}}^T$ required to maintain a constant scaled monomer concentration $c/c_{\text{pw}}^*$ increases with increasing simulation box size $L_{\text{box}}$, as does the maximum possible length of a linear wormlike micelle $L_{\text{max}} = n_{\text{pw}}^T \ell_{\text{pw}}$, which are both independent of the value of sticker energy $\epsilon_{st}$. This is demonstrated in the third and fourth columns of~\tref{table:finite_size_effect}, where $n_{\text{pw}}^T$ and $L_{\text{max}}$ are tabulated for various values of $L_{\text{box}}$ at a fixed value of scaled concentration $c/c_{\text{pw}}^* = 1$. When simulations are carried out at this value of $c/c_{\text{pw}}^*$, which corresponds to the semidilute regime, at two different values of $\epsilon_{st} = 7$ and $10$, it can be seen that for small box sizes $L_{\text{box}} \lesssim 10$ (for $\epsilon_{st} = 7$) to $15$ (for $\epsilon_{st} = 10$), the simulated value of mean length $\bar{L} \approx L_{\text{max}}$ and mean number of chains in a box (estimated from simulations) $\langle\mathcal{N}_{\text{wlm}}^{\text{lin}}\rangle \approx \mathcal{O} (1)$. It is also apparent from~\tref{table:finite_size_effect} that for these values of $L_{\text{box}}$, the simulated value of mean length $\bar{L}$ is significantly less than the value of mean length predicted by scaling theory at the given values of concentration and sticker energy, which is denoted here by $\bar{L}_{\text{inf}}$ (corresponding to an infinitely large box size), and given by the scaling expression, $\bar{L}_{\text{inf}}=A_s(c^{\text{eff}})^{0.6}\exp(\delta_s\epsilon_{st})$, with parameter values $A_s =22.11$ and $\delta_s=0.50$, which are valid in the semidilute regime. When the simulation box is relatively small, its finite size influences simulation predictions. Indeed, even the scaled effective concentration $c^{\text{eff}}/c_{\text{inf}}^*$ is seen to be dependent on system size, where $c_{\text{inf}}^*$ is the overlap concentration at the given value of  $\epsilon_{st}$, estimated for a sufficiently large box size such that it is independent of the box size. 

With increasing simulation box size, it can be seen from~\tref{table:finite_size_effect}, that the simulated value of mean length $\bar{L}$ comes closer to $\bar{L}_{\text{inf}}$, and becomes independent of box size, as does the scaled concentration $c^{\text{eff}}/c_{\text{inf}}^*$. The mean number of simulated chains in a box also increases and $\langle\mathcal{N}_{\text{wlm}}^{\text{lin}}\rangle$ becomes roughly of $\mathcal{O} (10)$. Scaling predictions can be expected to be obeyed for simulations carried out under these conditions, as can be seen from the plots displayed in~\frefs{fig21}. 

Figure~\ref{fig21}~(a) is a plot of the ratio $\bar{L}/\bar{L}_{\text{inf}}$ versus the ratio $\bar{L}_{\text{max}}/\bar{L}_{\text{inf}}$ at a fixed  value of $c/c_{\text{pw}}^* = 1$, for two different values of $\epsilon_{st}$. When the box size is small, and $\bar{L} \approx L_{\text{max}}$, both ratios are equal to each other and increase with system size with slope 1, since  $L_{\text{max}} \propto n_{\text{pw}}^T$. For sufficiently large box sizes, however, the ratio $\bar{L}/\bar{L}_{\text{inf}}$ becomes independent of box size and levels off to a constant value of one. In this regime, one expects finite-size effects to become negligible. It should be noted though that the size of the simulation box for which $\bar{L}$ approaches $\bar{L}_{\text{inf}}$ increases with increasing sticker energy $\epsilon_{st}$, thus making it harder to study systems at high sticker energy. 

As noted earlier, when finite-size effects are significant for small box sizes, the system is dominated by the presence of a single large chain with most of the monomers in it. This can be seen clearly in Figure~\ref{fig21}~(b), which is a plot of the scaled length distribution $p(x)$ versus the scaled length $x = L/\bar{L}$, for different box sizes at a fixed  value of $c/c_{\text{pw}}^* = 1$ and sticker energy $\epsilon_{st} = 7$. The exponential distribution is not valid for  $L_{\text{box}} \lesssim 24$, with a peak observable at $x = L_{\text{max}}/\bar{L} \sim \mathcal{O} (1)$, but it becomes an accurate representation of the scaled length distribution for larger box sizes, where finite-size effects can be neglected. 

\begin{figure*}[t]
\begin{center}
\begin{tabular}{cc}
\includegraphics[width=8.5cm]{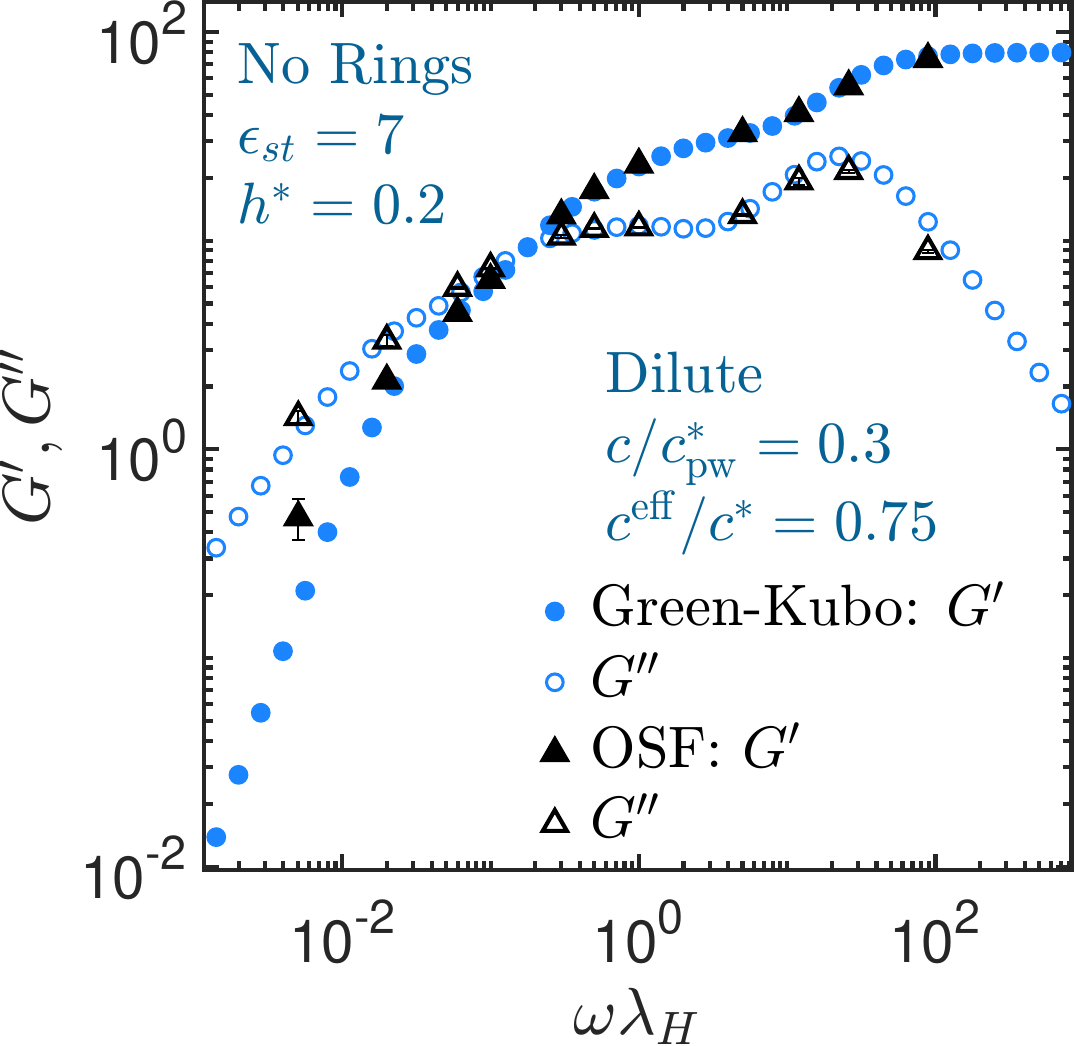} &
\includegraphics[width=8.5cm]{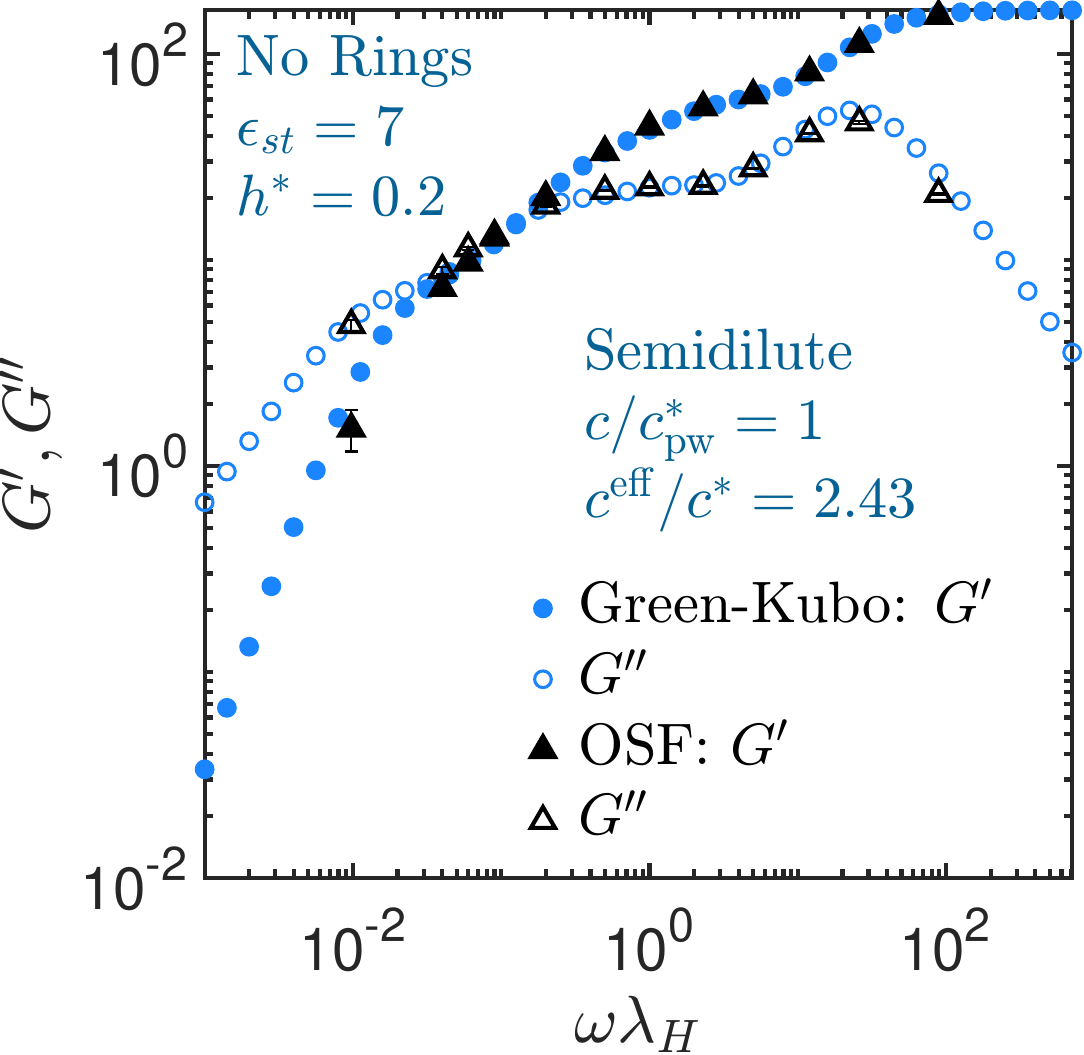} \\
(a)  & (b) 
\end{tabular}
\end{center}
\vspace{-7pt}
\caption{Comparison of the dynamic moduli of wormlike micelle solutions obtained using the Green-Kubo expression~\eref{Gprime} (circles), with simulation results of small amplitude oscillatory shear slow (OSF) (\eref{OSF}) (triangles). The comparison is carried out for (a) a dilute solution with $c^{\text{eff}}/c^* = 0.75$ and (b) a semidilute solution with $c^{\text{eff}}/c^* = 2.43$, both at a sticker energy of $\epsilon_{st} = 7$.   
\label{fig22}}
\end{figure*}

The final sign of finite-size effects is the interaction of chains with their periodic images, which is discussed here in the context of the end-to-end vector $\langle R_e \rangle$ of a linear wormlike micelle. The last column of~\tref{table:finite_size_effect} presents values of $\langle R_e \rangle$ for different box sizes, at a fixed  value of $c/c_{\text{pw}}^* = 1$, for two values of sticker energy $\epsilon_{st}$. It can be seen that for small box sizes $L_{\text{box}} \lesssim 15$ (for $\epsilon_{st} = 7$) and $L_{\text{box}} \lesssim 30$ (for $\epsilon_{st} = 10$), the magnitude of the end-to-end vector $\langle R_e \rangle$ is greater than the box size $L_{\text{box}}$. It is only for larger box sizes that $\langle R_e \rangle  \lesssim  L_{\text{box}}$. Typically, in molecular simulations of polymeric systems, the minimum box size is often chosen to be at least twice the size of the end-to-end vector in order to avoid finite-size effects. Here, a less conservative value is chosen for the box size since the results in~\frefs{fig21} and simulation results presented earlier suggest that scaling laws are obeyed and finite-size effects are negligible provided $\bar{L} \approx \bar{L}_{\text{inf}}$. 

With these considerations in mind, a box size of $L_{\text{box}} = 24$ for $\epsilon_{st} < 9$ and $L_{\text{box}} = 40$ for $\epsilon_{st} \geq 9$ have been used for all the simulation results reported here. These values lie in the asymptotic plateau region where $\bar{L}/\bar{L}_{\text{inf}} =1$, as shown in \fref{fig21}~(a). For both choices, the average number of wormlike micelles in the simulation box is sufficiently large, i.e. $\langle\mathcal{N}_{\text{wlm}}^{\text{lin}}\rangle \gtrsim 15$, to be able to establish a sufficiently wide distribution of lengths.

\subsection{\label{sec:gprime} Storage and loss moduli}

The linear viscoelastic behaviour of wormlike micellar solutions is examined here through the calculation of the storage and loss moduli, $G^\prime$ and $G^{\prime\prime}$. The purpose of this section is not to carry out a detailed study of these material functions, but is rather  meant to be a preliminary demonstration of the capabilities of the mesoscopic model introduced in this work. In particular, since the novel aspect of the algorithm used here is the inclusion of hydrodynamic interactions, the goal is to investigate briefly the influence of hydrodynamic interactions on the dependence of $G^\prime$ and $G^{\prime\prime}$ on the frequency of oscillation $\omega$. For simplicity, only the model that does not allow the formation of rings is considered, restricting the discusssion to solutions consisting  purely of linear wormlike micelles. A more thorough examination of the linear viscoelasticity of wormlike micellar solutions with and without rings will be presented in a future publication.

\subsubsection{\label{sec:gprimevalid} Green-Kubo versus small amplitude oscillatory shear flow}

It is common to estimate $G^\prime$ and $G^{\prime\prime}$ in simulations~\cite{Cruz2012} by emulating the experimental protocol of imposing an oscillatory shear strain on the system, $\gamma_{yx} (t) =  \gamma^0 \sin(\omega t)$, where $\gamma^0$ is the amplitude of oscillation (which is kept small to ensure that only linear behaviour is probed), and measuring the shear stress $\tau_{yx}^{\text{p}}$ that is developed in the fluid in response to the oscillatory strain. $\tau_{yx}^{\text{p}}$ oscillates with the same frequency, but is out of phase with the strain~\cite{Bird1987}, $\tau_{yx}^{\text{p}} = -A (\omega) \gamma^0 \sin(\omega t + \delta)$. By defining the storage and loss moduli through the expression,  $\tau_{yx}^{\text{p}} =  - G^\prime  \gamma^0 \sin(\omega t) - G^{\prime\prime} \gamma^0 \cos(\omega t)$, it follows that~\cite{Bird1987},
\begin{equation}
    G^\prime (\omega) = - A (\omega) \cos \delta \,; \quad 
    G^{\prime\prime} (\omega) = - A (\omega) \sin \delta    \label{OSF}
\end{equation}
The inconvenience of using this procedure is that the protocol must be repeated for each value of $\omega$ for which the values of $G^\prime$ and $G^{\prime\prime}$ are desired. An alternative procedure is to use the Green-Kubo expression and estimate the storage and loss moduli over the entire frequency range by carrying out equilibrium simulations, as outlined in~\sref{sec:propertydefs}. The disadvantage of this approach is that it is often computationally expensive since the shear relaxation modulus $G(t)$ has to be estimated with high accuracy for long times so that it can be fitted precisely and Fourier transformed to obtain $G^\prime$ and $G^{\prime\prime}$. The latter procedure is adopted here with a sufficiently large number of trajectories to obtain good statistics, and the results are displayed  in~\frefs{fig22} (filled and empty circles) for two concentrations, one in the dilute concentration regime and the other in the semidilute regime. The departure in the shape of the curves from the well known shapes for homopolymers  in the intermediate frequency regime is immediately apparent. For homopolymers, in this regime, $G^\prime$ and $G^{\prime\prime}$ scale with frequency as power laws, $\sim \omega^{1/2}$ in the Rouse model (without hydrodynamic interactions) and $\sim \omega^{1/3\nu}$ in the Zimm model (with hydrodynamic interactions)~\cite{doi-edwards,Rubinstein2003}, as will be discussed in more detail below. On the other hand, the curves for $G^\prime$ and $G^{\prime\prime}$ show many interesting features in addition to the power law regime. The microscopic origin of these additional features, and their connection to various physical phenomena occurring on the molecular scale will be discussed in a forthcoming publication. However, it is important to establish that they are not numerical artefacts of the procedure adopted here to obtain the dependence of $G^\prime$ and $G^{\prime\prime}$ on $\omega$. In order to do this, the storage and loss moduli were also calculated at various individual frequencies by carrying out the conventional small amplitude oscillatory shear flow protocol discussed above, with $\gamma^0 = 0.2$. Results obtained by this procedure are displayed  in~\frefs{fig22} by the filled and empty triangles. It is clear that the novel characteristics of the curves observed for wormlike micellar solutions are indeed a genuine reflection of underlying physics, and must be related to the fact that unlike in the case of homopolymers, wormlike micelles undergo scission and rejoining, and their solutions are highly polydisperse in nature.

\subsubsection{\label{sec:homopolymer} Comparison with monodisperse homopolymers}

\begin{figure}[b]
\vspace{-15pt}
\centering
\resizebox{0.90\columnwidth}{!}{\includegraphics{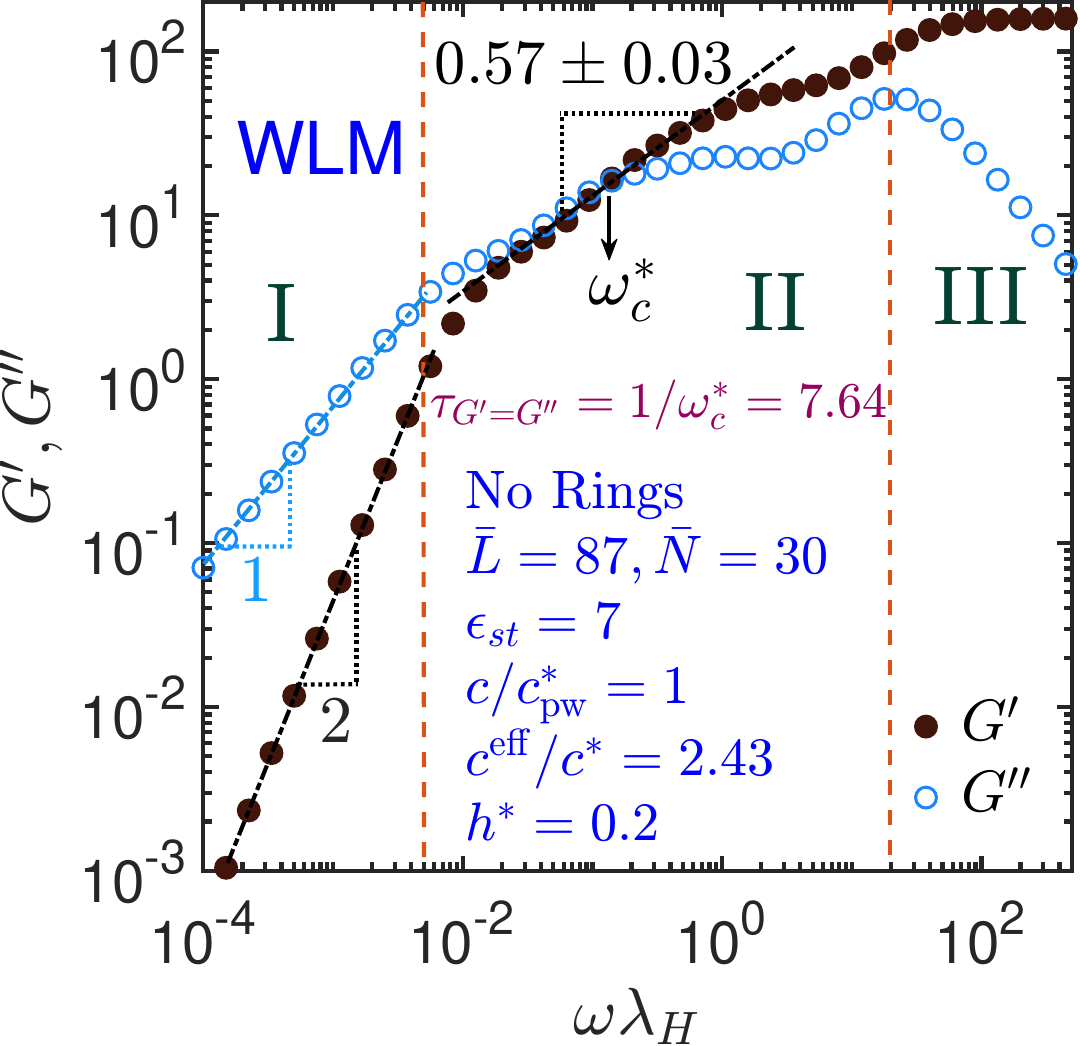}}\\
(a) \\[10pt]
\resizebox{0.93\columnwidth}{!}{\includegraphics{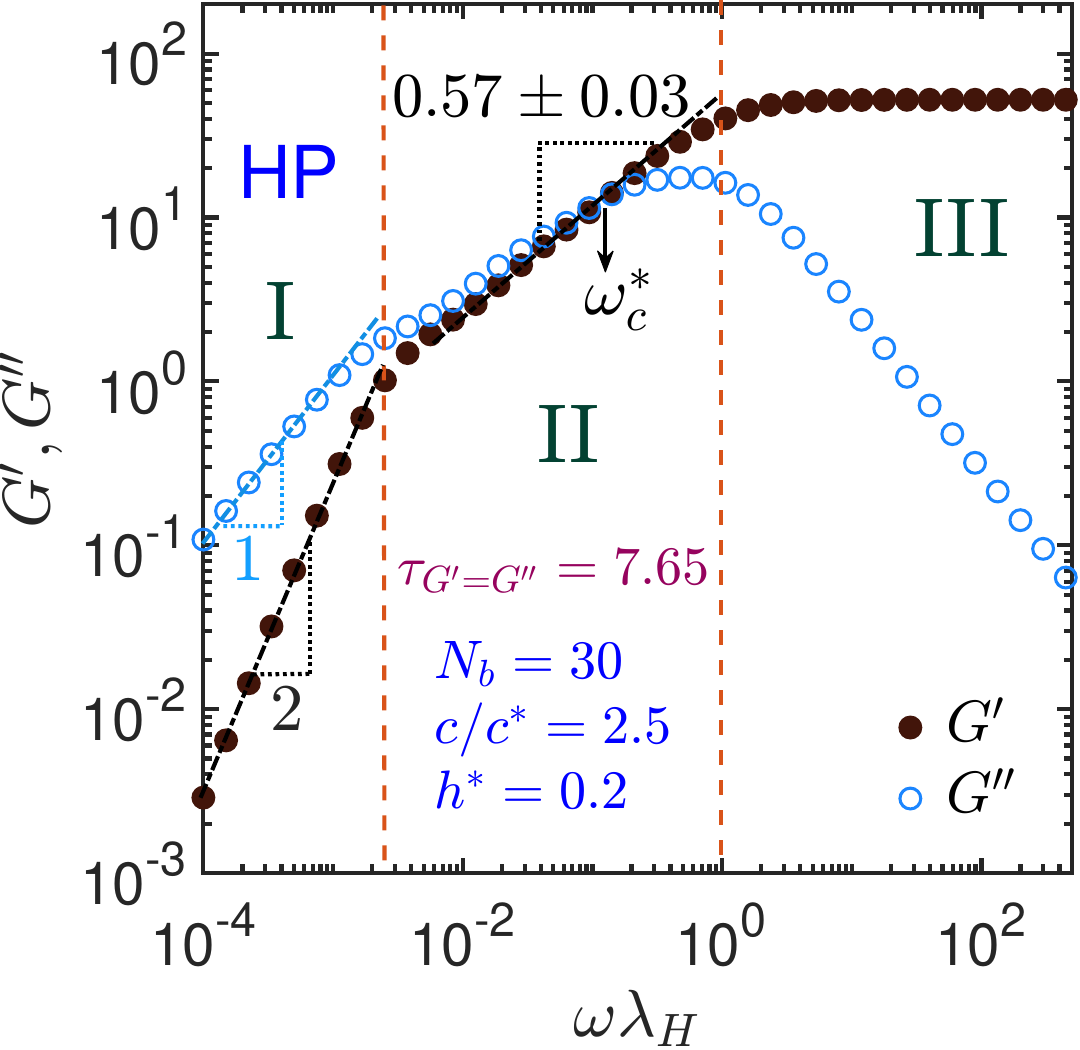}}\\
(b)
\vspace{-6pt}
\caption{Comparison of the dynamic response of (a) a polydisperse solution of linear wormlike micelles (WLM) with (b) a monodisperse homopolymer solutions (HP). The homopolymer solution has a scaled concentration of $c/c^* = 2.5$, which is close to the effective scaled concentration for the wormlike micellar solution ($c^{\mathrm{eff}}/c^*=2.43$). Each bead-spring chain in the homopolymer solution has ${N}_b = 30$ beads, which is set equal to the number of beads ($\bar{N}$) in a wormlike micelle of mean contour length $\bar{L}$, at the given concentration and sticker energy. 
\label{fig23}}
\end{figure}

A direct comparison of the dependence of the storage and loss moduli on frequency, for wormlike micellar and homopolymer solutions in the semidilute regime, is carried out in~\frefs{fig23}. In order to make the comparison meaningful, the homopolymer chain length and solution concentration are chosen to match the mean chain length and overlap concentration in the wormlike micellar solution. For a sticker energy $\epsilon_{st} = 7$ and persistent worm concentration $c/c_{\mathrm{pw}}^* = 1$, the scaled overlap concentration turns out to be $c^{\mathrm{eff}}/c^*=2.43$, with the mean contour length of wormlike micelles being $\bar{L} = 87$, which corresponds to a mean number of beads $\bar{N} = \bar{L}/b + 1 = 30$. For comparison, a monodisperse homopolymer solution of bead-spring chains, with $N_b = \bar{N} = 30$ beads in a chain, was simulated at a semidilute concentration of $c/c^* = 2.5$, under athermal solvent conditions in the presence of hydrodynamic interactions, with $h^* = 0.2$. It is clear from~\frefs{fig23}~(a) and~(b) that the behaviour of both solutions are similar to each other in the low frequency (indicated by I) and high frequency regimes  (indicated by III). In the former, the well know slopes of $G^{\prime} \sim \omega^2$ and $G^{\prime\prime} \sim \omega$ observed in the terminal regime for viscoelastic liquids~\cite{Bird1987} is captured, while at high frequencies, $G^{\prime}$ levels off to a constant value characteristic of elastic solids, and $G^{\prime\prime}$ goes to zero as the polymer no longer contributes to the dynamic viscosity. 

Interesting behaviour and a significant difference between the two solutions is observed in the intermediate frequency regime (indicated by II). In the presence of hydrodynamic interactions, the expected power law behaviour is observed for the homopolymer solution, with  $G^{\prime} \sim \omega^{0.57}$ (which corresponds to a value of $\nu$ between $0.58$ and $0.59$). It is harder to observe the power law for  $G^{\prime\prime}$ because of the rapid crossover to the high frequency regime. Much longer chains would have to be simulated to see the power law regime  over an extended frequency range for both material functions. A similar power law behaviour is also observed for a limited range of frequencies in the scaling of $G^{\prime}$ with $\omega$ for wormlike micellar solutions. As discussed above, there are additional features observed in these solutions, including a distinct upturn in the loss modulus $G^{\prime\prime}$, consistent with observations from previous studies~\cite{Zou2019}. The particularly interesting aspect of the comparison between the two solutions which is highlighted here, and that is worth noting, is that the longest relaxation time is identical in both cases.

The longest nondimensional relaxation time is determined here as the inverse of the frequency at which the storage and loss moduli curves intersect, denoted by $\tau_{G^{\prime}=G^{\prime\prime}} = 1/\omega_c^*$, where $\omega_c^*$ is the nondimensional crossover frequency. For the parameter values used to determine the simulation data in~\frefs{fig23}, $\omega_c^* = 0.13$ for both systems. As a result, at long times, monodisperse homopolymer solutions and polydisperse wormlike micellar solutions appear to relax on the same timescale when the homopolymer concentration matches the effective concentration of wormlike micelles, with the homopolymer chain length being the same as the mean length of the wormlike micelles. This dynamic similarity mirrors in some sense the static similarity seen earlier in~\sref{sec:radius_of_gyration}, where the mean radius of gyration for a  wormlike micellar solution was seen to be equivalent to that of a homopolymer solution with chains of length equal to the mean length $\bar{L}$. 

\subsubsection{\label{sec:HI} Role of hydrodynamic interactions}

The success of Zimm theory in describing the linear viscoelastic behaviour of dilute homopolymer solutions with quantitative accuracy is well known~\cite{Bird1987,Rubinstein2003}, and unequivocally demonstrates the importance of accounting for hydrodynamic interactions in the development of molecular theories to predict dynamic properties.  As discussed earlier, one of the most pronounced manifestations of the difference between Zimm and Rouse theories is in the predicted power law scaling in the intermediate frequency regime. Further, it is also well known that hydrodynamic interactions are screened with increasing concentration, and that the difference between Rouse and Zimm theories is expected to vanish for concentrated solutions. To our knowledge, the role of hydrodynamic interactions in determining the linear viscoelastic behaviour of wormlike micellar solutions has not been examined so far, neither in terms of its influence on the behaviour of $G^{\prime}$ and $G^{\prime\prime}$ in the intermediate frequency regime, nor in terms of the concentration beyond which it is screened. Here, the influence of hydrodynamic interactions on $G^{\prime}$ and $G^{\prime\prime}$ is studied systematically, with a view to address these questions. It is worth noting that the Zimm model, which is an analytical model, accounts for hydrodynamic interactions in a pre-averaged form, while here fluctuations are taken into account since Brownian dynamics simulations are an exact numerical solution of the governing equations.

\begin{figure*}[t]
\begin{center}
\begin{tabular}{cc}
\includegraphics[height=6.8cm]{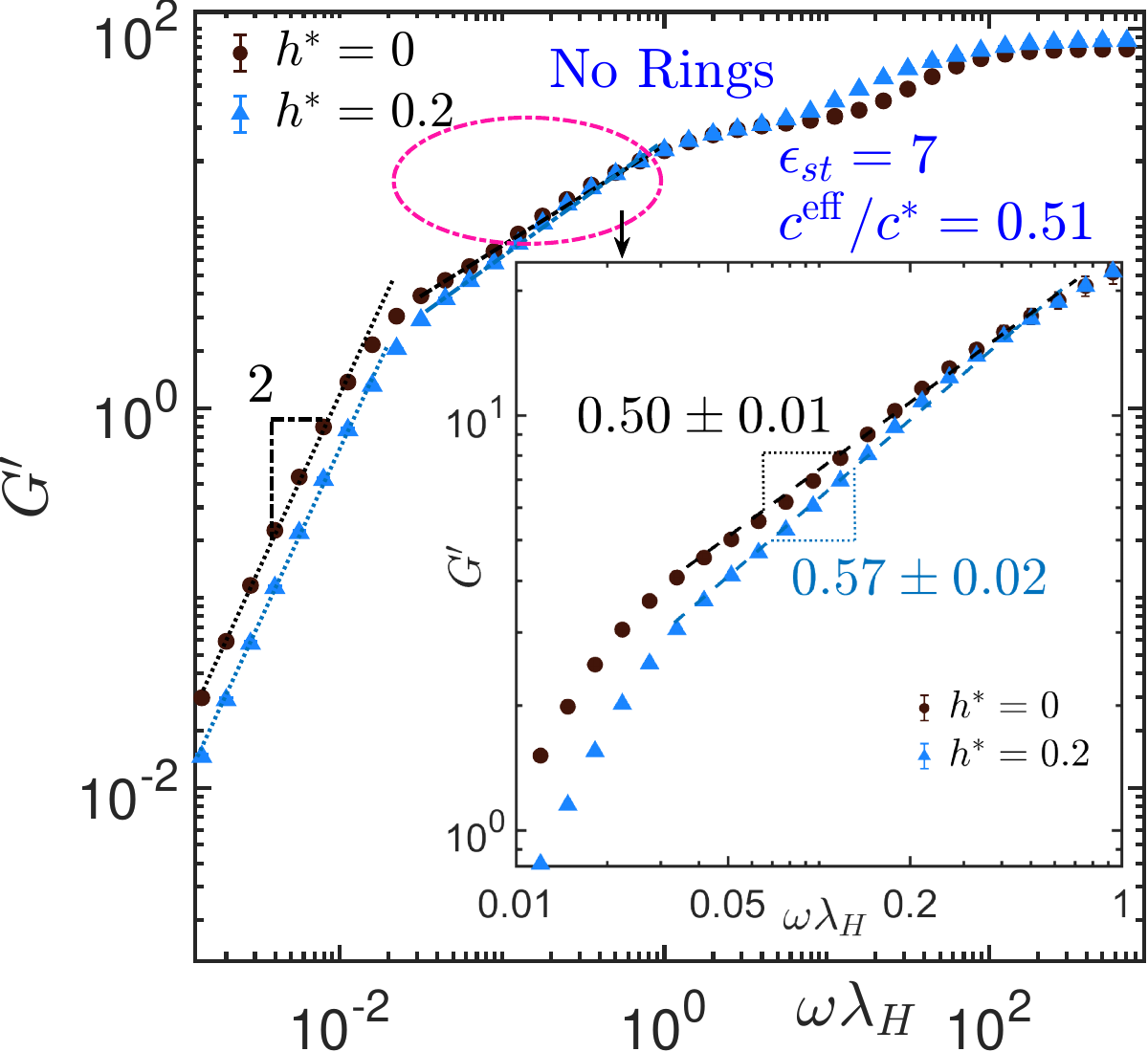} &
\includegraphics[height=6.8cm]{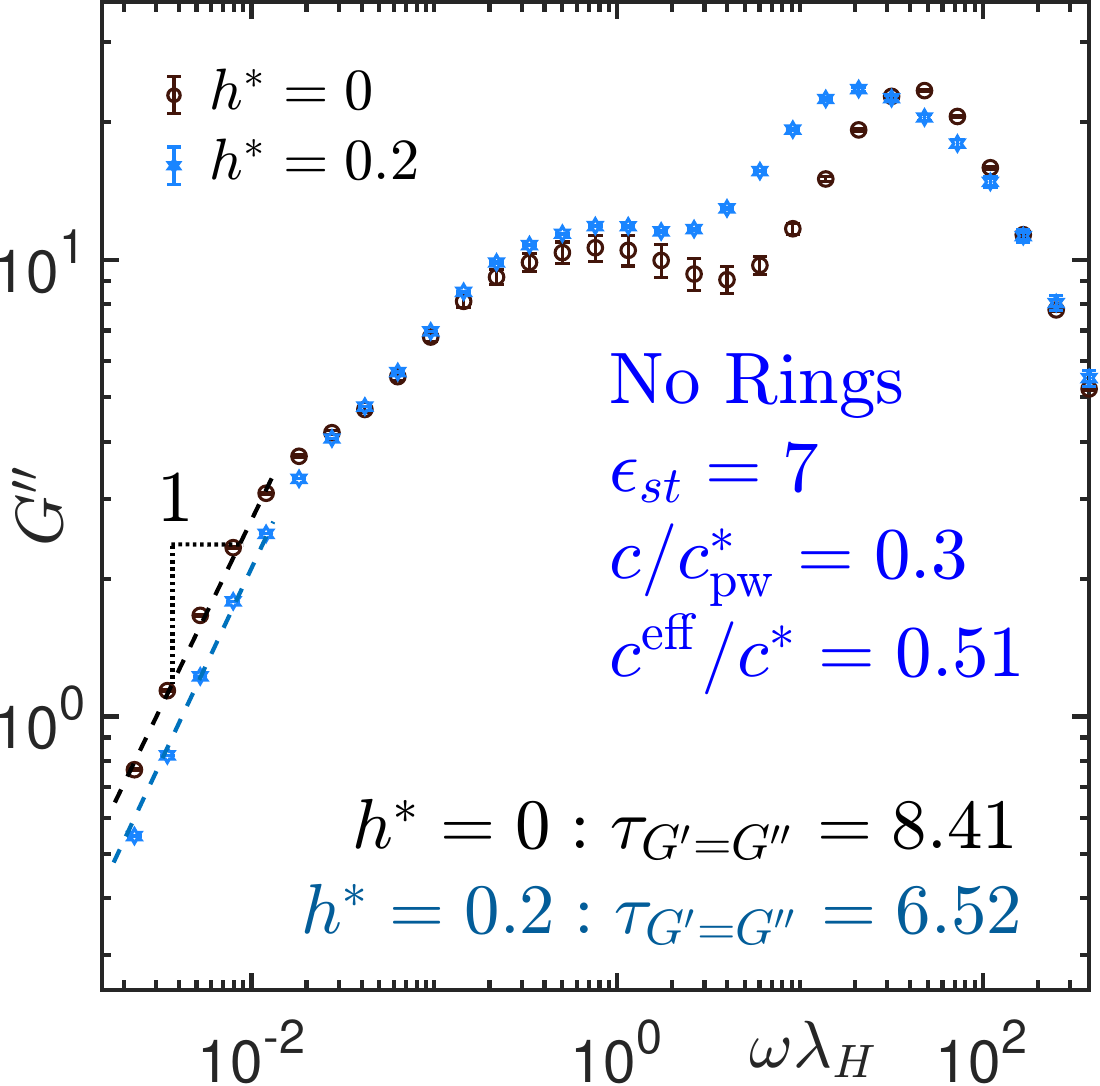} \\[-5pt] 
(a)  & (d) \\
\includegraphics[height=6.8cm]{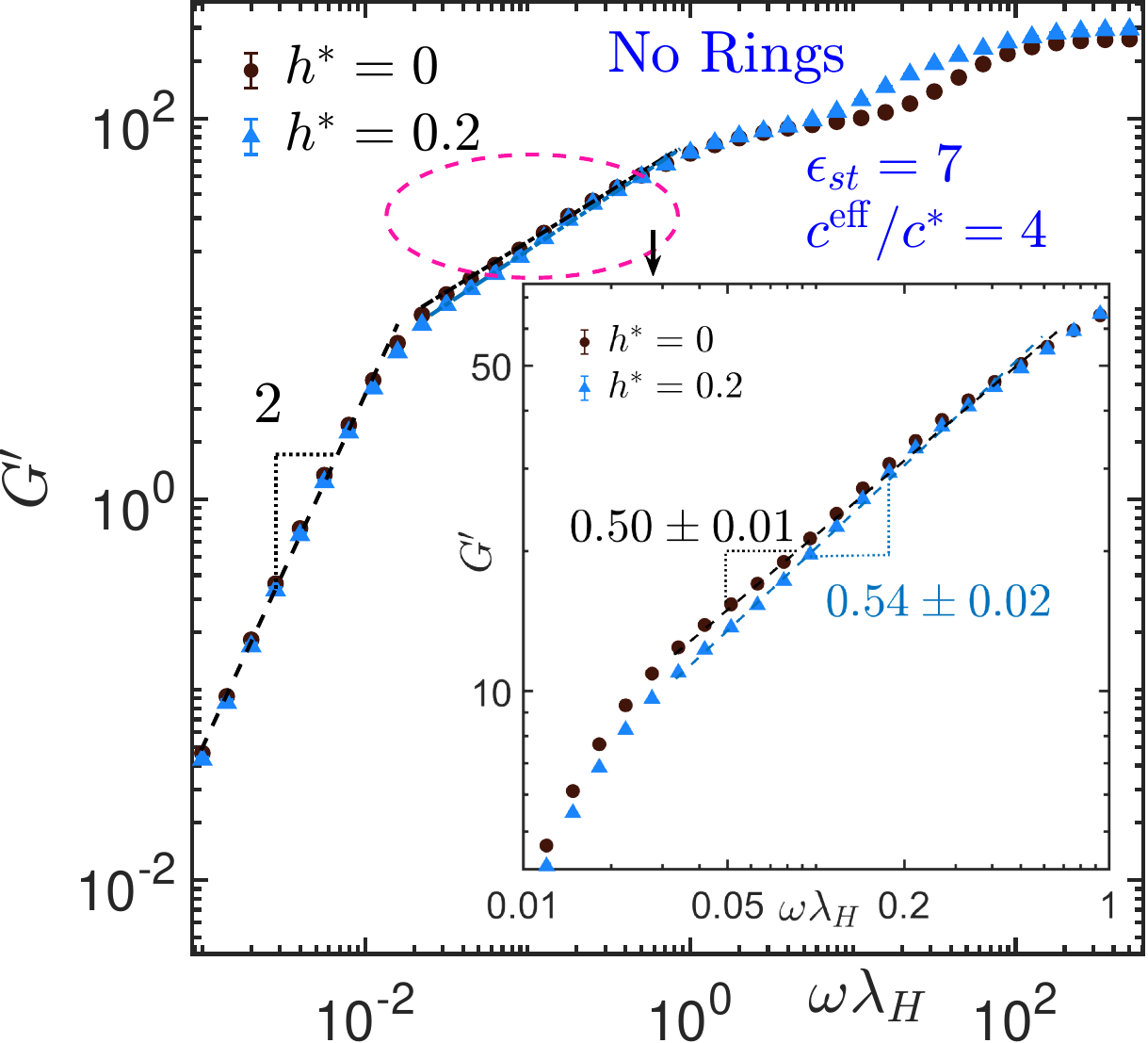} &
\includegraphics[height=6.9cm]{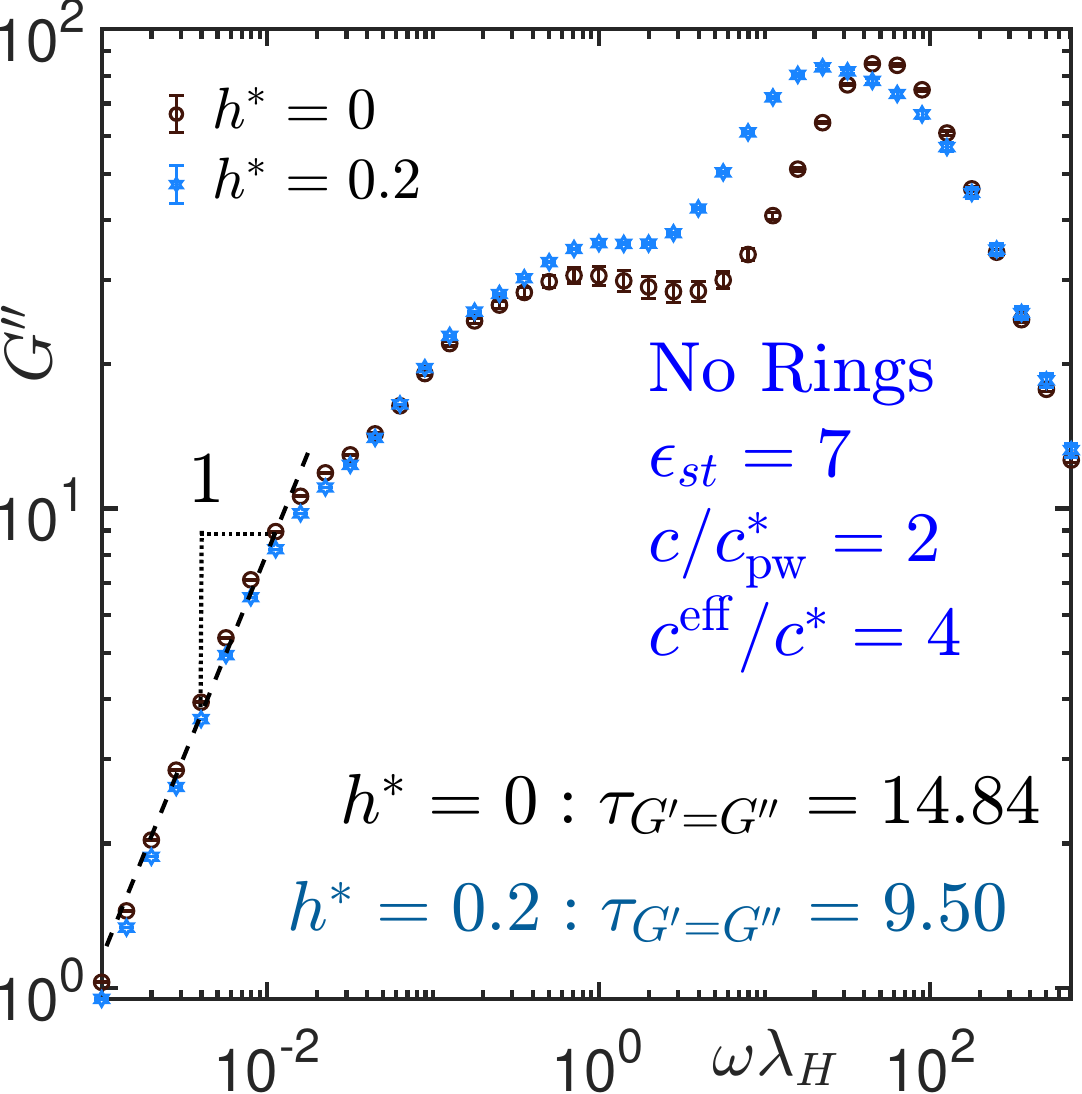}  \\[-5pt]   
(b)  & (e) \\
\includegraphics[height=6.8cm]{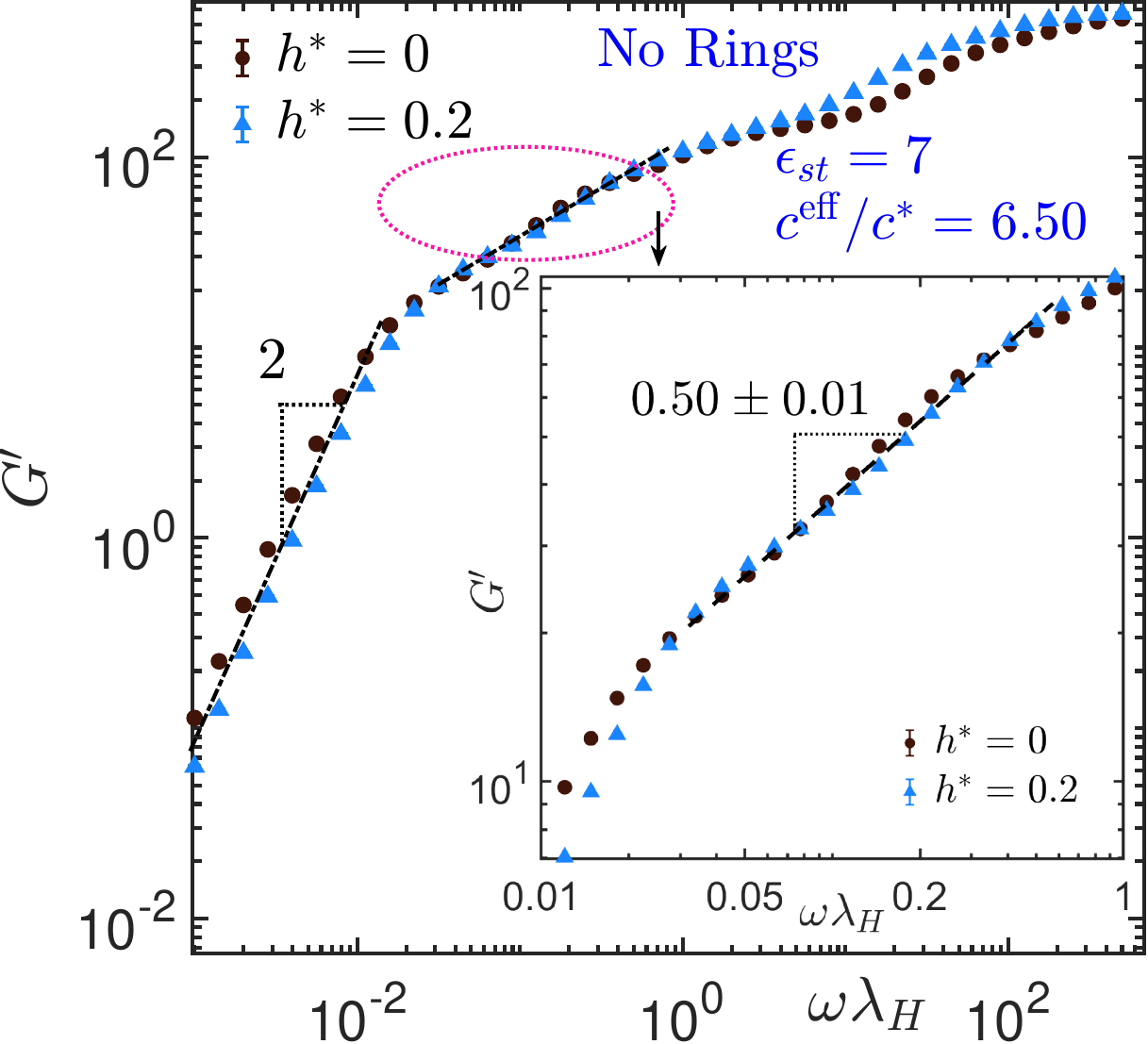} & 
\includegraphics[height=6.9cm]{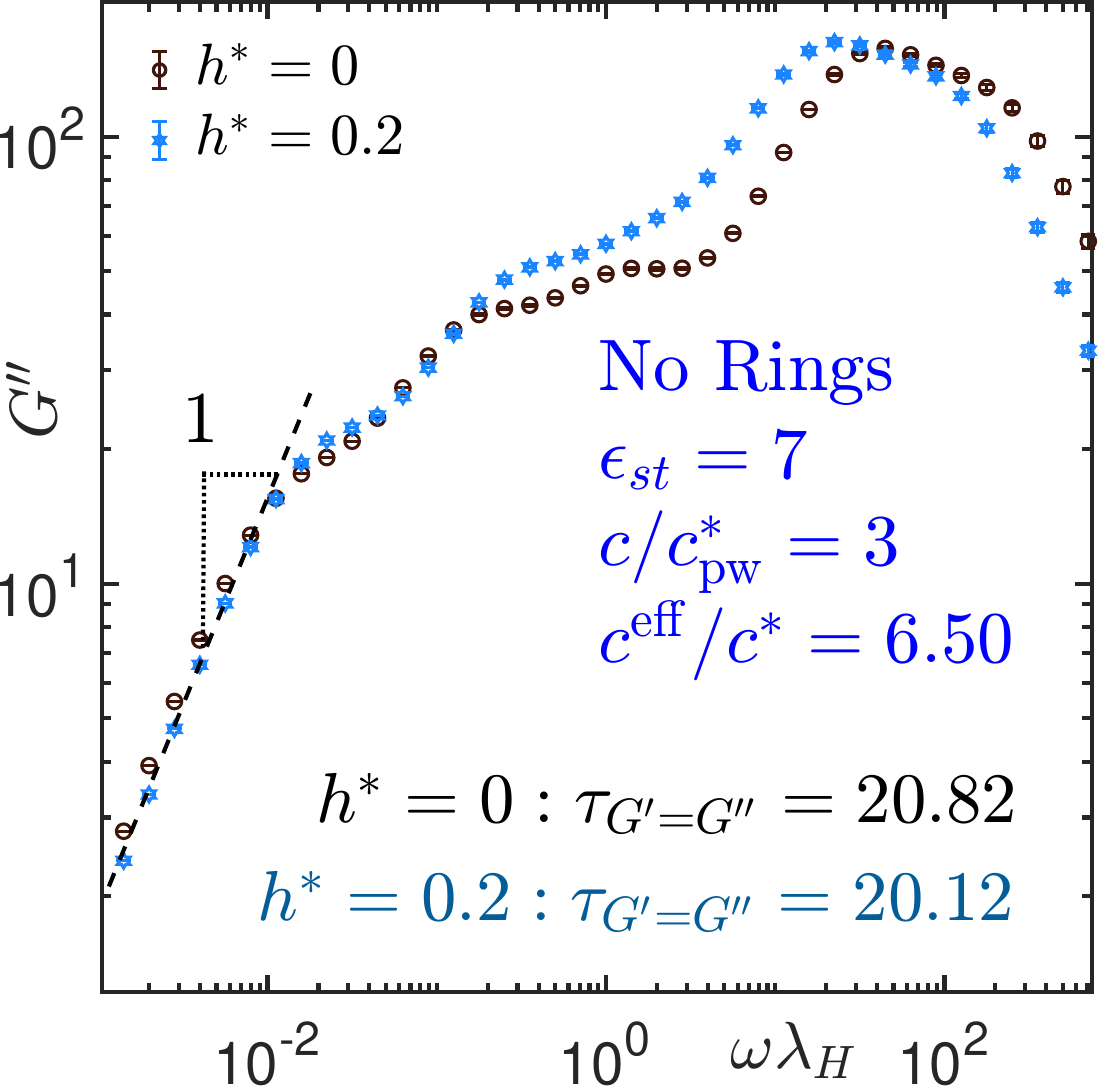}\\[-5pt] 
(c)  & (f) 
\end{tabular}
\vspace{-20pt}
\end{center}
\caption{Illustration of Rouse ($h^*=0$) and Zimm ($h^*=0.2$) dynamics in wormlike micelle solutions with changing concentration in the dilute and semidilute regimes, at a fixed sticker energy $\epsilon_{st}=7$. The first column presents the storage modulus $G^{\prime}$ at three concentrations: (a) $c^{\mathrm{eff}}/c^* = 0.51$, (b) $c^{\mathrm{eff}}/c^* = 4$, and (c) $c^{\mathrm{eff}}/c^* = 6.50$, while the second column (d) to (f), displays the loss modulus $G^{\prime\prime}$ at  the same three concentrations. In all these cases, the error bars in the simulation data are smaller than the symbol size.}
\label{fig24}
\end{figure*}

Figure~\ref{fig24} displays plots of $G^{\prime}$ and $G^{\prime\prime}$ as a function of frequency at three different concentrations in the dilute and semidilute regimes, for both the cases where hydrodynamic interactions are switched on or switched off. The first column [(a) to~(c)] presents the storage modulus $G^{\prime}$, while the second column [(d) to (f)] displays the loss modulus $G^{\prime\prime}$. The concentration is increased systematically, with $c^{\mathrm{eff}}/c^* = 0.51$ in the first row, $c^{\mathrm{eff}}/c^* = 4$ in the second row, and $c^{\mathrm{eff}}/c^* = 6.50$ in the third row. As mentioned earlier, the range of frequencies in the intermediate frequency regime where a power law slope in $G^{\prime}$ is observed is limited because of the appearance of signatures of various molecular phenomena in this regime. This is even more so in the case of $G^{\prime\prime}$. Nevertheless, from the limited range of frequencies where data can be fitted to a power law, it can be seen from the inset to~\fref{fig24}~(a) that the scaling with frequency is Rouse like ($ \sim \omega^{1/2}$) in the absence of hydrodynamic interactions, and Zimm like ($ \sim \omega^{1/3\nu}$, with $\nu \approx 0.6$) when they are taken into account. With increasing concentration, \frefs{fig24}~(a) to~(c) indicate that the screening of hydrodynamic interactions sets in, and the difference in the slopes of the power law regime between the two cases decreases, until it vanishes by a concentration of $c^{\mathrm{eff}}/c^* = 6.50$, at which point only Rouse scaling is observed. Interestingly, a difference in the curves  with and without hydrodynamic interactions persists for both $G^{\prime}$ and $G^{\prime\prime}$ at the high end of frequencies in regime II (before the crossover into regime III), regardless of the increase in concentration. 

For monodisperse homopolymer solutions, for chains of the same length, the magnitude of the longest relaxation time predicted by the Rouse model is always larger than that predicted by the Zimm model, and there is a  significant change in the exponent from 2 to 3/2 in the scaling of the relaxation time with chain length~\cite{Bird1987,Rubinstein2003}. It can be seen from~\frefs{fig24}~(d) to~(f), where the values of the longest relaxation times $\tau_{G^{\prime}=G^{\prime\prime}}$ are reported for the various cases, that accounting for hydrodynamic interactions decreases the magnitude for wormlike micellar solutions as well. While the relaxation time increases with concentration as expected, the presence of hydrodynamic interactions leads to a significant decrease of more than 30\% at all concentrations. It appears impossible to systematically examine through experimental observations, the dependence at a fixed concentration, of the longest relaxation time on chain length, for wormlike micellar solutions. It is definitely worth exploring if this can be done in simulations, but is outside the scope of the present work.

\section{\label{sec:conclusions} Conclusions}

A novel mesoscopic simulation framework has been introduced for modelling ``polyflexible'' and ``polydisperse" wormlike micellar solutions, in the dilute and unentangled semidilute concentration regimes, with the notion of ``persistent worms'' used to represent the shortest possible micelle length. The end beads of persistent worms are treated as sticky beads that can form transient bonds with other sticky beads through a Monte Carlo acceptance probability that depends both on the sticker energy \( \epsilon_{s t} \) (which models the scission energy), and the persistence length \( l_{p} \) (which captures the semiflexibility of wormlike micelles). Intra- and inter-persistent-worm bending potentials have been introduced to ensure a uniform bending stiffness along the backbone of micelles. Brownian dynamics simulations have been carried out, with hydrodynamic interactions incorporated, to compute the static and dynamics properties predicted by the mesoscopic model, with two versions of the algorithm -- one that permits the formation of rings, and another that doesn't. For simplicity, sticky beads are only allowed to associate in pairs, preventing the occurrence of branching. Static properties predicted by the simulations have been validated by comparison with classical results of mean-field and scaling theories, and the linear viscoelastic behaviour of wormlike micelle solutions in the presence of hydrodynamic interactions has been predicted for the first time. The key findings are summarised below.

\begin{enumerate}
  
\item Simulation data for the ratio of the mean length of wormlike micelles to the overlap length,  \(\bar{L} / L^{*}\), as a function of the scaled concentration \( c^{\mathrm{eff}} / c^{*} \), for different sticker energies \(\epsilon_{s t}\), collapse onto a universal master curve obeying \(\bar{L} / L^{*}=\left(c^{\text {eff }} / c^{*}\right)^{\alpha}\), with \(\alpha=0.46\) in the dilute limit and 0.6 in semidilute limit. A crossover occurs in the concentration range $0.7 \lesssim c^{\mathrm{eff}}/c^* \lesssim 1$ (see \fref{fig11}). The influence of sticker energy is captured through the overlap length \(L^{*}\) and the overlap concentration \(c^{*} \), both of which vary exponentially with \(\epsilon_{s t}\)  (see \fref{fig12}). The presence or absence of rings does not impact the scaling of linear chains, provided that the effective monomer concentration, \(c^{\mathrm{eff}}\), is computed only for linear wormlike micelles. Scaling behaviour is not observed at low sticker energies when \(c^{\mathrm{eff}} / c^{*}<0.2\), or if the linear micelles are too short, with \(\bar{L}<20)\).

\item  The scaled length, $x=L / \bar{L}$, of linear wormlike micelles follows a Schultz-Zimm distribution in the dilute limit, $p(x) \propto x^{\gamma-1} \exp (-x)$, with \(\gamma=\gamma_{\text {eff }} = 1.162 \pm 0.002 \) and an exponential distribution in the semi-dilute regime, with \( \gamma=\gamma_{\text {eff }} =1 \pm 0.02 \). A smooth crossover is observed in the value of \(\gamma_{\text {eff }}\) from the dilute to the semidilute regimes. The presence of rings does not affect \(p(x)\) for linear chains, but rather decreases the number of persistent worms available for the formation of linear chains (see \fref{fig13}).

  \item The length distribution of rings, \(\psi_{\mathrm{R}}(L)\), follows an algebraic decay with exponential damping, \(\psi_{\mathrm{R}}(L)=\lambda_{0} L^{-(1+3 \nu)} \exp (-\mu L) H\left(L-L_{\text {min }}^{R}\right)\), where \(L_{\text {min }}^{R}\) is the shortest possible length of rings, which corresponds here to the length of a single persistent worm \(\ell_{\mathrm{pw}}\)  (see \fref{fig15}). Note, the power-law scaling exponent depends on the value of the Flory exponent \(\nu \) in the different concentration regimes. Low effective concentrations \(c^{\mathrm{eff}}\) and high sticker energies are found to favour the formation of rings in the solution (see \frefs{fig16}).

\item The mean radius of gyration of all the linear wormlike micelles in a solution, at a given value of scaled concentration \(\left( c^{\text {eff }} / c^{*} \right) \), is found to obey the scaling law, \(R_{g}=R_{g_{L}}^{*}\left(\bar{N} / N^{*}\right)^{\nu}\), where $\nu$ is the Flory exponent with a value 0.59 in the dilute regime and 0.5 in the semidilute regime, independent of the sticker energy $\epsilon_{s t}$ (see \fref{fig19}~(b)). Here, \(R_{g_{L}}^{*}\) is the radius of gyration of a linear wormlike micelle with overlap length $L^{*}$ (with  $N^{*}$ beads) at $\epsilon_{s t}$, and $\bar{N}$ is the number of beads in a  linear wormlike micelle of length equal to the mean length $\bar{L}$, at the given values of \(\left( c^{\text {eff }} / c^{*} \right) \) and $\epsilon_{s t}$. Additionally, the ratio $R_{g} / R_{g0}$, where $R_{g0}$ is the size of a homopolymer chain with \( \bar{N} \) beads under athermal solvent conditions, is observed to obey the well know scaling law for homopolymers in semidilute solutions, $R_g^2/{R_{g0}^2} \sim \left ( c^{\text{eff}}/c^* \right)^{(2\nu-1)/(1-3\nu)}$, demonstrating the onset of Flory screening with increasing concentration  (see \frefs{fig20}).

\item The dynamic moduli of wormlike micellar solutions exhibit many additional features in the intermediate frequency regime compared to those observed for monodisperse homopolymer solutions. The molecular origin of these features, and their connection to the timescales of microscopic phenomena, will be explored in future studies. The longest relaxation time of polydisperse wormlike micelles, \(\tau_{G^{\prime}=G^{\prime \prime}}\), is found to be identical to that of a monodisperse homopolymer solution, provided that the homopolymer chain length and scaled solution concentration are equal to the mean length of the linear wormlike micelles, and the scaled effective concentration of the wormlike micellar solution, respectively (see \frefs{fig22}~and~\ref{fig23}).

\item Hydrodynamic interactions were observed to significantly affect the scaling with frequency of \( G^{\prime}\) and  \(G^{\prime \prime}\) for linear wormlike micelles, in the intermediate frequency regime. For dilute wormlike micellar solutions, \( G^{\prime}\) obeys Rouse-like scaling \(\left(\sim \omega^{1 / 2}\right)\) in the absence of hydrodynamic interactions, whereas it exhibits Zimm-like scaling (\(\sim \omega^{1 / 3 \nu}\))  in its presence. However, with increasing concentration, hydrodynamic interactions begin to get screened, with Rouse-like scaling observed in the intermediate frequency regime, regardless of the presence or absence of hydrodynamic interactions, for the scaled effective concentration \(c^{\mathrm{eff}} / c^{*} = 6.50\). Additionally, the longest relaxation time (\(\tau_{G^{\prime}=G^{\prime \prime}}\)) decreases significantly with the incorporation of hydrodynamic interactions (see \frefs{fig24}).
  
\end{enumerate}

To our knowledge, this is the first study to systematically examine Rouse and Zimm scaling in the linear viscoelastic behaviour of wormlike micellar solutions, and offers valuable new insights into the dynamics of living polymers. While the current work primarily focuses on flexible wormlike micelles, future investigations will explore the impact of bending stiffness on the static and dynamic properties of semiflexible wormlike micellar solutions.

\begin{acknowledgments}
This research was undertaken with the assistance of resources from the National Computational Infrastructure (NCI Australia), an NCRIS enabled capability supported by the Australian Government. This work was also employed computational facilities provided by Monash University through the DUG, MASSIVE and MonARCH systems. We also acknowledge the funding and general support received from the IITB-Monash Research Academy.
\end{acknowledgments}

\vspace{20pt}
\noindent
\textbf{AUTHOR DECLARATIONS}

\noindent
\textbf{Conflict of Interest}

\noindent
The authors have no conflicts to disclose.

\vspace{20pt}
\noindent
\textbf{DATA AVAILABILITY STATEMENT}

\noindent
The data that supports the findings of this study are available within the article. 

\bibliography{wlm}

\begin{thebibliography}{}
\newcommand{\enquote}[1]{``#1''}

\bibitem[Adams \emph{et~al.}(2018)Adams, Solomon and Larson]{Adams2018}
Adams, A.~A., M.~J. Solomon and R.~G. Larson, \enquote{A nonlinear
  kinetic-rheology model for reversible scission and deformation of unentangled
  wormlike micelles,} J. Rheol. \textbf{62}, 1419--1427 (2018).

\bibitem[Anderson \emph{et~al.}(2020)Anderson, Glaser and Glotzer]{HOOMD2020}
Anderson, J.~A., J.~Glaser and S.~C. Glotzer, \enquote{Hoomd-blue: A python
  package for high-performance molecular dynamics and hard particle monte carlo
  simulations,} Comput. Mater. Sci. \textbf{173}, 109363 (2020).

\bibitem[Bautista \emph{et~al.}(1999)Bautista, De~Santos, Puig and
  Manero]{Bautista1999}
Bautista, F., J.~De~Santos, J.~Puig and O.~Manero, \enquote{Understanding
  thixotropic and antithixotropic behavior of viscoelastic micellar solutions
  and liquid crystalline dispersions. i. {T}he model,} J. Non-Newton. Fluid
  Mech. \textbf{80}, 93--113 (1999).

\bibitem[Berret(2006)Berret]{Berret2006}
Berret, J.-F., \enquote{Rheology of wormlike micelles: Equilibrium properties
  and shear banding transitions,} in \emph{Molecular gels: materials with
  self-assembled fibrillar networks}, pp. 667--720, Springer (2006).

\bibitem[Bhardwaj \emph{et~al.}(2007)Bhardwaj, Miller and
  Rothstein]{Rothstein2007}
Bhardwaj, A., E.~Miller and J.~P. Rothstein, \enquote{Filament stretching and
  capillary breakup extensional rheometry measurements of viscoelastic wormlike
  micelle solutions,} J. Rheol. \textbf{51}, 693--719 (2007).

\bibitem[Bird \emph{et~al.}(1987)Bird, Curtiss, Armstrong and
  Hassager]{Bird1987}
Bird, R.~B., C.~F. Curtiss, R.~C. Armstrong and O.~Hassager, \emph{Dynamics of
  Polymeric Liquids - Volume 2 : Kinetic Theory}, John Wiley and Sons, New York
  (1987).

\bibitem[Boek \emph{et~al.}(2005)Boek, Padding, Anderson, Tardy, Crawshaw and
  Pearson]{Boek2005}
Boek, E., J.~Padding, V.~Anderson, P.~Tardy, J.~Crawshaw and J.~Pearson,
  \enquote{Constitutive equations for extensional flow of wormlike micelles:
  stability analysis of the {B}autista--{M}anero model,} J. Non-Newton. Fluid
  Mech. \textbf{126}, 39--46 (2005).

\bibitem[Carl \emph{et~al.}(1997)Carl, Makhloufi and Kr{\"o}ger]{Kroger1997}
Carl, W., R.~Makhloufi and M.~Kr{\"o}ger, \enquote{On the shape and rheology of
  linear micelles in dilute solutions,} J. Phys. II \textbf{7}, 931--946
  (1997).

\bibitem[Cates(1987)Cates]{Cates1987}
Cates, M., \enquote{Reptation of living polymers: dynamics of entangled
  polymers in the presence of reversible chain-scission reactions,}
  Macromolecules \textbf{20}, 2289--2296 (1987).

\bibitem[Cates and Candau(1990)Cates and Candau]{Cates1990}
Cates, M.~E. and S.~J. Candau, \enquote{Statics and dynamics of worm-like
  surfactant micelles,} J. Phys.: Condens. Matter \textbf{2}, 6869 (1990).

\bibitem[Cates and Fielding(2006)Cates and Fielding]{Cates2006}
Cates, M.~E. and S.~M. Fielding, \enquote{Rheology of giant micelles,} Adv.
  Phys. \textbf{55}, 799--879 (2006).

\bibitem[Chacko \emph{et~al.}(2018)Chacko, Mari, Cates and
  Fielding]{FieldingPRL2018}
Chacko, R.~N., R.~Mari, M.~E. Cates and S.~M. Fielding, \enquote{Dynamic
  vorticity banding in discontinuously shear thickening suspensions,} Phys.
  Rev. Lett. \textbf{121}, 108003 (2018).

\bibitem[Chu \emph{et~al.}(2013)Chu, Dreiss and Feng]{Dreiss2013}
Chu, Z., C.~A. Dreiss and Y.~Feng, \enquote{Smart wormlike micelles,} Chem.
  Soc. Rev. \textbf{42}, 7174--7203 (2013).

\bibitem[Clisby(2017)Clisby]{Clisby2017}
Clisby, N., \enquote{Scale-free {M}onte {C}arlo method for calculating the
  critical exponent $\gamma$ of self-avoiding walks,} J. Phys. A: Math. Theor.
  \textbf{50}, 264003 (2017).

\bibitem[Cruz \emph{et~al.}(2012)Cruz, Chinesta and Régnier]{Cruz2012}
Cruz, C., F.~Chinesta and G.~Régnier, \enquote{Review on the {B}rownian
  dynamics simulation of bead-rod-spring models encountered in computational
  rheology,} Arch. Comput. Methods E \textbf{19}, 227--259 (2012).

\bibitem[Decruppe \emph{et~al.}(1995)Decruppe, Cressely, Makhloufi and
  Cappelaere]{decruppe1995}
Decruppe, J., R.~Cressely, R.~Makhloufi and E.~Cappelaere, \enquote{Flow
  birefringence experiments showing a shear-banding structure in a ctab
  solution,} Colloid Polym. Sci. \textbf{273}, 346--351 (1995).

\bibitem[Dhont and Briels(2008)Dhont and Briels]{Dhont2008}
Dhont, J.~K. and W.~J. Briels, \enquote{Gradient and vorticity banding,} Rheol.
  Acta \textbf{47}, 257--281 (2008).

\bibitem[Doi and Edwards(1986)Doi and Edwards]{doi-edwards}
Doi, M. and S.~F. Edwards, \emph{The Theory of Polymer Dynamics}, Clarendon,
  Oxford (1986).

\bibitem[Dreiss(2007)Dreiss]{Dreiss2007}
Dreiss, C.~A., \enquote{Wormlike micelles: where do we stand? {R}ecent
  developments, linear rheology and scattering techniques,} Soft Matter
  \textbf{3}, 956--970 (2007).

\bibitem[Dutta and Graham(2018)Dutta and Graham]{Dutta2018}
Dutta, S. and M.~D. Graham, \enquote{Mechanistic constitutive model for
  wormlike micelle solutions with flow-induced structure formation,} J.
  Non-Newton. Fluid Mech. \textbf{251}, 97--106 (2018).

\bibitem[Ezrahi \emph{et~al.}(2007)Ezrahi, Tuval and Aserin]{Ezrahi2007}
Ezrahi, S., E.~Tuval and A.~Aserin, \enquote{Properties, main applications and
  perspectives of worm micelles,} Adv. Colloid Interface Sci. \textbf{128-130},
  77--102 (2007).

\bibitem[Fielding(2007a)Fielding]{Fielding2007}
Fielding, S.~M., \enquote{Complex dynamics of shear banded flows,} Soft Matter
  \textbf{3}, 1262--1279 (2007a).

\bibitem[Fielding(2007b)Fielding]{FieldingPRE2007}
Fielding, S.~M., \enquote{Vorticity structuring and velocity rolls triggered by
  gradient shear bands,} Phys. Rev. E \textbf{76}, 016311 (2007b).

\bibitem[Fiore \emph{et~al.}(2017)Fiore, Balboa~Usabiaga, Donev and
  Swan]{PSE2017}
Fiore, A.~M., F.~Balboa~Usabiaga, A.~Donev and J.~W. Swan, \enquote{Rapid
  sampling of stochastic displacements in brownian dynamics simulations,} J.
  Chem. Phys. \textbf{146}, 124116 (2017).

\bibitem[Gamez-Corrales \emph{et~al.}(1999)Gamez-Corrales, Berret, Walker and
  Oberdisse]{Berret1999}
Gamez-Corrales, R., J.-F. Berret, L.~Walker and J.~Oberdisse,
  \enquote{Shear-thickening dilute surfactant solutions: equilibrium structure
  as studied by small-angle neutron scattering,} Langmuir \textbf{15},
  6755--6763 (1999).

\bibitem[Hommel and Graham(2021)Hommel and Graham]{Graham2021}
Hommel, R. and M.~Graham, \enquote{Constitutive modeling of dilute wormlike
  micelle solutions: Shear-induced structure and transient dynamics,} J.
  Non-Newton. Fluid Mech. \textbf{295}, 104606 (2021).

\bibitem[Hommel and Graham(2024)Hommel and Graham]{Graham2024}
Hommel, R.~J. and M.~D. Graham, \enquote{Flow instabilities in circular couette
  flow of wormlike micelle solutions with a reentrant flow curve,} J.
  Non-Newton. Fluid Mech. \textbf{324}, 105183 (2024).

\bibitem[Howard \emph{et~al.}(2019)Howard, Statt, Madutsa, Truskett and
  Panagiotopoulos]{Nlist2019}
Howard, M.~P., A.~Statt, F.~Madutsa, T.~M. Truskett and A.~Z. Panagiotopoulos,
  \enquote{Quantized bounding volume hierarchies for neighbor search in
  molecular simulations on graphics processing units,} Comput. Mater. Sci.
  \textbf{164}, 139--146 (2019).

\bibitem[Hu \emph{et~al.}(1998)Hu, Boltenhagen and Pine]{Pine1998}
Hu, Y., P.~Boltenhagen and D.~Pine, \enquote{Shear thickening in
  low-concentration solutions of wormlike micelles. i. direct visualization of
  transient behavior and phase transitions,} J. Rheol. \textbf{42}, 1185--1208
  (1998).

\bibitem[Huang \emph{et~al.}(2006a)Huang, Xu, Crevel, Wittmer and
  Ryckaert]{Huang2006}
Huang, C.-C., H.~Xu, F.~Crevel, J.~Wittmer and J.-P. Ryckaert,
  \enquote{Reaction kinetics of coarse-grained equilibrium polymers: a brownian
  dynamics study,} in \emph{Computer Simulations in Condensed Matter Systems:
  From Materials to Chemical Biology Volume 2}, pp. 379--418, Springer (2006a).

\bibitem[Huang \emph{et~al.}(2006b)Huang, Xu and Ryckaert]{Ryckaert2006}
Huang, C.-C., H.~Xu and J.-P. Ryckaert, \enquote{Kinetics and dynamic
  properties of equilibrium polymers,} J. Chem. Phys. \textbf{125}, 094901
  (2006b).

\bibitem[Israelachvili(2011)Israelachvili]{Israel2011}
Israelachvili, J.~N., \emph{Intermolecular and surface forces}, Academic press
  (2011).

\bibitem[Jain \emph{et~al.}(2012)Jain, Sunthar, D{\"u}nweg and
  Prakash]{JainPRE2012}
Jain, A., P.~Sunthar, B.~D{\"u}nweg and J.~R. Prakash, \enquote{Optimization of
  a {B}rownian dynamics algorithm for semidilute polymer solutions,} Phys. Rev.
  E \textbf{85}, 066703 (2012).

\bibitem[Kroeger(1998)Kroeger]{Kroger1998}
Kroeger, M., \enquote{Micro/mesoscopic approaches to the ring formation in
  linear wormlike micellar systems,} Macromol. Symp. \textbf{133}, 101--112
  (1998).

\bibitem[Kroeger(2004)Kroeger]{Kroger2004}
Kroeger, M., \enquote{Simple models for complex nonequilibrium fluids,} Phys.
  Rep. \textbf{390}, 453--551 (2004).

\bibitem[Kr\"oger and Makhloufi(1996)Kr\"oger and Makhloufi]{Kroger1995}
Kr\"oger, M. and R.~Makhloufi, \enquote{Wormlike micelles under shear flow: A
  microscopic model studied by nonequilibrium-molecular-dynamics computer
  simulations,} Phys. Rev. E \textbf{53}, 2531--2536 (1996).

\bibitem[Kumar and Saha~Dalal(2022)Kumar and Saha~Dalal]{Indranil2022}
Kumar, P. and I.~Saha~Dalal, \enquote{Fraenkel springs as an efficient
  approximation to rods for brownian dynamics simulations and modeling of
  polymer chains,} Macromol. Theory Simul. \textbf{31}, 2200008 (2022).

\bibitem[Land{\'a}zuri \emph{et~al.}(2016)Land{\'a}zuri, Mac{\'\i}as,
  Garc{\'\i}a-Sandoval, Hern{\'a}ndez, Manero, Puig and
  Bautista]{Landazuri2016}
Land{\'a}zuri, G., E.~R. Mac{\'\i}as, J.~P. Garc{\'\i}a-Sandoval,
  E.~Hern{\'a}ndez, O.~Manero, J.~E. Puig and F.~Bautista, \enquote{On the
  modelling of the shear thickening behavior in micellar solutions,} Rheol.
  Acta \textbf{55}, 547--558 (2016).

\bibitem[Lerouge and Berret(2010)Lerouge and Berret]{Lerouge2010}
Lerouge, S. and J.-F. Berret, \enquote{Shear-induced transitions and
  instabilities in surfactant wormlike micelles,} in \emph{Polymer
  Characterization: Rheology, Laser Interferometry, Electrooptics}, eds.
  K.~Dusek and J.-F. Joanny, pp. 1--71, Springer Berlin Heidelberg, Berlin,
  Heidelberg (2010).

\bibitem[Liu and Pine(1996)Liu and Pine]{Pine1996}
Liu, C.-h. and D.~Pine, \enquote{Shear-induced gelation and fracture in
  micellar solutions,} Phys. Rev. Lett. \textbf{77}, 2121 (1996).

\bibitem[López-Aguilar \emph{et~al.}(2018)López-Aguilar, Webster,
  Tamaddon-Jahromi and Manero]{Manero2018}
López-Aguilar, J., M.~Webster, H.~Tamaddon-Jahromi and O.~Manero,
  \enquote{Predictions for circular contraction-expansion flows with
  viscoelastoplastic and thixotropic fluids,} J. Non-Newton. Fluid Mech.
  \textbf{261}, 188--210 (2018).

\bibitem[López-Aguilar \emph{et~al.}(2022)López-Aguilar, Resendiz-Tolentino,
  Tamaddon-Jahromi, Ellero and Manero]{Manero2022}
López-Aguilar, J.~E., O.~Resendiz-Tolentino, H.~R. Tamaddon-Jahromi, M.~Ellero
  and O.~Manero, \enquote{Flow past a sphere: Numerical predictions of
  thixo-viscoelastoplastic wormlike micellar solutions,} J. Non-Newton. Fluid
  Mech. \textbf{309}, 104902 (2022).

\bibitem[May and Ben-Shaul(2007)May and Ben-Shaul]{May2007}
May, S. and A.~Ben-Shaul, \enquote{Molecular packing in cylindrical micelles,}
  in \emph{Giant Micelles}, pp. 41--80, CRC Press (2007).

\bibitem[Milchev \emph{et~al.}(2000)Milchev, Wittmer and Landau]{Milchev2000}
Milchev, A., J.~Wittmer and D.~Landau, \enquote{Dynamical {M}onte {C}arlo study
  of equilibrium polymers: {E}ffects of high density and ring formation,} Phys.
  Rev. E \textbf{61}, 2959 (2000).

\bibitem[Moorcroft and Fielding(2014)Moorcroft and Fielding]{Moorcroft2014}
Moorcroft, R.~L. and S.~M. Fielding, \enquote{Shear banding in time-dependent
  flows of polymers and wormlike micelles,} J. Rheol. \textbf{58}, 103--147
  (2014).

\bibitem[Nagarajan(2007)Nagarajan]{Nagarajan2007}
Nagarajan, R., \enquote{Molecular thermodynamics of giant micelles,} in
  \emph{Giant Micelles}, pp. 1--40, CRC Press (2007).

\bibitem[Nicolas and Morozov(2012)Nicolas and Morozov]{Morozov2012}
Nicolas, A. and A.~Morozov, \enquote{Nonaxisymmetric instability of
  shear-banded taylor-couette flow,} Phys. Rev. Lett. \textbf{108}, 088302
  (2012).

\bibitem[Nicolas-Morgantini(2007)Nicolas-Morgantini]{Nicolas2007}
Nicolas-Morgantini, L., \enquote{Giant micelles and shampoos,} in \emph{Giant
  Micelles}, pp. 493--514, CRC Press (2007).

\bibitem[Olmsted(2008)Olmsted]{olmsted2008}
Olmsted, P.~D., \enquote{Perspectives on shear banding in complex fluids,}
  Rheol. Acta \textbf{47}, 283--300 (2008).

\bibitem[Padding and Boek(2004)Padding and Boek]{PaddingPRE2004}
Padding, J.~T. and E.~S. Boek, \enquote{Influence of shear flow on the
  formation of rings in wormlike micelles: A nonequilibrium molecular dynamics
  study,} Phys. Rev. E \textbf{70}, 031502 (2004).

\bibitem[Padding \emph{et~al.}(2005)Padding, Boek and Briels]{Padding2005}
Padding, J.~T., E.~S. Boek and W.~J. Briels, \enquote{Rheology of wormlike
  micellar fluids from brownian and molecular dynamics simulations,} J. Phys.:
  Condens. Matter \textbf{17}, S3347 (2005).

\bibitem[Padding \emph{et~al.}(2008)Padding, Boek and Briels]{Padding2008}
Padding, J.~T., E.~S. Boek and W.~J. Briels, \enquote{Dynamics and rheology of
  wormlike micelles emerging from particulate computer simulations,} J. Chem.
  Phys. \textbf{129}, 074903 (2008).

\bibitem[Peterson and Cates(2021)Peterson and Cates]{Peterson2021}
Peterson, J.~D. and M.~E. Cates, \enquote{Constitutive models for
  well-entangled living polymers beyond the fast-breaking limit,} J. Rheol.
  \textbf{65}, 633--662 (2021).

\bibitem[Peterson \emph{et~al.}(2023)Peterson, Zou, Larson and
  Cates]{Peterson2023}
Peterson, J.~D., W.~Zou, R.~G. Larson and M.~E. Cates, \enquote{Wormlike
  micelles revisited: A comparison of models for linear rheology,} J.
  Non-Newton. Fluid Mech. \textbf{322}, 105149 (2023).

\bibitem[Pincus \emph{et~al.}(2023)Pincus, Rodger and Ravi~Prakash]{Isaac2023}
Pincus, I., A.~Rodger and J.~Ravi~Prakash, \enquote{{Dilute polymer solutions
  under shear flow: Comprehensive qualitative analysis using a bead-spring
  chain model with a FENE-Fraenkel spring},} J. Rheol. \textbf{67}, 373--402
  (2023).

\bibitem[Prakash(2019)Prakash]{Prakash2019}
Prakash, J.~R., \enquote{{Universal dynamics of dilute and semidilute solutions
  of flexible linear polymers},} Curr. Opin. Colloid Interface Sci.
  \textbf{43}, 63--79 (2019).

\bibitem[{Quintero F.} \emph{et~al.}(2022){Quintero F.}, Zhou and
  Cook]{Cook2020}
{Quintero F.}, L., L.~Zhou and L.~Cook, \enquote{Transiently linked nonlinear
  fene two species dynamics,} J. Non-Newton. Fluid Mech. \textbf{301}, 104720
  (2022).

\bibitem[Raghavan and Feng(2017)Raghavan and Feng]{Raghavan2017}
Raghavan, S.~R. and Y.~Feng, \enquote{Wormlike micelles: Solutions, gels, or
  both?} in \emph{Wormlike Micelles: Advances in Systems, Characterisation and
  Applications}, The Royal Society of Chemistry (2017).

\bibitem[Robe \emph{et~al.}(2024)Robe, Santra, McKinley and Prakash]{Robe2024}
Robe, D., A.~Santra, G.~H. McKinley and J.~R. Prakash, \enquote{Evanescent
  gels: Competition between sticker dynamics and single-chain relaxation,}
  Macromolecules \textbf{57}, 4220--4235 (2024).

\bibitem[Rouault(1999)Rouault]{Rouault1999}
Rouault, Y., \enquote{Off-lattice {B}rownian dynamics simulation of wormlike
  micelles: {T}he dependence of the mean contour length on concentration,} J.
  Chem. Phys. \textbf{111}, 9859--9863 (1999).

\bibitem[Rouault and Milchev(1995)Rouault and Milchev]{Rouault1995}
Rouault, Y. and A.~Milchev, \enquote{Monte {C}arlo study of living polymers
  with the bond-fluctuation method,} Phys. Rev. E \textbf{51}, 5905 (1995).

\bibitem[Rubinstein and Colby(2003)Rubinstein and Colby]{Rubinstein2003}
Rubinstein, M. and R.~H. Colby, \emph{Polymer Physics}, Oxford University Press
  (2003).

\bibitem[Saadat and Khomami(2016)Saadat and Khomami]{Khomami2016}
Saadat, A. and B.~Khomami, \enquote{{A new bead-spring model for simulation of
  semi-flexible macromolecules},} J. Chem. Phys. \textbf{145}, 204902 (2016).

\bibitem[Salipante \emph{et~al.}(2024a)Salipante, Cromer and
  Hudson]{Hudson2024a}
Salipante, P.~F., M.~Cromer and S.~D. Hudson, \enquote{Two-species model for
  nonlinear flow of wormlike micelle solutions. part i: Model,} J. Rheol.
  \textbf{68}, 873--894 (2024a).

\bibitem[Salipante \emph{et~al.}(2024b)Salipante, Cromer and
  Hudson]{Hudson2024b}
Salipante, P.~F., M.~Cromer and S.~D. Hudson, \enquote{Two-species model for
  nonlinear flow of wormlike micelle solutions. part ii: Experiment,} J. Rheol.
  \textbf{68}, 895--911 (2024b).

\bibitem[Santra \emph{et~al.}(2019)Santra, Kumari, Padinhateeri, Dünweg and
  Prakash]{Santra2019}
Santra, A., K.~Kumari, R.~Padinhateeri, B.~Dünweg and J.~R. Prakash,
  \enquote{Universality of the collapse transition of sticky polymers,} Soft
  Matter \textbf{15}, 7876--7887 (2019).

\bibitem[Sato and Larson(2022)Sato and Larson]{Sato2022}
Sato, T. and R.~G. Larson, \enquote{Nonlinear rheology of entangled wormlike
  micellar solutions predicted by a micelle-slip-spring model,} J. Rheol.
  \textbf{66}, 639--656 (2022).

\bibitem[Shibaev \emph{et~al.}(2015)Shibaev, Molchanov and
  Philippova]{Shibaev2015}
Shibaev, A.~V., V.~S. Molchanov and O.~E. Philippova, \enquote{Rheological
  behavior of oil-swollen wormlike surfactant micelles,} J. Phys. Chem. B
  \textbf{119}, 15938--15946 (2015).

\bibitem[Soddemann \emph{et~al.}(2001)Soddemann, Duenweg and Kremer]{SDK2001}
Soddemann, T., B.~Duenweg and K.~Kremer, \enquote{A generic computer model for
  amphiphilic systems,} Eur. Phys. J. E \textbf{6} (2001).

\bibitem[Sullivan \emph{et~al.}(2007)Sullivan, Nelson, Anderson and
  Hughes]{Sullivan2007}
Sullivan, P., E.~B. Nelson, V.~Anderson and T.~Hughes, \enquote{Oilfield
  applications of giant micelles,} in \emph{Giant Micelles}, pp. 453--472, CRC
  Press (2007).

\bibitem[Tamano \emph{et~al.}(2020)Tamano, Hamanaka, Nakano, Morinishi and
  Yamada]{Yamada2020}
Tamano, S., S.~Hamanaka, Y.~Nakano, Y.~Morinishi and T.~Yamada,
  \enquote{Rheological modeling of both shear-thickening and thinning behaviors
  through constitutive equations,} J. Non-Newton. Fluid Mech. \textbf{283},
  104339 (2020).

\bibitem[Turner and Cates(1992)Turner and Cates]{Turner1992}
Turner, M.~S. and M.~E. Cates, \enquote{Flow-induced phase transitions in
  rod-like micelles,} J. Phys.: Condens. Matter \textbf{4}, 3719 (1992).

\bibitem[Varchanis \emph{et~al.}(2022)Varchanis, Haward, Hopkins, Tsamopoulos
  and Shen]{Shen2022}
Varchanis, S., S.~J. Haward, C.~C. Hopkins, J.~Tsamopoulos and A.~Q. Shen,
  \enquote{Evaluation of constitutive models for shear-banding wormlike
  micellar solutions in simple and complex flows,} J. Non-Newton. Fluid Mech.
  \textbf{307}, 104855 (2022).

\bibitem[Vasquez \emph{et~al.}(2007)Vasquez, McKinley and {Pamela
  Cook}]{Vasquez2007}
Vasquez, P.~A., G.~H. McKinley and L.~{Pamela Cook}, \enquote{A network
  scission model for wormlike micellar solutions: I. {M}odel formulation and
  viscometric flow predictions,} J. Non-Newton. Fluid Mech. \textbf{144},
  122--139 (2007).

\bibitem[Wittmer \emph{et~al.}(1998)Wittmer, Milchev and Cates]{Wittmer1998}
Wittmer, J., A.~Milchev and M.~Cates, \enquote{Dynamical {M}onte {C}arlo study
  of equilibrium polymers: Static properties,} J. Chem. Phys. \textbf{109},
  834--845 (1998).

\bibitem[Wittmer \emph{et~al.}(2000)Wittmer, van~der Schoot, Milchev and
  Barrat]{Wittmer2000}
Wittmer, J., P.~van~der Schoot, A.~Milchev and J.~Barrat, \enquote{Dynamical
  {M}onte {C}arlo study of equilibrium polymers. ii. {T}he role of rings,} J.
  Chem. Phys. \textbf{113}, 6992--7005 (2000).

\bibitem[Yamakawa and Yoshizaki(2016)Yamakawa and Yoshizaki]{yamakawa2016}
Yamakawa, H. and T.~Yoshizaki, \emph{Helical wormlike chains in polymer
  solutions}, Springer (2016).

\bibitem[Yerushalmi \emph{et~al.}(1970)Yerushalmi, Katz and
  Shinnar]{Shinnar1970}
Yerushalmi, J., S.~Katz and R.~Shinnar, \enquote{The stability of steady shear
  flows of some viscoelastic fluids,} Chem. Eng. Sci. \textbf{25}, 1891--1902
  (1970).

\bibitem[Zakin \emph{et~al.}(2007)Zakin, Zhang and Ge]{Zakin2007}
Zakin, J.~L., Y.~Zhang and W.~Ge, \enquote{Drag reduction by surfactant giant
  micelles,} Giant Micelles pp. 473--492 (2007).

\bibitem[Zhou \emph{et~al.}(2014)Zhou, McKinley and Cook]{Zhou2014}
Zhou, L., G.~H. McKinley and L.~P. Cook, \enquote{Wormlike micellar solutions:
  Iii. vcm model predictions in steady and transient shearing flows,} J.
  Non-Newton. Fluid Mech. \textbf{211}, 70--83 (2014).

\bibitem[Zou \emph{et~al.}(2019)Zou, Tan, Jiang, Vogtt, Weaver, Koenig,
  Beaucage and Larson]{Zou2019}
Zou, W., G.~Tan, H.~Jiang, K.~Vogtt, M.~Weaver, P.~Koenig, G.~Beaucage and
  R.~G. Larson, \enquote{From well-entangled to partially-entangled wormlike
  micelles,} Soft {Ma}tter \textbf{15}, 642--655 (2019).

\end{thebibliography}
\bibliographystyle{JORnat}
\end{document}